\documentclass[journal]{IEEEtran}
\usepackage[numbers,compress]{natbib}
\usepackage[inline]{enumitem}
\usepackage{amsfonts,amssymb,mathrsfs,mathdots,mathtools,amsthm}
\usepackage{dsfont}
\usepackage{algorithmic}
\usepackage{graphicx}
\usepackage{textcomp}
\usepackage{xcolor}
\usepackage{dirtytalk}
\usepackage{color, soul}
\usepackage{balance}

\ifCLASSOPTIONcompsoc
\usepackage[caption=false,font=normalsize,labelfont=sf,textfont=sf]{subfig}
\else
\usepackage[caption=false,font=footnotesize]{subfig}
\fi
\usepackage{hyperref}

\DeclarePairedDelimiter\abs{\lvert}{\rvert}%
\DeclarePairedDelimiter\ceil{\lceil}{\rceil}
\DeclarePairedDelimiter\floor{\lfloor}{\rfloor}
\DeclareMathOperator*{\argmax}{arg\,max}
\DeclareMathOperator*{\argmin}{arg\,min}

\newtheorem{theorem}{Theorem}
\newtheorem{example}{Example}
\newtheorem{definition}{Definition}

\theoremstyle{remark}
\newtheorem{corollary}{Corollary}
\newtheorem{remark}{Remark}
\newtheorem{lemma}{Lemma}
\newtheorem{prop}{Proposition}

\newcounter{relctr} 
\everydisplay\expandafter{\the\everydisplay\setcounter{relctr}{0}} 

\AtBeginDocument{} 

\allowdisplaybreaks
\begin{document}

\title{Pointwise Maximal Leakage}

\author{Sara~Saeidian,~\IEEEmembership{Member,~IEEE,}
        Giulia~Cervia,~\IEEEmembership{Member,~IEEE,}
        Tobias~J.~Oechtering,~\IEEEmembership{Senior~Member,~IEEE} 
        Mikael~Skoglund,~\IEEEmembership{Fellow,~IEEE}
\thanks{This work has been supported by the Strategic Research Agenda Program, Information and Communication Technology – The Next Generation (SRA ICT – TNG) funded by the Swedish Government and Digital Futures center within the project DataLEASH and in part by the Swedish
Research Council (VR) under grant 2019-03606. This article was presented in part at the 2022 IEEE International Symposium on Information Theory.}
\thanks{Sara~Saeidian, Tobias~J.~Oechtering, and Mikael~Skoglund are with the Division of Information Science and Engineering, School of Electrical Engineering and Computer Science, KTH Royal Institute of Technology, 100~44 Stockholm, Sweden (e-mail: saeidian@kth.se; oech@kth.se; skoglund@kth.se).}
\thanks{Giulia Cervia is with IMT Nord Europe, Institut Mines-Télécom, Univ. Lille, Centre for Digital Systems, F-59000 Lille, France (e-mail: giulia.cervia@imt-nord-europe.fr)}
}

\maketitle

\begin{abstract}
We introduce a privacy measure called \emph{pointwise maximal leakage}, generalizing the pre-existing notion of \emph{maximal leakage}, which quantifies the amount of information leaking about a secret $X$ by disclosing a single outcome of a (randomized) function calculated on $X$. Pointwise maximal leakage is a robust and operationally meaningful privacy measure that captures the largest amount of information leaking about $X$ to adversaries seeking to guess arbitrary (possibly randomized) functions of $X$, or equivalently, aiming to maximize arbitrary gain functions. We study several properties of pointwise maximal leakage, e.g., how it composes over multiple outcomes, how it is affected by pre- and post-processing, etc. Furthermore, we propose to view information leakage as a \emph{random variable} which, in turn, allows us to regard privacy \emph{guarantees} as requirements imposed on different statistical properties of the information leakage random variable. We define several privacy guarantees and study how they behave under pre-processing, post-processing and composition. Finally, we examine the relationship between pointwise maximal leakage and other privacy notions such as local differential privacy, local information privacy, $f$-information, and so on. Overall, our paper constructs a robust and flexible framework for privacy risk assessment whose central notion has a strong operational meaning which can be adapted to a variety of applications and practical scenarios.
\end{abstract}

\section{Introduction}

Suppose $X$ is a random variable representing some data containing sensitive information. As we aim to remain general, we intentionally keep $X$ abstract. For example, $X$ may be a single data entry collected from an individual (the \emph{local} setting), $X$ may represent an entire database containing sensitive information (the \emph{centralized} setting), or $X$ may be a secret such as a password that must be kept confidential (the \emph{side-channel} setting). 

Further, suppose $Y$ is the output of a (randomized) function with input $X$. For example, in the local setting, $Y$ denotes the perturbed version of a single user's data which is collected by a data curator. In the centralized setting, $Y$ may be some aggregate statistic calculated from a database, and in the side-channel setting, $Y$ is the output of a side-channel with input $X$, for instance, the inter-keystroke delays when typing in a password. In all of these scenarios, we are interested in answering the following question: How much information is $Y$ leaking about $X$? 

The above question has been studied and answered in different contexts using a myriad of different privacy measures. For instance, differential privacy (DP)~\cite{dwork2014algorithmic, dwork2006differential} was introduced in a centralized setting within the context of database privacy in order to ensure that no single individual's participation can be revealed from the output of a function calculated on a database (i.e., ensuring \emph{membership privacy}). Later on, borrowing ideas from survey privacy~\cite{warner1965randomized}, the adaptation of differential privacy to the local setting led to the concept of local differential privacy (LDP)~\cite{evfimievski2003limiting, kasiviswanathan2011can, duchi2013local}, where, roughly speaking, the goal is to provide plausible deniability for all possible values of $X$. Parallel to these developments, in the information theory literature, a wide range of privacy measures have been proposed and studied that aim to measure the dependence between two random variables $X$ and $Y$. These include, for example, mutual information~\cite{asoodeh2016information, asoodeh2015maximal, wang2016relation, makhdoumi2014information}, and its generalizations~\cite{verdu2015alpha, liao2019tunable}, divergence-based measures (e.g., metrics based on $f$-divergence~\cite{diaz2019robustness, rassouli2019optimal}), probability of correctly guessing~\cite{asoodeh2018estimation}, information privacy~\cite{du2012privacy,jiang2020local,jiang2021context} and indistinguishability~\cite{wang2016relation}. The two recent surveys by~\citet{wagner2018technical} and~\citet{bloch2021overview} contain an extensive list of various privacy measures. 

In much of the literature, the prevalent approach to tackling privacy problems has been to start with a particular definition of privacy, study the properties that follow from the definition, and design/optimize mechanisms that guarantee a certain level of privacy and utility. An alternative approach is to start from a \emph{threat model} in which an adversary with explicitly-described capabilities is pursuing a specific objective and study the system's vulnerability as a result of this adversarial model. This approach has several advantages. First, it encourages us to make our assumptions about the capabilities of the adversary (e.g., in terms of computational power or prior knowledge of the system) and her objectives explicit. Second, the privacy definition obtained by studying a threat model is operationally meaningful and easier to interpret. Third, the discussions around the advantages and limitations of different privacy measures become more transparent and objective. Note that certain privacy measures such as differential privacy and differential indistinguishability~\cite{lee2012differential} have been interpreted using powerful adversarial scenarios. For example, \citet{wasserman2010statistical} and~\citet{kairouz2015composition} have shown that differential privacy imposes a tradeoff between Type I and Type II error probabilities of a hypothesis testing adversary who wishes to determine whether or not a data point is included in a database. However, here, we should make a subtle but essential distinction: While these measures have been \emph{interpreted} using adversarial models, their definitions do not follow from any such model. In fact, not making this distinction may lead to misconceptions/disagreements about what a privacy definition does or does not promise. An example of this is the long ongoing debate about whether or not differential privacy (implicitly) requires assumptions about the data generating (prior) distribution, such as the assumption that the entries in a database are drawn independently~\cite{kifer2011no, tschantz2020differential}. This disagreement may have been avoided had the assumptions about the adversarial model been made explicit in the definition. 

The threat-model approach to privacy has been adopted in a line of work termed \emph{quantitative information flow}~\cite{smith2009foundations, braun2009quantitative, espinoza2013min, alvim2012measuring, alvim2014additive, alvim2020science}, in which several notions of information leakage are motivated, defined and studied. One such notion is \emph{min-entropy leakage}~\cite{smith2009foundations, braun2009quantitative} which is defined in a setup where a passive but computationally-unbounded adversary is trying to guess the value of the secret $X$ in one try. Min-entropy leakage, then, quantifies the increase in the probability of correctly guessing $X$ having observed the output $Y$, compared to guessing the secret with no observations. Naturally, min-entropy leakage depends both on the prior distribution of $X$, denoted by $P_X$, and the channel from $X$ to $Y$, denoted by $P_{Y \mid X}$. Therefore, to obtain a privacy measure that depends only on the channel $P_{Y \mid X}$,~\citet{braun2009quantitative} maximize the min-entropy leakage over all possible priors, which leads to a quantity (later) called \emph{maximal leakage}, and expressed as 
\begin{equation}
\label{eq:intro_maximal_leakage}
    \mathcal L(P_{Y \mid X}) = \log \sum_{y} \max_{x : P_X(x) > 0} P_{Y \mid X=x}(y).
\end{equation}
Interestingly, the worst-case prior in this problem is the uniform distribution~\cite{braun2009quantitative}. 

Subsequent works in this area have extended the above adversarial model~\cite{espinoza2013min, alvim2012measuring, alvim2014additive, alvim2020science}, developed maximal leakage into a practical tool for quantifying privacy in learning applications~\cite{saeidian2021quantifying}, and considered maximal leakage as the privacy constraint in privacy-utility tradeoff problems~\cite{liao2017hypothesis, wu2020optimal, saeidian2021optimal}. We find two such works particularly interesting: the $g$-leakage framework introduced by~\citet{alvim2012measuring} and the maximal leakage definition of~\citet{issa2019operational}. The $g$-leakage~\cite{alvim2012measuring} formulation generalizes the setup by considering an adversary aiming to construct a guess of $X$ that maximizes a certain \emph{gain} function and constitutes a useful tool for modeling a variety of adversarial goals, such as guessing the secret $X$ in $k \geq 1$ attempts or approximately guessing the secret~\cite{alvim2012measuring}. Moreover, it has been shown that for all prior distributions, maximizing $g$-leakage over all possible gain functions yields a quantity that is equal to maximal leakage~\cite{alvim2014additive}. The setup put forward by~\citet{issa2019operational}, on the other hand, considers adversaries who are interested in guessing a (possibly randomized) discrete function of $X$, called $U$. Then, taking the supremum over all such $U$'s, the resulting quantity is once again equal to maximal leakage.  

The above results lead us to believe that maximal leakage is a powerful privacy measure with a robust definition that is relevant in a multitude of different scenarios. However, one apparent limitation of the definition given in~\eqref{eq:intro_maximal_leakage} is that maximal leakage is defined for the \emph{average outcome} $Y$. Hence, a privacy guarantee given in terms of maximal leakage does not allow us to distinguish between individual outcomes based on how much information they leak. To see why this can be problematic, suppose $X$ is a uniformly distributed ternary random variable, and consider the following channels from $X$ to $Y$: 
\begin{equation*}
    P_{Y \mid X} = \begin{bmatrix}
    1 & 0 & 0\\[0.5em]
    \frac{1}{2} & \frac{1}{2} & 0\\[0.5em]
    0 & \frac{1}{2} & \frac{1}{2}
    \end{bmatrix}, \qquad
    Q_{Y \mid X} = \begin{bmatrix}
    \frac{2}{3} & \frac{1}{6} & \frac{1}{6}\\[0.5em]
    \frac{1}{6} & \frac{2}{3} & \frac{1}{6}\\[0.5em]
    \frac{1}{6} & \frac{1}{6} & \frac{2}{3}
    \end{bmatrix}.
\end{equation*}
It is easy to see that both channels have equal maximal leakage $\mathcal{L}(P_{Y \mid X}) = \mathcal{L}(Q_{Y \mid X}) = \log 2$; however, they are qualitatively different. For example, the third outcome in $P_{Y \mid X}$, can be considered to be far less private compared to the other two as it completely reveals the value of the secret. On the other hand, due to the symmetry in $Q_{Y \mid X}$, we may expect that all outcomes leak the same amount of information. One justification for this issue is that maximal leakage remains small as long as highly revealing outputs occur with small probability~\cite{issa2019operational}. Nevertheless, the average-case guarantee provided by maximal leakage may be deemed insufficient in  privacy-critical applications with strict requirements (for example, in the healthcare sector), which motivates the search for alternative privacy measures that capture the \emph{distribution} of the leakage over the outcomes. Using the leakage distribution not only do we get a precise description of privacy in a given system, but we will also have the flexibility to adapt our definition of a private system to each specific application. For example, system designers will be able to decide whether or not highly-revealing but low-probability outcomes pose a privacy risk on a per-application basis.
 
It is also worth mentioning that a distributional view of privacy has already been adopted in the differential privacy literature, where the log-likelihood ratio is referred to as the \emph{privacy loss random variable}~\cite{mironov2017renyi, dwork2016concentrated, bun2016concentrated,sommer2018privacy,zhu2022optimal}. The importance of the privacy loss random variable is that it allows a sharper characterization of the privacy budget, which has led to highly practical tools such as the \emph{moments accountant}~\cite{abadi2016deep}. The framework we develop in this paper differs from these works mainly in that here we let $X$ represent any type of data containing sensitive information, while the works on differential privacy target specifically the centralized setting, where the notion of neighboring databases (that is, databases that differ in a single entry) plays a central role in all definitions. On top of that, the privacy measure introduced in this paper has a clear and useful operational meaning and is obtained by analyzing specific threat models (more on this below). 

\subsection{Overview and Contributions}
\label{ssec:contributions}
\subsubsection{Introducing pointwise maximal leakage} In this paper, our main goal is to define a robust and operationally meaningful privacy measure that describes the amount of information leaking about the secret $X$ due to disclosing a single outcome $Y=y$. We consider the same adversarial models that were used to obtain maximal leakage as a privacy measure (e.g.,~\cite{issa2019operational, alvim2012measuring}), but redirect our attention from the \say{average outcome} characterization of the previous works to individual outcomes, and obtain a new privacy definition which we will call \emph{pointwise maximal leakage} (PML), denoted by $\ell_{P_{XY}}(X \to y)$, and expressed as 
\begin{equation}
\label{eq:intro_pml}
    \ell_{P_{XY}}(X \to y) = \log \max_{x: P_X(x) > 0} \frac{P_{X \mid Y=y}(x)}{P_X(x)}.
\end{equation}
In Section~\ref{ssec:def_randomized}, we start from the threat model of~\cite{issa2019operational}, in which an adversary attempts to guess the outcome of a randomized function of $X$, denoted by $U$. More concretely, we define a quantity $\ell_U(X\to y)$ as the logarithm of the ratio of the probability of correctly guessing $U$ having observed an outcome $y$, and the probability of correctly guessing $U$ with no observations. Then, we define pointwise maximal leakage as the supremum of $\ell_U(X\to y)$ over all $U$, that is 
\begin{equation*}
    \ell_{P_{XY}}(X\to y) \coloneqq \sup_{P_{U \mid X}} \ell_U(X\to y),
\end{equation*}
and obtain expression~\eqref{eq:intro_pml}. We will call this approach the \emph{randomized function view} of leakage. Afterwards, in Section~\ref{ssec:def_gain}, we consider the threat model of the $g$-leakage framework~\cite{alvim2012measuring}, in which an adversary is trying to maximize a certain gain function $g$. More precisely, we define a quantity $\ell_g(X \to y)$ as the logarithm of the ratio of the expected gain having observed an outcome $y$, and the expected gain with no observations. We will call this approach the \emph{gain function view} of leakage, and show that it is in fact equivalent to the randomized function view of leakage. Specifically, we show that for all joint distributions over $X$ and $Y$ and for each randomized function of $X$, denoted by $U$, there exists a gain function $g_{_U}$ such that $\ell_U(X \to y)  = \ell_{g_{_U}} (X \to y)$. Conversely, we show that for each gain function $g$, there exists a randomized function of $X$, denoted by $U_g$, such that $\ell_g(X \to y) = \ell_{U_g} (X \to y)$. It follows that PML can alternatively be defined as
\begin{equation*}
    \ell_{P_{XY}}(X \to y) \coloneqq \sup_{g} \ell_g(X \to y).
\end{equation*}
This result not only unifies two seemingly different ways of defining PML, but also signifies the \emph{robustness} of PML as a privacy measure against a large class of adversaries with different objectives. Once we have established the definition of PML, in Section~\ref{ssec:properties}, we study several of its properties, e.g., how it composes when several outcomes are observed, how the leakage is affected by pre- and post-processing, and so on.  

\subsubsection{Defining privacy guarantees based on PML}
Our second objective in this work is to argue in favor of viewing information leakage as a \emph{random variable}. The idea behind this is simple: The amount of information leaked due to disclosing an outcome $Y=y$ is equal to $\ell_{P_{XY}}(X \to y)$ which is a function of $y$. Since $Y$ is a random variable distributed according to $P_Y$ (i.e., the output distribution induced by $P_X$ and $P_{Y \mid X}$), this in turn allows us to define a random variable $\ell_{P_{XY}}(X \to Y)$ whose distribution is induced by $P_Y$. Adopting this view, a \emph{privacy guarantee} is essentially a requirement we impose on some statistical property of $\ell_{P_{XY}}(X \to Y)$; thus, we have the flexibility to define different types of guarantees depending on how strict we may want to be. For example, we may require small information leakage \emph{with probability one}. Less stringently, we can define guarantees that bound either the \emph{tail} of the information leakage random variable, or its \emph{expectation}.\footnote{By considering the expectation of information leakage, we retrieve the original definition of maximal leakage.} These privacy guarantees, which are the subject of Section~\ref{sec:guarantees}, can be informally expressed as follows: Given an arbitrary but fixed prior $P_X$, we say that  
\begin{itemize}
    \item $P_{Y \mid X}$ satisfies $\epsilon$-PML with $\epsilon \geq 0$ if $\ell_{P_{XY}}(X \to Y)$ is bounded by $\epsilon$ with probability one;
    \item $P_{Y \mid X}$ satisfies $(\epsilon, \delta)$-PML with $\epsilon \geq 0$ and $0 \leq \delta \leq 1$ if $\ell_{P_{XY}}(X \to Y)$ is bounded by $\epsilon$ with probability at least $1 - \delta$; and 
    \item $P_{Y \mid X}$ satisfies $\mathcal L(P_{Y \mid X}) \leq \epsilon$ with $\epsilon \geq 0$, if the expectation of $\exp \Big( \ell_{P_{XY}}(X \to Y) \Big)$ is bounded by $\exp(\epsilon)$, where $\mathcal L(\cdot)$ denotes maximal leakage.
\end{itemize}

In the rest of Section~\ref{sec:guarantees}, we study the \emph{data-processing} and \emph{composition} properties of the privacy guarantees introduced in this paper. Specifically, in Section~\ref{ssec:data_processing}, we study how $\epsilon$-PML and $(\epsilon, \delta)$-PML are affected by pre- and post-processing. Interestingly, it turns out that $(\epsilon, \delta)$-PML is not closed under post-processing, that is, given a privacy mechanism $P_{Y \mid X}$ that satisfies $(\epsilon, \delta)$-PML, we may be able to come up with a post-processing mechanism $P_{Z \mid Y}$ such that the overall mechanism $P_{Z \mid X}$ does not satisfy $(\epsilon, \delta)$-PML. In response to this observation, we introduce another privacy guarantee, called $(\epsilon, \delta)$-\emph{event maximal leakage} (EML) with $\epsilon \geq 0$ and $0 \leq \delta \leq 1$, which resembles $(\epsilon, \delta)$-PML but is closed under post-processing. Informally, a mechanism $P_{Y \mid X}$ satisfies $(\epsilon, \delta)$-EML if the pointwise leakage for all post-processed outcomes of $P_{Y \mid X}$ with probability at least $\delta$ is bounded by $\epsilon$. 

Next, in Section~\ref{ssec:composition}, we study how different privacy guarantees change as a result of composing privacy mechanisms. More concretely, we are interested to find out what types of guarantees we can get for a mechanism $P_{YZ \mid X}$ which is obtained by \emph{adaptively} composing two mechanisms $P_{Y \mid X}$ and $P_{Z \mid XY}$. Naturally, one can formulate different problems by making various assumptions about the involved mechanisms $P_{Y \mid X}$ and $P_{Z \mid XY}$. We present several such problem formulations and their corresponding results.

\subsubsection{Comparing PML with other privacy notions}
In Section~\ref{sec:comparisons}, we study how pointwise maximal leakage relates to several other privacy/statistical notions, namely, max-information~\cite{dwork2015generalization,rogers2016max}, local differential privacy~\cite{du2012privacy}, local information privacy~\cite{jiang2021context}, local differential identifiability~\cite{wang2016relation}, mutual information, $f$-information~\cite{diaz2019robustness} and total-variation privacy~\cite{rassouli2019optimal}. We derive bounds between the different notions and discuss their implications.

As a final note, in Section~\ref{ssec:dynamic_model} we discuss a privacy framework called the \emph{dynamic consumption of secrecy}~\cite{espinoza2013min} which, in the same spirit as our work, attempts to quantify the information leakage due to disclosing a single outcome of the random variable $Y$. Somewhat surprisingly,~\cite{espinoza2013min} argues that the privacy definition resulting from this dynamic view suffers from limitations that convince the authors against pursuing this line of research. In Section~\ref{ssec:dynamic_model}, we discuss what these limitations are, and explain why they do \emph{not} apply to pointwise maximal leakage.

\subsection{Applications}
The versatility, robustness, and explainability of PML render it a suitable privacy notion to be imposed in a variety of practical applications, both in centralized settings (e.g., extracting aggregate information from healthcare databases~\cite{hassan2019differential}) and in local settings (e.g., collecting data from browsers~\cite{erlingsson2014rappor} and wearable health monitors~\cite{saifuzzaman2022systematic}, or location-based services~\cite{andres2013geo}). This is true especially in light of a recent trend that has given new significance to the so-called \emph{context-aware} privacy measures~\cite{jiang2021context, jiang2020local, li2013membership, yang2015bayesian,kifer2014pufferfish,he2014blowfish}. Context-aware privacy measures embed adversarial assumptions into the problem formulation, usually in the form of a prior distribution, which is in contrast with traditional differential privacy-based definitions that depend only on the privacy mechanism. Adding context to a problem usually serves one of two purposes: to ensure that conclusions drawn from a privacy analysis apply to certain specific adversaries, or to relax the setting of differential privacy (especially the local version) to enable higher utility. The former goal is a response to the idea that differential privacy may not provide sufficient protection when the dataset in question contains highly correlated values~\cite{kifer2011no} and has led to frameworks such as Pufferfish privacy~\cite{kifer2014pufferfish} and Bayesian differential privacy~\cite{yang2015bayesian}. The latter goal, on the other hand, aims to address concerns that achieving local differential privacy with suitably small parameters significantly reduces utility~\cite{tramer2015differential}, which is why in practical implementations sometimes the resulting privacy parameters are very large~\cite{fredrikson2014privacy}. For example,~\citet{tang2017privacy} report that while Apple's deployment of local differential privacy uses a small per-datum privacy budget, the overall daily privacy budget can be significantly larger, as high as 16 per day. As a result, context-aware privacy measures such as local information privacy~\cite{jiang2021context,jiang2020local} are proposed to achieve better privacy-utility tradeoffs. 

PML is a context-aware privacy measure that can be deployed to simultaneously achieve both of the aforementioned goals. Specifically, the gain function view of leakage can be used to explicitly model a large variety of adversarial goals, e.g., membership privacy attacks (Example~\ref{ex:membership}). Moreover, PML can lead to better privacy-utility tradeoffs compared to context-free notions such as local differential privacy (Example~\ref{ex:randomized_response}). Note that while assessing PML requires assumptions about the prior distribution, in many applications, priors are either publicly known or can be estimated. For instance, population density maps can supply priors in geopositioning applications.

\subsection{Notation}
In this work, we restrict our attention to finite random variables, therefore, all sets are assumed to be finite. We use uppercase letters to refer to random variables, e.g., $X$. Sets are represented by uppercase calligraphic letters, for example, the alphabet of $X$ will be denoted by $\mathcal {X}$. Let $\mathcal E \subseteq \mathcal X$. We will use both $P_X(\mathcal E)$ and $\mathbb P_{X \sim P_X}[\mathcal E]$ to describe the probability of an event $\mathcal E$ according to distribution $P_X$. Similarly, we will use $\mathbb E_{X \sim P_X}[\cdot]$ to represent expectation with respect to $P_X$. The notation $\mathrm{supp}(P_X) \coloneqq \{x \in \mathcal{X}: P_X(x) > 0\}$ will be used to refer to the support set of distribution $P_X$. Given probability distributions $P_X$ and $Q_X$ on $\mathcal X$, we write $P_X \ll Q_X$ to imply that $P_X$ is absolutely continuous with respect to $Q_X$. 

Let $n$ be a positive integer. We use $[n] \coloneqq \{1, \ldots, n\}$ to denote the set of all positive integers smaller than or equal to $n$. Suppose $X$ is a random variable whose alphabet has cardinality $\abs{\mathcal X} = n$, and $Y$ is a random variable induced by a channel $P_{Y \mid X}$, whose alphabet has cardinality $\abs{\mathcal Y} =m$. Then, the channel $P_{Y \mid X} \in [0,1]^{n \times m}$ is a row-stochastic matrix with elements $(P_{Y \mid X})_{ij} = P_{Y \mid X = x_i}(y_j)$ for $i \in [n]$ and $j \in [m]$. We say that a channel $P_{Y \mid X}$ is \emph{deterministic} if it consists only of zeros and ones. Similarly, we say that an outcome $y_j$ with $j \in [m]$ is deterministic if its corresponding column in the matrix $P_{Y \mid X}$ consists only of zeros and ones. Suppose the Markov chain $X-Y-Z$ holds. We write $P_{Z \mid X} = P_{Z \mid Y} \circ P_{Y \mid X}$ to denote marginalization over $Y$, that is, $P_{Z \mid X=x}(z) = \sum_{y \in \mathcal Y} P_{Z \mid Y=y}(z) P_{Y \mid X=x}(y)$ for $x \in \mathcal X$ and $z \in \mathcal Z$.  Finally, we use $\log(\cdot)$ to denote the natural logarithm and $\mathds{1}[\cdot]$ to denote the indicator function.

\section{Definition, interpretations and properties}
\subsection{Randomized Function View of Leakage}
\label{ssec:def_randomized}
We begin by describing our first threat model, which is a pointwise adaptation of the model described in~\cite{issa2019operational}. Suppose $X$ is a random variable defined over a finite alphabet $\mathcal X$. We use $X$ to represent some data containing sensitive information. Further, suppose $Y$ is a random variable taking values in a finite alphabet $\mathcal{Y}$ which is the output of a channel (i.e., kernel) $P_{Y \mid X}$ with input $X$. We will also refer to the channel $P_{Y \mid X}$ as a \emph{privacy mechanism}. Consider an adversary who is interested in guessing the value of a possibly randomized function of $X$, called $U$, characterized by $P_{U \mid X}$. The adversary, who is computationally unbounded, observes an outcome $y \in \mathrm{supp}(P_Y)$ (where $P_Y = P_{Y \mid X} \circ P_X$ is the output distribution) and constructs a guess of $U$ called $\hat{U}$ according to a kernel $P_{\hat{U} \mid Y}$. The adversary is passive in the sense that she cannot affect the outcomes of the system, but can verify if her guess is correct. Furthermore, the adversary knows the joint distribution $P_{UXY}$, and therefore, can optimize her choice of guessing kernel $P_{\hat{U} \mid Y}$ to maximize her chances of correctly guessing $U$. 

To measure the information leakage of a disclosed outcome $y$, the system designer considers the ratio of the probability of correctly guessing $U$ having observed $y$, and the probability of correctly guessing $U$ with no observations (in this case, the best guess is the most probable outcome according to $P_U$). Accordingly, we define the pointwise \emph{$U$-leakage} of $X$ as follows:
\begin{equation}
\label{eq:u-leakage}
    \ell_U(X\to y) \coloneqq \log \frac{\sup_{P_{\hat U \mid Y}} \mathbb P \left[U=\hat U \mid Y=y \right]}{\max_{u 
    \in \mathcal{U}} P_U(u)},
\end{equation}
where $\mathcal{U}$ denotes the alphabet of $U$. As the system designer may not know what $U$ the adversary is interested in, or different adversaries may be interested in guessing different $U$'s, we investigate the worst-case scenario by taking the supremum of~\eqref{eq:u-leakage} over all possible randomized functions of $X$. Considering this setup, we define pointwise maximal leakage (PML) denoted by $\ell_{P_{XY}}(X \to y)$ as follows.  

\begin{definition}[Pointwise maximal leakage]
Let $P_{XY}$ denote the joint distribution of $X$ and $Y$. The pointwise maximal leakage from $X$ to $y \in \mathrm{supp}(P_Y)$, $\ell_{P_{XY}}(X \to y)$, is defined as 
\begin{align}
\label{eq:privacy_rv_def}
    \ell_{P_{XY}}(X\to y) &\coloneqq \sup_{P_{U \mid X}} \ell_U(X\to y) \\
    &= \log \sup_{P_{U \mid X}} \frac{\sup_{P_{\hat U \mid Y}} \mathbb P \left[U=\hat U \mid Y=y \right]}{\max_{u 
    \in \mathcal{U}} P_U(u)}.\nonumber
\end{align}
\end{definition}
\begin{figure}
    \centering
    \includegraphics[scale=0.105]{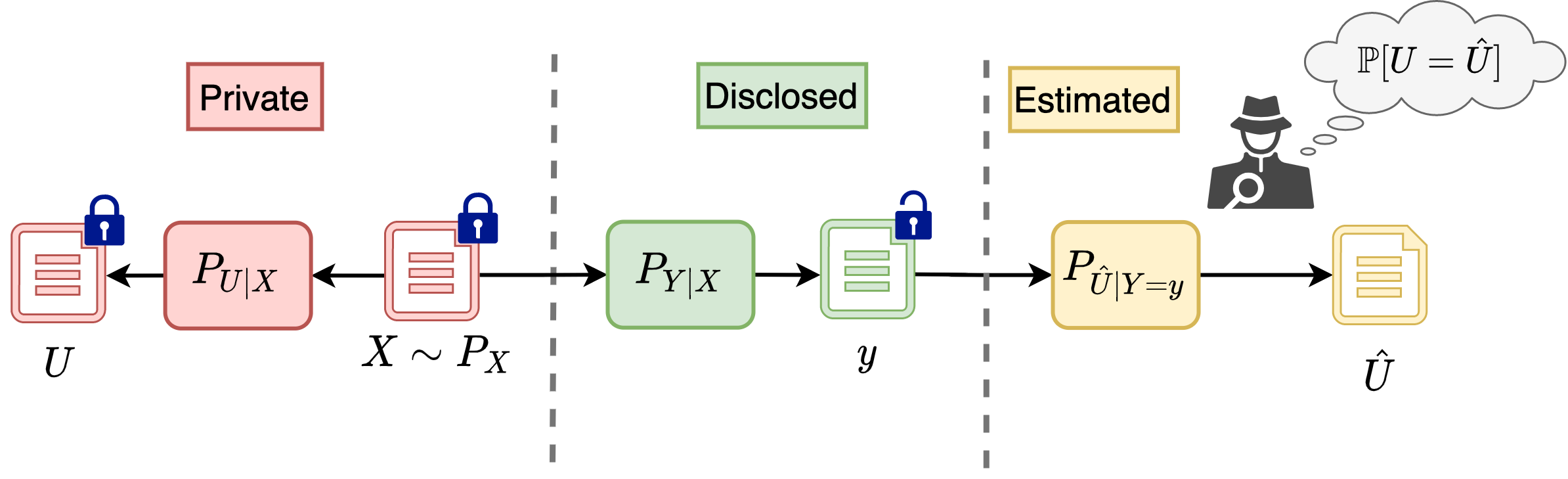}
    \caption{System model for the randomized function view of leakage: An adversary observes an outcome $y$ of the channel $P_{Y \mid X}$, and tries to guess the value of a randomized function of $X$, denoted by $U$.}
    \label{fig:randomized_func}
\end{figure}
Below, we show that $\ell_{P_{XY}}(X \to y)$ can be simplified to~\eqref{eq:intro_pml}. 

\begin{theorem}
\label{thm:privacy_leakage_rv}
Given a joint distribution $P_{XY}$ on $\mathcal X \times \mathcal Y$, the pointwise maximal leakage from $X$ to $y \in \mathrm{supp}(P_Y)$ can be expressed as 
\begin{equation}
\label{eq:privacy_rv}
    \ell_{P_{XY}}(X \to y) = \log \max_{x \in \mathrm{supp}\left(P_X \right)} \frac{P_{X \mid Y=y}(x)}{P_X(x)}.
\end{equation}
\end{theorem}
\begin{IEEEproof}
Fix an arbitrary random variable $U$ satisfying the Markov chain $U-X-Y$. The numerator of~\eqref{eq:u-leakage} can be written as 
\begin{align*}
    &\sup_{P_{\hat U \mid Y}} \mathbb P \left[U=\hat U \mid Y=y \right] = \sup_{P_{\hat U \mid Y}} \sum_{u, \hat u} \mathds{1}[u=\hat u] P_{U\hat U \mid Y=y}(u,\hat u)\\
    &= \sup_{P_{\hat U \mid Y}} \; \sum_{u, \hat u} \mathds{1}[u=\hat u] \; P_{U \mid Y=y}(u) \; P_{\hat U \mid Y=y}(\hat u)\\
    &= \sup_{P_{\hat U \mid Y}} \; \sum_{u} \;  P_{U \mid Y=y}(u) \;P_{\hat U \mid Y=y}(u)\\
    &= \max_{u \in \mathcal U} P_{U \mid Y=y}(u),
\end{align*}
where the last equality follows from the fact that the optimal estimator $P^*_{\hat U \mid Y}$ in the above problem satisfies
\begin{equation*}
    P^*_{\hat U \mid Y=y}(u) = \begin{cases}
    1 & \mathrm{for\;some} \; u \in \argmax\limits_{u \in \mathcal{U}} P_{U \mid Y=y}(u), \\
    0 & \mathrm{otherwise}.
    \end{cases}
\end{equation*}
Thus, we can write
\begin{subequations}
\begin{align}
    &\exp \big(\ell_U(X \to y) \big) = \frac{\sup_{P_{\hat U \mid Y}} \mathbb P \left[U=\hat U \mid Y=y \right]}{\max_{u' \in \mathcal{U}} P_U(u')}\nonumber\\
    &= \frac{\max_{u \in \mathcal U} P_{U \mid Y=y}(u)}{\max_{u' \in \mathcal{U}} P_U(u')} \nonumber\\
    &= \frac{\max_{u \in \mathcal U} \sum_{x \in \mathrm{supp}(P_X)} P_{UX \mid Y=y}(u,x)}{\max_{u' \in \mathcal{U}} P_U(u')} \nonumber\\
    &= \frac{\max_{u \in \mathcal U} \sum_{x \in \mathrm{supp}(P_X)} P_{X \mid Y=y}(x) \; P_{U \mid X=x}(u)}{\max_{u' \in \mathcal{U}} P_U(u')}\nonumber\\
    &= \max_{u \in \mathcal U}  \sum_{x \in \mathrm{supp}(P_X)}  \frac{P_{X \mid Y=y}(x)}{P_X(x)} P_{X \mid U=u}(x)\;  \frac{P_U(u)}{\max_{u' \in \mathcal{U}} P_U(u')}\nonumber\\
    &\leq \max_{u \in \mathcal U} \sum_{x \in \mathrm{supp}(P_X)} \frac{P_{X \mid Y=y}(x)}{P_X(x)} \; P_{X \mid U=u}(x)\label{subeq:uniform_u}\\
    &\leq \max_{x \in \mathrm{supp}(P_X)} \frac{P_{X \mid Y=y}(x)}{P_X(x)}.\label{subeq:map_u}
\end{align}
\end{subequations}

Taking the supremum over all $U$'s satisfying $U - X - Y$ we obtain
\begin{equation}
\label{eq:ub_privacy_rv}
    \ell_{P_{XY}}(X\to y) \leq  \log \max_{x \in \mathrm{supp}(P_X)} \frac{P_{X \mid Y=y}(x)}{P_X(x)}.
\end{equation}
To prove the reverse inequality, we construct a $U$ achieving the bound in~\eqref{eq:ub_privacy_rv}. Note that inequality~\eqref{subeq:map_u} holds with equality if there exists $u^* \in \mathcal U$ such that 
\begin{align}
\begin{split}
\label{eq:opt_rv_cond1}
    &P_{X \mid U=u^*}(x) \\
    &= \begin{cases}
    1 & \mathrm{for\;some} \; x \in \argmax\limits_{x \in \mathrm{supp}(P_X)} \frac{P_{X \mid Y=y}(x)}{P_X(x)},\\
    0 & \mathrm{otherwise}.
    \end{cases}
\end{split}
\end{align}
Furthermore, $u^*$ will also satisfy~\eqref{subeq:uniform_u} with equality if
\begin{equation}
\label{eq:opt_rv_cond2}
    P_U(u^*) = \max_{u \in \mathcal{U}} P_U(u). 
\end{equation}
An example of $U$ satisfying both of the above conditions can be obtained through the \say{shattering} channel defined in the proof of~\cite[Thm. 1]{issa2019operational}. Roughly speaking, the shattering channel breaks down each $x \in \mathrm{supp} (P_X)$ with probability $P_X(x)$ into $k(x)$ corresponding elements with probability $\min_{x \in \mathrm{supp}(P_X)} P(x)$, thus creating a random variable $U_\mathrm{S}$ with an (almost) uniform distribution. We recall the definition of the shattering channel $P_{U_\mathrm{S} \mid X}$ for completeness.
\begin{definition}[Shattering channel~\cite{issa2019operational}]
\label{def:shattering}
Let $p^* \coloneqq \min\limits_{x \in \mathrm{supp}(P_X)} P_X(x)$. Given $x \in \mathrm{supp}(P_X)$, let $k(x) \coloneqq \frac{P_X(x)}{p^*}$, and let $\mathcal{U}_\mathrm{S} = \bigcup_{x \in \mathrm{supp}(P_X)} \{(x,1), \ldots, (x, \ceil{k(x)})\}$, where $\ceil{k(x)}$ denotes the smallest integer greater than or equal to $k(x)$. The shattering channel $P_{U_\mathrm{S} \mid X}$ is defined as 
\begin{align*}
    P_{U_\mathrm{S} \mid X=x} &(i_u, j_u)\\
    &= \begin{cases}
    \frac{p^*}{P_X(x)} & \mathrm{if} \; i_u = x, \;j_u = 1, \ldots, \floor{k(x)},\\
    1 - \frac{\floor{k(x)} p^*}{P_X(x)} & \mathrm{if} \; i_u = x, \;j_u = \ceil{k(x)},\\
    0 & \mathrm{otherwise}, 
    \end{cases}
\end{align*}
with $u=(i_u, j_u) \in \mathcal{U}_\mathrm{S}$ and $x \in \mathrm{supp}(P_X)$, where $\floor{k(x)}$ denotes the largest integer smaller than or equal to $k(x)$. 
\end{definition}
The shattering channel induces the joint distribution $P_{U_\mathrm{S} X}$: 
\begin{align*}
    &P_{U_\mathrm{S} X}((i_u, j_u), x) \\
    &= \begin{cases}
    p^* & \mathrm{if} \; i_u = x, \;j_u = 1, \ldots, \floor{k(x)},\\
    P_X(x) - \floor{k(x)} p^* & \mathrm{if} \; i_u = x, \;j_u = \ceil{k(x)},\\
    0 & \mathrm{otherwise}, 
    \end{cases}
\end{align*}
and $P_{U_\mathrm{S}}$ is obtained as 
\begin{align*}
    &P_{U_\mathrm{S}}(i_u, j_u)\\
    &= \begin{cases}
    p^* & \mathrm{if} \; i_u = x, \;j_u = 1, \ldots, \floor{k(x)},\\
    P_X(x) - \floor{k(x)} p^* & \mathrm{if} \; i_u = x, \;j_u = \ceil{k(x)},
    \end{cases}
\end{align*}
for $(i_u, j_u) \in \mathcal U_\mathrm{S}$. Clearly, each $u$ is mapped to exactly one $x$, so condition~\eqref{eq:opt_rv_cond1} holds. Furthermore, each $x$ corresponds to at least one $u$ with probability $P_{U_\mathrm{S}}(u)=\max_{u'} P_{U_\mathrm{S}}(u') = p^*$, so condition~\eqref{eq:opt_rv_cond2} also holds. Consequently, the random variable $U_\mathrm{S}$ obtained through the shattering channel satisfies both~\eqref{eq:opt_rv_cond1} and~\eqref{eq:opt_rv_cond2}, and attains the bound in~\eqref{eq:ub_privacy_rv}.
\end{IEEEproof}

\begin{remark}
PML can alternatively be written as
\begin{align*}
    \ell_{P_{XY}}(X \to y) &= \max_{x \in \mathrm{supp}\left(P_X \right)} \log \frac{P_{X \mid Y=y}(x)}{P_X(x)}\\[0.5em]
    &= \max_{x \in \mathrm{supp}\left(P_X \right)} \log \frac{P_{Y \mid X=x}(y)}{P_Y(y)}\\[0.5em]
    &= D_\infty \left(P_{X \mid Y=y} \Vert P_X \right)\\[0.5em]
    &= \max_{x \in \mathrm{supp}(P_X)} i_{P_{XY}}(x;y), 
\end{align*}
for $y \in \mathrm{supp}(P_Y)$, where $D_\infty\left(P_{X \mid Y=y} \Vert P_X \right)$ denotes the Rényi divergence of order infinity~\cite{renyi1961measures, van2014renyi} between the posterior $P_{X \mid Y=y}$ and the prior $P_X$, while $i_{P_{XY}}(x;y) \coloneqq \log \frac{P_{XY}(x,y)}{P_X(x) P_Y(y)}$ denotes (the value of) the information density of $P_{XY}$ at $x$ and $y$. The above is also related to a recent result by~\citet{kurri2022variational}, where a variational formula for Rényi divergence of order infinity is derived as the ratio of the expected gains for guessing a randomized function of $X$. 
\end{remark}

\subsection{Gain Function View of Leakage}
\label{ssec:def_gain}
The threat model assumed in Theorem~\ref{thm:privacy_leakage_rv} considers an adversary who is interested in guessing the value of a randomized function of $X$. In this section, we argue that pointwise maximal leakage can be obtained using an alternative threat model based on (a pointwise adaptation of) the $g$-leakage framework introduced in~\cite{alvim2012measuring}. First, we describe this alternative threat model. 

Suppose a passive and computationally unbounded adversary observes $y \in \mathrm{supp}(P_Y)$, an outcome of the channel $P_{Y \mid X}$, and constructs a guess $\hat X$ of $X$ using a kernel $P_{\hat X \mid Y}$ in order to maximize her expected \emph{gain}. The adversary selects her guess from a non-empty finite set $\hat{\mathcal X}$ (not necessarily equal to $\mathcal X$), and her gain is captured by a function $g$ of the form $g: \mathcal{X} \times \hat{\mathcal{X}} \to \mathbb{R}^+$. In order to measure the amount of information leaking from $y$, the system designer considers the ratio of the expected adversarial gain having observed $y$, and the expected adversarial gain with no observations. As such, we define the pointwise \emph{$g$-leakage} of $X$ as follows: 
\begin{equation}
\label{eq:g-leakage}
    \ell_g(X \to y) \coloneqq \log \frac{\sup_{P_{\hat{X} \mid Y}} \mathbb{E} \left[g(X,\hat{X}) \mid Y=y \right]}{\max_{\hat{x} \in \hat{\mathcal X}} \mathbb E\left[g(X, \hat{x})\right]}.
\end{equation}

\begin{figure}
    \centering
    \includegraphics[scale=0.12]{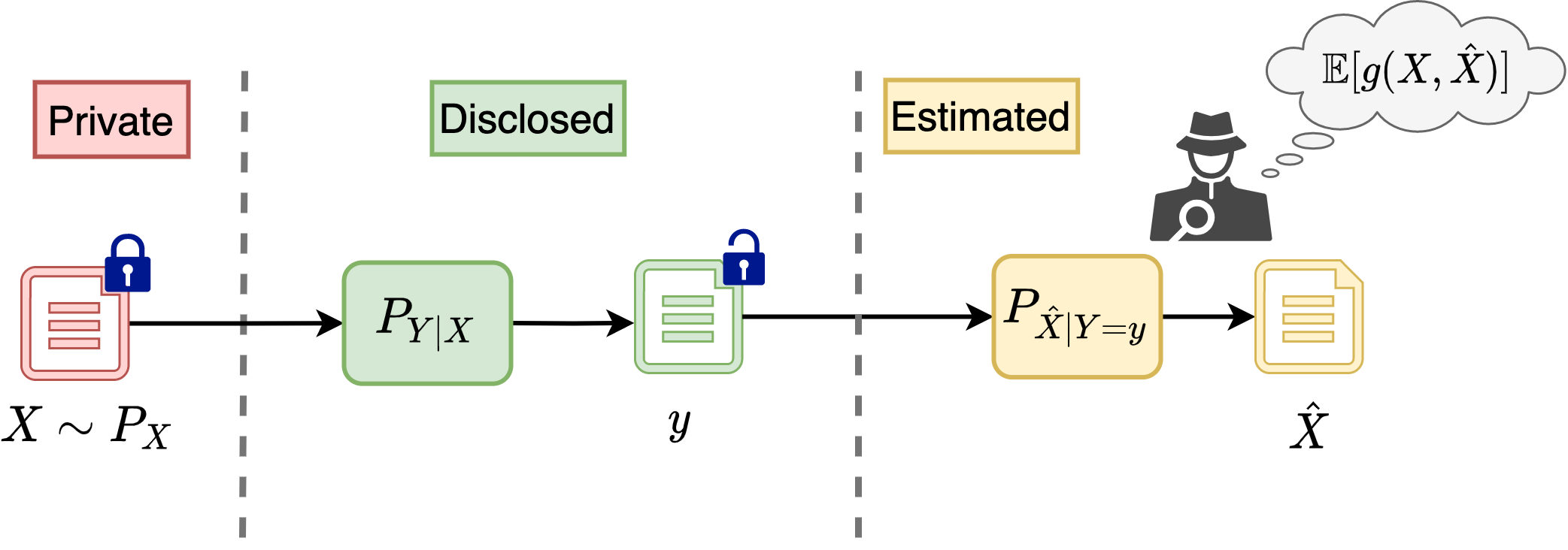}
    \caption{System model for the gain function view of leakage: An adversary observes an outcome $y$ of the channel $P_{Y \mid X}$, and tries to construct a guess $\hat{X}$ of $X$ in order to maximize a gain function $g$.}
    \label{fig:gain_func}
\end{figure}

In Theorem~\ref{thm:gain_u_equivalence}, we will show that the randomized function view and the gain function view of leakage are equivalent in the sense that for every gain function $g$, there exists a corresponding randomized function of $X$, $U_g$, such that $\ell_g(X \to y) = \ell_{U_g}(X \to y)$, and conversely, for every randomized function of $X$, $U$, there exists a corresponding gain function $g_{_U}$ such that $\ell_U(X \to y)  = \ell_{g_{_U}} (X \to y)$. Before presenting this result, let us demonstrate through a few examples how gain functions can be used to model different adversarial objectives. 

\begin{example}[The identity gain function~{\cite[Def. 3.5]{alvim2012measuring}}]
\label{ex:identity_gain}
The simplest type of gain function is the identity gain which models an adversary interested in guessing the secret $X$ itself, who is only rewarded for correct guesses. Here, the guessing space of the adversary is $\hat{\mathcal{X}} = \mathcal{X}$, and her gain function is given by $g^{\mathrm{identity}}(x,\hat{x}) = \mathds{1} [x=\hat{x}]$. The $g$-leakage for the identity gain is 
\begin{equation}
\label{eq:identity_gain}
    \ell_{g^{\mathrm{identity}}}(X \to y) = \frac{\max_{x \in \mathcal{X}} P_{X \mid Y=y}(x)}{\max_{x \in \mathcal{X}} P_X(x)},
\end{equation}
which is equal to the dynamic min-entropy leakage defined in~\cite[Def. 3]{espinoza2013min}. We will further discuss the identity gain and its associated $g$-leakage in Section~\ref{ssec:dynamic_model}.  
\end{example}

\begin{example}[Membership/group privacy]
\label{ex:membership}
Consider a centralized setting in which each $x \in \mathcal{X}$ represents a database whose rows constitute data collected from individuals, and $X$ describes the random selection of a database according to some distribution $P_X$. Suppose an adversary is interested in guessing whether or not  Alice's data is included in $X$, and is rewarded with a binary gain depending on whether or not her guess is correct. We can model this problem as follows: Let $\mathcal{X}_1 = \{x \in \mathcal{X} \colon x \text{ contains Alice's data}\}$ denote the set of databases that contain Alice's data, and let $\mathcal{X}_0 = \mathcal{X} \setminus \mathcal{X}_1$ be the set of databases that do not contain Alice's data. An adversary who is interested in finding out Alice's membership makes a binary guess from $\hat{\mathcal{X}} = \{0,1\}$, and is rewarded according to $g(x,x') = \mathds{1}[\hat{x} = i]$ with $i \in \{0,1\}$ and $x \in \mathcal{X}_i$.

More generally, suppose the adversary has a list of $k$ individuals and is interested in guessing if any one of their data points is included in $X$. Further, suppose the adversary is rewarded based on the number of correct guesses that she makes. To model this problem, we bi-partition the set of all databases in $k$ different ways, one for each individual on the list. Let $\mathcal{X}_{j_1} = \{x \in \mathcal{X} \colon x \text{ contains the $j$-th individual's data}\}$ and $\mathcal{X}_{j_{0}} = \mathcal{X} \setminus \mathcal{X}_{j_{1}}$ for $j = 1, \ldots, k$. Then, $\hat{\mathcal{X}} = \{0,1\}^k$ is the guessing space of the adversary, and $g(x,\hat{x}) = \sum_{j=1}^k \mathds{1}[\hat{x}_j = j_i]$ with $x \in \mathcal{X}_{j_i}$ and $i \in \{0,1\}$ is her gain function. This example can be easily extended to model cases where different individuals signify different gains for the adversary.  
\end{example}

As a side note, we should point out that in the membership privacy example above (or more generally, in our consideration of the centralized setting) we are \emph{not} assuming that the adversary is \emph{informed}~\cite{dwork2006calibrating} (an informed adversary knows all the entries in the database except for a single entry which may be Alice's). In our setup, we assume that the adversary knows the joint distribution $P_{X Y}$ (and the spaces $\mathcal{X}$ and $\mathcal{Y}$), while any other side information should be explicitly modeled as such. The concept of an informed adversary was originally proposed as a model for a very powerful adversary. However, it has been argued that more side information does not necessarily make an adversary more effective~\cite{kifer2011no,yang2015bayesian}. For example,~\cite{kifer2011no} provides three definitions of privacy against adversaries that either 
\begin{enumerate*}[label=(\roman*)]
    \item know all the entries in a database except for a single entry,
    \item know all the attributes in a database except for a single attribute of a single entry,
    \item know all the bits in a database except for a single bit of a single entry.\label{itm:bit_dp}
\end{enumerate*}
Then, it is shown that the privacy definition that seeks to limit the inference of the more knowledgeable adversary (i.e., the third adversary) may actually leak more sensitive information to the less knowledgeable adversaries.

In Section~\ref{ssec:properties}, we will define a conditional form of pointwise maximal leakage that can be used to model an adversary who possesses some side information about the secret. There, we will see that side information can both increase and decrease the information leakage due to observing an outcome (more on this in Remark~\ref{rem:example_side_info}). 

\begin{example}[Multiple guesses (the $k$-tries gain function in~\cite{alvim2012measuring})] Consider a side-channel setting in which $X$ represents a password and $Y$ represents some information leaking about the password, for example, through the inter-keystroke delays. Suppose an adversary is allowed $k \geq 1$ attempts at guessing the password correctly before getting cut off from the system. Let $\mathcal X$ be the set of all possible passwords and let $\hat{\mathcal{X}}= \{\hat{x} \subset \mathcal{X} \colon \abs{\hat{x}} \leq k\}$ denote the collection of subsets of $\mathcal X$ containing $k$ or less passwords. Then, we can model the adversary's gain through the function $g(x,\hat{x}) = \mathds{1}[x \in \hat{x}]$, where $\hat{x}$ denotes the set of $k$ or fewer attempts that the adversary makes at guessing the correct password $x$. 
\end{example}

\begin{example}[Metric spaces~\cite{alvim2012measuring}]
Suppose $(\mathcal{X},\rho)$ is a metric space, where $\mathcal X$ is a finite set, and $\rho$ is a metric on $\mathcal X$. Suppose the goal of the adversary is to construct a guess $\hat{x}$ of $x$ that minimizes $\rho(x,\hat{x})$. This scenario can be modeled by taking $\hat{\mathcal{X}} = \mathcal{X}$ and some non-negative gain function that is decreasing in $\rho(x,\hat x)$, for example, $g(x,\hat{x}) = \exp(-\rho(x,\hat{x}))$. Many problems can be modeled as metric spaces. A simple example is in geo-location applications where the goal of an adversary may be to locate a user as accurately as possible based on partial or noisy measurements. 
\end{example}

We have now seen how a variety of adversarial objectives can be modeled using gain functions. In the following result, which is one of the main contributions of this paper, we show that the definition of $g$-leakage given in~\eqref{eq:g-leakage} is equivalent to the definition of $U$-leakage in~\eqref{eq:u-leakage}. Thus, we unify two seemingly different ways of defining (pointwise) maximal leakage. The proof of the theorem is deferred to Appendix~\ref{sec:eqvlnc_thm_proof}.
\begin{theorem}
\label{thm:gain_u_equivalence}
For all joint distributions $P_{XY}$ on $\mathcal{X} \times \mathcal{Y}$, the randomized function view and the gain function view of leakage are equivalent. That is, for every randomized function of $X$, denoted by $U$, there exists a space $\hat{\mathcal{X}}_U$ and a gain function $g_{_U} : \mathcal{X} \times \hat{\mathcal X}_{_U} \to \mathbb R^+$ such that $\ell_U(X \to y)  = \ell_{g_{_U}} (X \to y)$. Conversely, for every gain function $g : \mathcal X \times \hat{\mathcal X} \to \mathbb R^+$, there exists a randomized function of $X$, denoted by $U_g$, such that $\ell_g(X \to y) = \ell_{U_g}(X \to y)$.
\end{theorem}

Note that while the above result establishes the equivalence of the gain function view and the randomized function view for pointwise leakages, it generalizes readily to the average-case leakages of~\cite{alvim2014additive} and~\cite{issa2019operational}. Furthermore, we have the following corollary which provides an alternative definition of PML. 
\begin{corollary}
Pointwise maximal leakage can also be defined as
\begin{align*}
\begin{split}
    \ell_{P_{XY}}(X \to y) &\coloneqq \sup_{g} \ell_g(X \to y)\\
    &= \log \, \sup_{g} \frac{\sup_{P_{\hat{X} \mid Y}} \mathbb{E} \left[g(X,\hat{X}) \mid Y=y \right]}{\max_{\hat{x} \in \hat{\mathcal X}} \mathbb E\left[g(X, \hat{x})\right]},
\end{split}
\end{align*}
where the supremum is over all gain functions with non-negative and finite ranges. 
\end{corollary}

\begin{remark}
Unlike privacy measures such as maximal leakage and (local) differential privacy that depend only on the mechanism $P_{Y \mid X}$, PML depends both on the mechanism $P_{Y \mid X}$ and the prior $P_X$, i.e., it is a property of the joint distribution $P_{XY}$. Thus, in the rest of this paper, we often assume that the prior $P_X$ is arbitrary but fixed, and study PML as a function of the mechanism $P_{Y \mid X}$. 
\end{remark}

\subsection{Properties}
\label{ssec:properties}
In this section, we recount several useful properties of $\ell_{P_{XY}}(X \to y)$. For instance, we discuss how pointwise maximal leakage composes over multiple outcomes, how it is affected by pre- and post-processing, and so on. Before we discuss these properties, let us first define a conditional form of pointwise maximal leakage which allows us to model adversaries who possess some side information about the secret $X$. 

\begin{definition}[Conditional pointwise maximal leakage]
Let $P_{XYZ}$ denote the joint distribution of $X$, $Y$, and $Z$. Given $z \in \mathrm{supp}(P_Z)$, the conditional pointwise maximal leakage from $X$ to $y \in \mathrm{supp}(P_{Y \mid Z=z})$ is defined as 
\begin{align*}
    &\ell_{P_{XY \mid Z}}(X \to y \mid z) \\
    &\coloneqq \log \sup_{P_{U \mid X,Z}} \frac{\sup_{P_{\hat U \mid Y,Z}} \mathbb P \left[U=\hat U \mid Y=y, Z=z \right]}{\sup_{P_{\Tilde U \mid Z}} \mathbb P \left[U=\Tilde U \mid Z=z \right]}.
\end{align*}
\end{definition}
To obtain a simpler expression for $\ell_{P_{XY \mid Z}}(X \to y \mid z)$, we may condition all the distributions in the proof of Theorem~\ref{thm:privacy_leakage_rv} on $Z=z$ and get
\begin{align*}
    \ell_{P_{XY \mid Z}}(X \to y \mid z) &= \log \max_{x \in \mathrm{supp}(P_{X \mid Z=z})} \frac{P_{X \mid Y=y,Z=z}(x)}{P_{X \mid Z=z}(x)}\\[0.7em]
    &= \log \max_{x \in \mathrm{supp}(P_{X \mid Z=z})} \frac{P_{Y \mid X=x,Z=z}(y)}{P_{Y \mid Z=z}(y)}\\[0.7em]
    &= D_\infty \left(P_{X \mid Y=y,Z=z} \Vert P_{X \mid Z=z} \right)\\[0.7em]
    &= \max_{x \in \mathrm{supp}(P_{X \mid Z=z})} i_{P_{XY \mid Z}}(x;y \mid z), 
\end{align*}
where 
\begin{equation*}
    i_{P_{XY \mid Z}}(x;y \mid z) \coloneqq \log \frac{P_{XY \mid Z=z}(x,y)}{P_{X \mid Z=z}(x) P_{Y \mid Z=z}(y)},
\end{equation*}
denotes the (value of the) \emph{conditional information density}. In the remainder of this paper, when the joint distribution used to calculate pointwise maximal leakage or information density is clear from the context, we do not specify it as a subscript.

The following lemma collects several useful properties of PML. The proof of the lemma is provided in Appendix~\ref{sec:porperties_lemma_proof}. 
\begin{lemma}
\label{lemma:properties}
Pointwise maximal leakage satisfies the following properties:  
\begin{enumerate}
\item (Upper/lower bounds). Given an arbitrary but fixed prior $P_X$, for all mechanisms $P_{Y \mid X}$ and all $y \in \mathrm{supp}(P_Y)$ it holds that
\begin{equation*}
    0 \leq \ell(X\to y) \leq - \log \Big(\min_{x \in \mathrm{supp}(P_X)} P_X(x)\Big),
\end{equation*}
where the left-hand side inequality holds with equality if and only if $P_{Y \mid X=x}(y) = P_{Y \mid X=x'}(y)$ for all $x,x' \in \mathrm{supp}(P_X)$, and the right-hand side inequality holds with equality if and only if $P_{X \mid Y=y}(x^*) = 1$ for some $x^* \in \argmin_{x \in \mathrm{supp}(P_X)} P_X(x)$. 

\item (Independence/deterministic mappings). If $X$ and $Y$ are independent random variables, then $\ell(X \to y) = 0$ for all $y \in \mathrm{supp}(P_Y)$. If $Y$ is the output of a deterministic mapping with input $X$, then $\ell(X \to y) = - \log P_Y(y)$. 

\item (Pre-processing). Suppose the Markov chain $X - Y - Z$ holds, where $Y$ represents some processing of the secret $X$, and $Z$ denotes the observable outcome of a channel $P_{Z \mid Y}$ with input $Y$. For all $z \in \mathrm{supp}(P_Z)$ we have
\begin{equation*}
    \ell(X \to z) \leq \ell(Y \to z),
\end{equation*}
with equality if and only if there exists $x^* \in \argmax_{x \in \mathrm{supp}(P_X)} i(x;z)$ such that $i(y;z) = i(y';z) = \ell(Y \to z)$ for all $y, y' \in \mathrm{supp}(P_{Y \mid X=x^*})$.

\item (Post-processing). Suppose the Markov chain $X - Y - Z$ holds, where $Y$ represents the observable outcome of a channel with input $X$, and $Z$ denotes some post-processing of $Y$. For all $z \in \mathrm{supp}(P_Z)$ we have
\begin{equation*}
    \ell(X \to z) \leq \max_{y \in \mathrm{supp}(P_Y)} \ell(X \to y),
\end{equation*}
with equality if and only if one of the following conditions is satisfied: 
\begin{enumerate}
    \item $X$ and $Y$ are independent, or 
    \item there exists $y_z \in \mathrm{supp}(P_Y)$ such that $P_{Y \mid Z=z}(y_z) =1$ and $\ell(X \to y_z) = \max_{y \in \mathrm{supp}(P_Y)} \ell(X \to y)$.
\end{enumerate}

\item (Conditionally-independent side information). If the Markov chain $Z - X - Y$ holds, where $Z$ represents some side information about $X$, then,
\begin{equation*}
    \ell(X \to y \mid z) = \ell(X \to y) - i(y;z),
\end{equation*}
for all $z \in \mathrm{supp}(P_Z)$ and $y \in \mathrm{supp}(P_{Y \mid Z=z})$.

\item (Composition). Given a prior $P_X$ and a mechanism $P_{YZ \mid X}$, for all $(y,z) \in \mathrm{supp}(P_{YZ})$ it holds that
\begin{align*}
    \ell(X \to y,z) &= \max_{x \in \mathrm{supp}(P_X)} i(x;y,z)\\
    &\leq \ell(X \to z) + \ell(X \to y \mid z),
\end{align*}
with equality if and only if the sets $\argmax\limits_{x \in \mathrm{supp}(P_X)} i(x ; y \mid z)$ and $\argmax\limits_{x \in \mathrm{supp}(P_X)} i(x ; z)$ have non-empty intersection.
\end{enumerate}
\end{lemma}

\begin{remark}
A few points are worth emphasizing about the above properties:
\begin{enumerate}
    \item The second property in the lemma describes the information leakage of deterministic mechanisms. Surprisingly, not all deterministic outcomes leak the same amount of information, and outcomes with lower probabilities have higher leakage. This is because pointwise maximal leakage is a relative privacy measure in which, roughly speaking, the information leaked to an adversary scales depending on how consistent the observed outcome is with the adversary's prior beliefs (captured by the joint distribution $P_{XY}$). As such, deterministic outcomes with smaller probabilities leak more information since an adversary would be \say{more surprised} by observing them. 

    \item Concerning the post-processing property, one may hope for the stronger statement $\ell(X \to z) \leq \ell(X \to y)$ for all $y \in \mathrm{supp}(P_Y)$. To see why this statement is not valid, suppose $Z = Y$ (or $Z$ is a deterministic mapping of $Y$). In this case, the best bound we can have is indeed $\ell(X \to y) \leq \max_y \ell(X \to y)$. 

    \item \label{rem:example_side_info}
    In general, side information can both increase and decrease information leakage. As an example, suppose $X$, $Y$, $Z$ are binary random variables where $X$ is uniformly distributed and the joint distribution $P_{XYZ}$ is induced by the following channels: 
    \begin{gather*}
        P_{Z \mid X = x}(0) = \begin{cases}
            \frac{2}{5} & \mathrm{if} \; x=0,\\
            \frac{3}{5} & \mathrm{if} \; x=1,
        \end{cases}\\
        P_{Y \mid X=x,Z=z}(0) =   \begin{cases}
            \frac{1}{2} & \mathrm{if} \; x=0,z=0\\
            \frac{1}{3} & \mathrm{if} \; x=0,z=1\\
            \frac{2}{3} & \mathrm{if} \; x=1,z=0\\
            \frac{1}{2} & \mathrm{if} \; x=1,z=1,
        \end{cases}
    \end{gather*}
    with $P_{Z \mid X=x}(1) = 1 - P_{Z \mid X=x}(0)$ and $P_{Y \mid X=x, Z=z}(1) = 1 - P_{Y \mid X=x,Z=z}(0)$. Then, it can be verified that $\ell_{P_{XY \mid Z}}(X \to 0 \mid 0) = \log \frac{10}{9}$, $\ell_{P_{XY \mid Z}}(X \to 0 \mid 1) = \log \frac{5}{4}$ and $\ell_{P_{XY}}(X \to 0) = \log \frac{6}{5}$. Therefore, 
    \begin{align*}
        \ell_{P_{XY \mid Z}}(X \to 0 \mid 0) &< \ell_{P_{XY}}(X \to 0)\\
        &< \ell_{P_{XY \mid Z}}(X \to 0 \mid 1).
    \end{align*}
\end{enumerate}
\end{remark}

\subsection{Dynamic Consumption of Secrecy}
\label{ssec:dynamic_model}
In the last part of this section, we discuss a notion of privacy introduced in~\cite[Section 2.2 ]{espinoza2013min} that, similar to our work, aims to measure the information leakage associated with individual observations. In \cite{espinoza2013min}, the \emph{dynamic min-entropy leakage} of an outcome $y \in \mathrm{supp}(P_Y)$ is defined as:
\begin{equation*}
    \ell^{\text{ dynamic}}(X \to y) \coloneqq \log \frac{\max_{x \in \mathcal{X}} P_{X \mid Y=y}(x)}{\max_{x \in \mathcal X} P_X(x)}, 
\end{equation*}
which is equal to the $g$-leakage associated with the identity gain in~\eqref{eq:identity_gain}. That is, the dynamic leakage is derived under the assumption that the adversary is trying to guess the secret $X$ itself, but does not consider other gain functions.

The authors of~\cite{espinoza2013min} withdraw from further developing the idea of measuring the pointwise information leakage based on $\ell^{\text{ dynamic}}(X \to y)$ as they believe the above privacy measure suffers from two drawbacks. First, they argue that the above definition cannot be axiomatically justified as it is shown that $\ell^{\text{ dynamic}}(X \to y)$ may be negative (see~\cite[Example 4]{espinoza2013min}), that is, the adversary's certainty about the secret may actually \emph{decrease} by observing an outcome $y$. Second, they believe that dynamic policy enforcement based on individual outcomes (for example, discarding high-leakage outcomes) may reveal information about the secret $X$. 

Note that the first issue mentioned above does not apply to $\ell(X \to y)$. It is easy to see that $\ell^{\text{ dynamic}}(X \to y) \leq \ell(X \to y)$,
and we have shown in Lemma~\ref{lemma:properties} that $\ell(X \to y) \geq 0$, which implies that in our current setup, observations can never decrease certainty about a secret $X$. This is because $\ell(X \to y)$ is defined by considering all possible gain functions an adversary may be interested in, while $\ell^{\text{ dynamic}}(X \to y)$ is defined only for the identity gain of Example~\ref{ex:identity_gain}.

Furthermore, while it is true that some policies defined based on individual outcomes, such as discarding high-leakage outcomes, may reveal information about the secret, we believe that this is not sufficient reason for abandoning the subject area altogether. In fact, contrary to~\cite{espinoza2013min}, we believe that effective policy enforcement depends crucially on the ability to quantify the information leaking from individual outcomes as this allows us to treat information leakage as a \emph{random variable}. Viewing information leakage as a random variable, we have the flexibility to define different types of privacy guarantees by specifying requirements on the statistical properties of information leakage. The resulting framework is then versatile enough to be applied to a wide range of problems. We develop this idea in the next section. 

\section{Privacy guarantees}
\label{sec:guarantees}
In Theorem~\ref{thm:privacy_leakage_rv}, we showed that $\ell(X \to y)$ can be written as a function of $y$. Since $Y$ is a random variable distributed according to $P_Y$, this in turn allows us to define a random variable $\ell(X \to Y)$ with a distribution induced by $P_Y$. From this point of view, a privacy guarantee is essentially a requirement we impose on some statistical property of $\ell(X \to Y)$; thus, we have the flexibility to define different types of privacy guarantees depending on how strict privacy requirements we need to meet. This section contains several examples of such guarantees: The \emph{almost-sure} guarantee, which bounds the information leakage with probability one, the \emph{tail-bound} guarantee, which bounds the leakage with high probability, and the \emph{average-case} guarantee, which bounds maximal leakage. We start by defining the almost-sure guarantee. 

\begin{definition}[Almost-sure guarantee]
Given an arbitrary but fixed prior $P_X$, we say that a privacy mechanism $P_{Y \mid X}$ satisfies $\epsilon$-PML with $\epsilon \geq 0$ if
\begin{equation}
\label{eq:strict_1}
    \mathbb P_{Y \sim P_Y}[\ell(X\to Y) \leq \epsilon] = 1.
\end{equation}
\end{definition}
As we are assuming that the random variables $X$ and $Y$ are finite, the above condition can also be expressed as  
\begin{equation}
\label{eq:strict_2}
\max_{(x,y) \in \mathrm{supp}(P_{XY})} i_{P_{XY}}(x;y) \leq \epsilon. 
\end{equation}
The expression $\max_{x,y} i_{P_{XY}}(x;y)$ coincides with the definition of maximal realizable leakage~\cite[Def. 8]{issa2019operational} and also max-information~\cite{dwork2015generalization}. Moreover, expression~\eqref{eq:strict_2} is also related to the notion of information privacy leakage~\cite{du2012privacy,jiang2021context}. We will discuss the relationship between our privacy definitions and pre-existing notions from the literature in the next section. 

The following lemma establishes some basic facts about $\epsilon$-PML guarantees.
\begin{lemma}
Given an arbitrary but fixed prior $P_X$, we have:
\begin{enumerate}
    \item All privacy mechanisms $P_{Y \mid X}$ satisfy $\epsilon_{\mathrm{max}}$-PML, where $\epsilon_{\mathrm{max}} \coloneqq -\log \min_{x \in \mathrm{supp}(P_X)} P_X(x)$ and $\epsilon_\mathrm{max} \geq \log 2$. Furthermore, we have
    \begin{equation*}
        \inf \{\epsilon \geq 0 \colon \mathbb P_{Y \sim P_Y}[\ell(X\to Y) \leq \epsilon] = 1\} = \epsilon_{\mathrm{max}},
    \end{equation*}
    if and only if there exists $y \in \mathrm{supp}(P_Y)$ and $x^* \in \argmin_x P_X(x)$ such that $P_{X \mid Y=y}(x^*) = 1$, or equivalently, $P_{Y \mid X=x}(y) = 0$ for all $x \neq x^*$. 
    \item A privacy mechanism $P_{Y \mid X}$ satisfies $\epsilon$-PML with $\epsilon=0$ if and only if $X$ and $Y$ are independent random variables. 
\end{enumerate}
\end{lemma}
\begin{IEEEproof}
\begin{enumerate}
\item For all $P_{Y \mid X}$ and all $y \in \mathrm{supp}(P_Y)$ we have 
\begin{align*}
    \log \max_{x \in \mathrm{supp}(P_X)} \frac{P_{X \mid Y=y}(x)}{P_X(x)} &\leq \log \max_{x \in \mathrm{supp}(P_X)} \frac{1}{P_X(x)}\\
    &= \log \frac{1}{\min\limits_{x \in \mathrm{supp}(P_X)} P_X(x)}. 
\end{align*}
Note that $\min_{x \in \mathrm{supp}(P_X)} P_X(x) \leq \frac{1}{2}$ which implies that $\epsilon_\mathrm{max} \geq \log 2$. The second half of the statement is clear from the above inequality.

\item If $X$ and $Y$ are independent, then $P_{X \mid Y=y}(x) = P_X(x)$ for all $x,y$, thus the mechanism $P_{Y \mid X}$ satisfies $\epsilon$-PML with $\epsilon=0$. Conversely, if $P_{Y \mid X}$ satisfies $\epsilon$-PML with $\epsilon=0$ this implies that $P_{X \mid Y=y}(x) = P_X(x)$ for all $x,y$ which means that $X$ and $Y$ are independent.
\end{enumerate}
\end{IEEEproof}

\begin{example}[Binary symmetric channel]
Suppose $P_{Y \mid X}$ is a binary symmetric channel with crossover probability $0 \leq \alpha \leq 1$. The input distribution is described by $P_X(0) = 1 - P_X(1) = q$, where $0 < q \leq 0.5$. Then, $P_{Y \mid X}$ satisfies $\epsilon$-PML with 
\begin{align*}
    \epsilon &= \begin{cases}
        \log \, {\displaystyle \frac{1 - \alpha}{q (1 - 2\alpha) + \alpha}} & \mathrm{if} \; \alpha \leq \frac{1}{2}, \\[1em]
        \log \, {\displaystyle \frac{\alpha}{q (2\alpha - 1) + 1 -\alpha}} & \mathrm{if} \; \alpha > \frac{1}{2}, \\
    \end{cases}\\[0.5em]
    &= \log \; {\displaystyle \frac{\abs{\alpha - \frac{1}{2}} + \frac{1}{2}}{\frac{1}{2} - \abs{\alpha - \frac{1}{2}} (1 - 2q)}}.
\end{align*}
\figurename~\ref{fig:BSC} depicts the leakage of the binary symmetric channel as a function of $\alpha$ for different values of $q$.  
\begin{figure}
    \centering
    \includegraphics[scale=0.2]{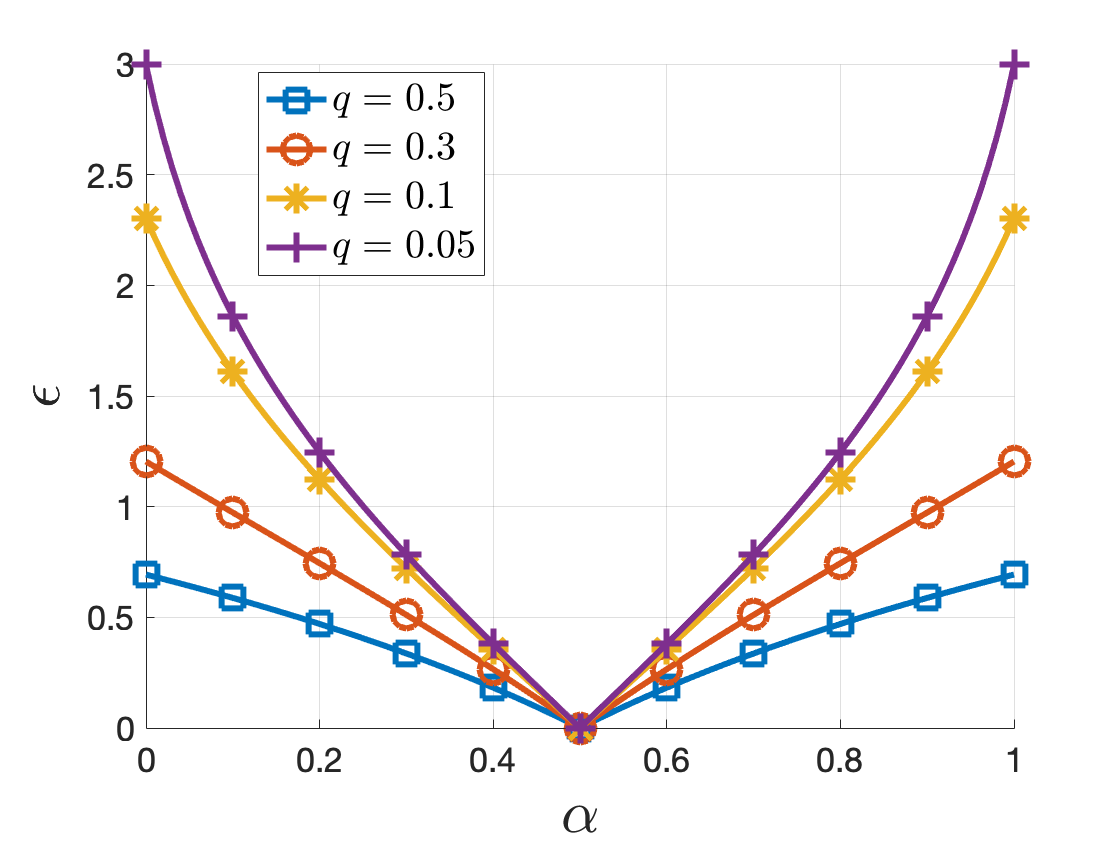}
    \caption{Leakage of the binary symmetric channel with $q \in \{0.05, 0.1, 0.3, 0.5\}$.}
    \label{fig:BSC}
\end{figure}
As expected, at $\alpha = \frac{1}{2}$ no information is leaked because $X$ and $Y$ are independent. It is also interesting to note that assuming a fixed $\alpha$, $\epsilon$ is decreasing in $q$ which implies that skewed priors leak more information. 
\end{example}

In order for an $\epsilon$-PML guarantee to hold, all $y \in \mathrm{supp}(P_Y)$ must satisfy $\ell(X \to y) \leq \epsilon$. As this condition may prove to be too restrictive in practice, in what follows we define two possible relaxations of the almost-sure guarantee: We either bound the information leakage by $\epsilon$ with high probability, or we bound the expected leakage (i.e., maximal leakage) by $\epsilon$. 

\begin{definition}[Tail-bound guarantee]
\label{def:tail_bound_guarantee}
Given an arbitrary but fixed prior $P_X$, we say that a mechanism $P_{Y \mid X}$ satisfies $(\epsilon,\delta)$-PML with $\epsilon \geq 0$ and $0 \leq \delta \leq 1$ if
\begin{equation}
    \mathbb{P}_{Y \sim P_Y}[\ell(X\to Y) \leq \epsilon] \geq 1-\delta.
\end{equation}
\end{definition}
Clearly, $\epsilon$-PML and $(\epsilon, 0)$-PML are equivalent. Also, note that given an arbitrary but fixed prior $P_X$, if a channel $P_{Y \mid X}$ satisfies $(\epsilon, \delta)$-PML, then it also satisfies $(\epsilon', \delta')$-PML for all $\epsilon \leq \epsilon'$ and all $\delta \leq \delta' \leq 1$. 
\begin{definition}[Average-case guarantee]
Given an arbitrary but fixed prior $P_X$, we say that the expected information leakage of a mechanism $P_{Y \mid X}$ is bounded by $\epsilon \geq 0$ if 
\begin{equation*}
    \mathbb E_{Y \sim P_Y}\Big[\exp\big(\ell(X\to Y)\big)\Big] \leq e^\epsilon,
\end{equation*}
or equivalently, if $\mathcal{L}(P_{Y \mid X}) \leq \epsilon$, where $\mathcal{L}(P_{Y \mid X})$ denotes maximal leakage as defined in~\cite[Thm. 1]{issa2019operational}.
\end{definition}
Note that here we denote maximal leakage by $\mathcal{L}(P_{Y \mid X})$ instead of $\mathcal{L}(X \to Y)$ used in~\cite{issa2019operational} to emphasize that maximal leakage is a property of the channel $P_{Y \mid X}$ and does not depend on the prior $P_X$.\footnote{Technically, maximal leakage depends on the support set of the prior $P_X$, but we can without loss of generality assume that $P_X$ has full support.}

\subsection{Data-processing Properties}
\label{ssec:data_processing}
Data-processing inequalities are often used while analyzing the end-to-end information leakage in larger systems. While the properties presented in Lemma~\ref{lemma:properties} allow us to assess pointwise maximal leakage for the outcomes of a pre- or post-processed random variable, it is also of practical benefit to understand how different privacy guarantees are affected by pre- and post-processing. This type of characterization is useful when we do not have access to the distribution of the leakage over the outcomes, but know that a privacy mechanism satisfies a certain privacy guarantee.

What we are specifically interested in is to understand whether or not different privacy guarantees are \emph{closed}\footnote{Suppose the channel $P_{Y \mid X}$ satisfies Property $A$. Given a Markov chain $X-Y-Z$, we say that Property $A$ is \emph{closed under post-processing} if the fact that $P_{Y \mid X}$ satisfies Property $A$ implies that $P_{Z \mid X}$ also satisfies Property $A$. Closedness under pre-processing is defined similarly.} under pre- and post-processing (in~\cite{wang2017privacy}, a privacy guarantee that is closed under pre-processing is said to satisfy the \emph{linkage} inequality). Suppose the channel $P_{Y \mid X}$ satisfies some privacy guarantee, say, $\epsilon$-PML. If the $\epsilon$-PML guarantee is closed under post-processing, then we can rest assured that for all post-processing channels $P_{Z \mid Y}$, the overall channel $P_{Z \mid X}$ also satisfies $\epsilon$-PML, that is, there exists no channel $P_{Z \mid Y}$ that an adversary could use to undermine the original guarantee. Similarly, if the $\epsilon$-PML guarantee is closed under pre-processing, then all (randomized) functions of $X$ would be at least as well-protected as $X$. 

The following lemma collects the data-processing properties satisfied by the privacy guarantees defined above. Part~\ref{itm:ml_data_proc} of the result concerning maximal leakage was shown in~\cite[Lemma 1]{issa2019operational}, and we re-state it here for completeness. 

\begin{prop}
\label{prop:data_processing_1}
Suppose the three random variables $X$, $Y$, and $Z$ form the Markov chain $X - Y - Z$ and that the prior $P_X$ is arbitrary but fixed. Given $\epsilon \geq 0$ and $0 \leq \delta \leq 1$, we have
\begin{enumerate}
    \item If $P_{Z \mid Y}$ satisfies $\epsilon$-PML, then $P_{Z \mid X}$ also satisfies $\epsilon$-PML.
    
    \item If $P_{Y \mid X}$ satisfies $\epsilon$-PML, then $P_{Z \mid X}$ also satisfies $\epsilon$-PML. 
    
    \item If $P_{Z \mid Y}$ satisfies $(\epsilon, \delta)$-PML, then $P_{Z \mid X}$ also satisfies $(\epsilon, \delta)$-PML.
    
    \item $\mathcal{L}(P_{Z \mid X}) \leq \min \{\mathcal{L}(P_{Y \mid X}), \mathcal{L}(P_{Z \mid Y})\}$.\label{itm:ml_data_proc} 
\end{enumerate}
\end{prop}
\begin{IEEEproof}
\begin{enumerate}
    \item By the pre-processing property of Lemma~\ref{lemma:properties}, if $P_{Z \mid Y}$ satisfies $\epsilon$-PML then for all $z \in \mathrm{supp}(P_Z)$ we have $\ell(Y \to z) \leq \epsilon$. Hence, 
    \begin{equation*}
        \max_{z \in \mathrm{supp}(P_Z)} \ell(X \to z) \leq \max_{z \in \mathrm{supp}(P_Z)} \ell(Y \to z) \leq \epsilon,
    \end{equation*}
    and $P_{Z \mid X}$ satisfies $\epsilon$-PML. 
    
    \item By the post-processing property of Lemma~\ref{lemma:properties},
    \begin{equation*}
        \max_{z \in \mathrm{supp}(P_Z)} \ell(X \to z) \leq \max_{y \in \mathrm{supp}(P_Y)} \ell(X \to y) \leq \epsilon,
    \end{equation*}
    so $P_{Z \mid X}$ satisfies $\epsilon$-PML. 
    
    \item Similarly to the above, the pre-processing property of Lemma~\ref{lemma:properties} yields
    \begin{align*}
        \mathbb P_{Z \sim P_Z} \big[\ell(X \to Z) > \epsilon \big] \leq \mathbb P_{Z \sim P_Z} \big[\ell(Y \to Z) > \epsilon \big] \leq \delta,
    \end{align*}
    hence, $P_{Z \mid X}$ satisfies $(\epsilon, \delta)$-PML.
\end{enumerate}
\end{IEEEproof}

\begin{example}[Truncated geometric mechanism~\cite{kairouz2014extremal,ghosh2009universally}]
A common perturbation method that satisfies $\alpha$-local differential privacy (LDP) adds geometrically distributed noise to the finite-valued input $X$, and then truncates the outcome to maintain the support of $X$. Suppose $\mathcal X = [k]$ with $k \geq 3$, $\mathcal Y = \{\ldots, -1, 0, 1, \ldots \}$, and let $Y$ be the output of the geometric mechanism described by 
\begin{equation*}
    P_{Y \mid X=x}(y) = \frac{1 - \exp(-\frac{\alpha}{k-1})}{1 + \exp(-\frac{\alpha}{k-1})} \cdot \exp(-\frac{\alpha \abs{y-x}}{k-1}),
\end{equation*}
where $x \in \mathcal X$, $y \in \mathcal Y$, and $\alpha \geq 0$. Let $Z$ denote the truncated version of $Y$, that is,
\begin{equation*}
    Z = \begin{cases} 
        Y & \mathrm{if} \; Y \in \mathcal X,\\
        1 & \mathrm{if} \; Y<1,\\
        k & \mathrm{if} \; Y>k.\\
    \end{cases}
\end{equation*}
The truncated geometric mechanism behaves interestingly under PML because the truncation, which is a form of post-processing, does not further decrease the leakage. To see why, note that 
\begin{equation*}
    \ell_{P_{XY}}(X \to y) = \begin{cases}
    - \log \, \mathbb E[\exp(-\frac{\alpha \abs{y-X}}{k-1})] & \mathrm{if} \; y \in \mathcal X, \\[0.5em]
     - \log \, \mathbb E[\exp(-\frac{\alpha (k-X)}{k-1})] & \mathrm{if} \; y > k, \\[0.5em]
     - \log \, \mathbb E[\exp(-\frac{\alpha (X-1)}{k-1})] & \mathrm{if} \; y <1,
    \end{cases}
\end{equation*}
that is, PML does not depend on the actual value of $y$ when $y >k$ or $y < 1$. Furthermore, $\ell_{P_{XZ}}(X \to i) = \ell_{P_{XY}}(X \to i)$ for $i \in \{2, \ldots, k-1\}$, 
\begin{align*}
    \ell_{P_{XZ}}(X \to 1) &= \log \; \frac{\max_x \sum_{y = -\infty}^{1} P_{Y \mid X=x}(y)}{\sum_{y = -\infty}^{1} P_Y(y)}\\
    &= \log \frac{\sum_{y = -\infty}^{1} P_{Y \mid X=1}(y)}{\sum_{y = -\infty}^{1} P_Y(y)}\\
    &= \ell_{P_{XY}}(X \to 1),
\end{align*}
and
\begin{align*} 
    \ell_{P_{XZ}}(X \to k) &= \log \; \frac{\max_x \sum_{y = k}^{\infty} P_{Y \mid X=x}(y)}{\sum_{y = k}^{\infty} P_Y(y)}\\
    &= \log \frac{\sum_{y = k}^{\infty} P_{Y \mid X=k}(y)}{\sum_{y = k}^{\infty} P_Y(y)}\\
    &= \ell_{P_{XY}}(X \to k).
\end{align*}
Now, since $\abs{y - x} \leq \max \{y-1, k-y\}$ for all $x,y \in \mathcal X$, both $P_{Y \mid X}$ and $P_{Z \mid X}$ satisfy $\epsilon$-PML with
\begin{align*}
    \epsilon &= \max_{y \in \mathcal Y} \, \ell_{P_{XY}}(X \to y)\\
    &= \max_{z \in \mathcal Z} \, \ell_{P_{XZ}}(X \to z)\\
    &= \max \Bigg\{ - \log \, \mathbb E \left [\exp(-\frac{\alpha (k-X)}{k-1}) \right], \\
    &\hspace{6em}- \log \, \mathbb E \left[\exp(-\frac{\alpha (X-1)}{k-1} \right] \Bigg\}\\
    &\leq \frac{\alpha \cdot \, \max \Big\{k - \mathbb E[X], \mathbb E[X] - 1)\Big\}}{k-1},
\end{align*}
where the last line is due to Jensen's inequality. 

When the distribution $P_X$ is highly skewed with either $\mathbb E[X] \to 1$ or $\mathbb E[X] \to k$ the gap between $\epsilon$ and $\alpha$ vanishes and $\epsilon \to \alpha$ since Jensen's inequality holds with equality for degenerate distributions. On the other hand, when $X$ is uniformly distributed, the bound reduces to $\epsilon \leq \frac{\alpha}{2}$ and the gap between $\epsilon$ and $\alpha$ is maximized. Thus, incorporating information about the prior is useful as it may lead to smaller privacy parameters under favorable conditions. 
\end{example}

Conspicuous by its absence in Proposition~\ref{prop:data_processing_1} is the post-processing property for the $(\epsilon, \delta)$-PML privacy guarantee. It turns out that, in general, $(\epsilon, \delta)$-PML is not closed under post-processing. To understand why, let us consider the following example.

\begin{example}
\label{ex:eps_delta_post_proc}
Suppose $X$ is a uniformly distributed random variable defined over an alphabet with four elements, and that the Markov chain $X-Y-Z$ holds. Suppose the channels $P_{Y \mid X}$ and $P_{Z \mid Y}$ are defined as 
\begin{align*}
    P_{Y \mid X} = \begin{bmatrix}
    0 & 0 & \frac{1}{2} & \frac{1}{2}\\[0.5em]
    0 & 0 & \frac{1}{2} & \frac{1}{2}\\[0.5em]
    0 & \frac{1}{3} & \frac{1}{3} & \frac{1}{3}\\[0.5em]
    \frac{1}{3} & 0 & \frac{1}{3} & \frac{1}{3}
    \end{bmatrix}, \qquad
    P_{Z \mid Y} = \begin{bmatrix}
    1 & 0\\[0.5em]
    0 & 1\\[0.5em]
    1 & 0 \\[0.5em]
    0 & 1 
    \end{bmatrix}.
\end{align*}
It can be easily verified that $\ell(X \to y_1) = \ell(X \to y_2) = \log 4$, and $\ell(X \to y_3) = \ell(X \to y_4) = \log \frac{6}{5}$. Since $P_Y(y_1) = P_Y(y_2) = \frac{1}{12}$, $P_{Y \mid X}$ satisfies $(\epsilon_1, \delta_1)$-PML with $\epsilon_1 = \log \frac{6}{5}$ and $\delta_1 = \frac{1}{6}$. On the other hand, one may also verify that $P_Z(z_1) = P_Z(z_2) = \frac{1}{2}$ and $\ell(X \to z_1) = \ell(X \to z_2) = \log \frac{4}{3}$. Hence, $P_{Z \mid X}$ does not satisfy $(\epsilon_1, \delta_1)$-PML; instead, it satisfies $\epsilon_2$-PML with $\epsilon_2 = \log \frac{4}{3} > \epsilon_1$ (and $\delta_2=0$). Note that the outcome $z_1$ is equivalent to the event $\{y_1, y_3\}$, $z_2$ is equivalent to the event $\{y_2, y_4\}$, and both outcomes have probability greater than $\delta_1$. 
\end{example}

Informally speaking, when we say that a mechanism $P_{Y \mid X}$ satisfies $(\epsilon, \delta)$-PML this implies that $\mathrm{supp}(P_Y)$ can be partitioned into two sets: a set of \say{good} $y$'s with probability at least $1- \delta$ whose members satisfy $\ell(X \to y) \leq \epsilon$, and a set of \say{bad} $y$'s with probability at most $\delta$ and $\ell(X \to y) > \epsilon$ (see Definition~\ref{def:tail_bound_guarantee}). However, through a post-processing channel $P_{Z \mid Y}$, we may define new outcomes as a combination of the members of the good and bad sets of $y$ (as in Example~\ref{ex:eps_delta_post_proc}). As a result, the probability of the set whose members satisfy $\ell(X \to z) > \epsilon$ (that is, the set of \say{bad} $z$'s) may no longer be bounded by $\delta$. Also, note that while in Example~\ref{ex:eps_delta_post_proc} we have $\epsilon_2 > \epsilon_1$ and $\delta_2 < \delta_1$, this need not always be the case; one may come up with examples where both $\epsilon$ and $\delta$ increase by post-processing.
\begin{remark}
Interestingly, a similar behavior has been observed with differential privacy. Specifically, it has been shown that \emph{probabilistic} DP, that is, a type of privacy guarantee where we require pure DP to hold with probability at least $1-\delta$~\cite{machanavajjhala2008privacy,meiser2018approximate}, is not closed under post-processing~\cite{kifer2012axiomatic, meiser2018approximate}. Differential privacy resolves this problem by introducing \emph{approximate} DP (i.e., $(\epsilon, \delta)$-DP defined based on an \emph{additive} parameter $\delta$~\cite{dwork2006our}) which is closed under post-processing. Note that approximate DP is a strictly weaker guarantee compared to probabilistic DP in the sense that probabilistic DP implies approximate DP but the reverse direction does not necessarily hold~\cite{meiser2018approximate}. In our current work with PML, we take a different approach to solving this issue and come up with a new privacy guarantee that maintains its probabilistic flavor.   
\end{remark}

Now, we define a new probabilistic privacy guarantee that is similar to $(\epsilon, \delta)$-PML, but is closed under post-processing. Drawing on Example~\ref{ex:eps_delta_post_proc}, our new definition ensures that all post-processed outcomes with probability at least $\delta$ have their PML bounded by $\epsilon$. We provide two alternative formulations of our new privacy guarantee: The first one in Definition~\ref{def:stable_post_proc} describes a somewhat technical condition, so we re-state it in a more intuitive form in Definition~\ref{def:eps_delta_eml}.

\begin{definition}
\label{def:stable_post_proc}
Given an arbitrary but fixed prior $P_X$, we say that a privacy mechanism $P_{Y \mid X}$ satisfies $(\epsilon, \delta)$-closedness with $\epsilon \geq 0$ and $0 \leq \delta \leq 1$, if for all post-processing channels $P_{Z \mid Y}$ and all $z \in \mathrm{supp}(P_Z)$, $P_Z(z) \geq \delta$ implies $\ell_{P_{XZ}}(X \to z) \leq \epsilon$.
\end{definition}

Based on Definition~\ref{def:stable_post_proc}, to check whether or not a certain mechanism $P_{Y \mid X}$ satisfies the desired closedness property, one needs to examine all possible post-processing channels $P_{Z \mid Y}$. This raises the question of whether it is possible to come up with a definition equivalent to Definition~\ref{def:stable_post_proc}, which can be stated as a property of the channel $P_{Y \mid X}$ itself. In what follows, we show that this is indeed possible, but we need a few other ingredients before we are ready to state this alternative definition. First, we recall two concepts from~\cite{mciver2014abstract}.

\begin{definition}[Similar outcomes~{\cite{mciver2014abstract}}]
Given a channel $P_{Y \mid X}$, we say that the outcomes $y,y' \in \mathrm{supp}(P_Y)$ are similar if their corresponding columns in the matrix of $P_{Y \mid X}$ are scalar multiples of each other, or equivalently, if $P_{X \mid Y=y}(x) = P_{X \mid Y=y'}(x)$ for all $x \in \mathrm{supp}(P_X)$.
\end{definition}

\begin{remark}
Note that if the outcomes $y,y' \in \mathrm{supp}(P_Y)$ are similar, then $i(x;y) = i(x;y')$ for all $x \in \mathrm{supp}(P_X)$ and $\ell(X \to y) = \ell(X \to y')$.
\end{remark}

\begin{definition}[Reduced channel~{\cite[Def. 3]{mciver2014abstract}}]
Given a channel $P_{Y \mid X}$, its reduced channel denoted by $P_{Y_r \mid X}$ is formed by removing all-zero columns from $P_{Y \mid X}$, and merging (i.e., adding) the columns corresponding to similar outcomes.
\end{definition}
Let $P_{Y_r \mid X}$ denote the reduced channel of the mechanism $P_{Y \mid X}$. We can define an equivalence relation $P_{\bar Y \mid X} \sim P_{Y \mid X}$ if $P_{\bar Y \mid X}$ has $P_{Y_r \mid X}$ as its reduced channel. Then, the equivalence class of $P_{Y \mid X}$, denoted by $\mathcal C(P_{Y \mid X})$, is the collection of all mechanisms whose reduced channel is $P_{Y_r \mid X}$. Suppose $P_{\bar Y \mid X} \in \mathcal C(P_{Y \mid X})$. We will use $\bar Y$ to denote the (output) random variable induced by the channel $P_{\bar Y \mid X}$ whose alphabet is represented by $\bar{\mathcal Y}$, and whose marginal distribution is denoted by $P_{\bar Y}$. 

Similar outcomes lead to the same posterior distribution, information density, and PML. Thus, the channels in a class $\mathcal C(P_{Y \mid X})$ behave identically with respect to information measures that are defined based on the information density, such as mutual information and maximal leakage. In the following, we show that if $P_{Y \mid X}$ satisfies $(\epsilon, \delta)$-closedness, then all $P_{\bar Y \mid X} \in \mathcal C(P_{Y \mid X})$ also satisfy $(\epsilon, \delta)$-closedness. 

\begin{prop}
\label{prop:one_for_all}
Given an arbitrary but fixed prior $P_X$ and an $(\epsilon, \delta)$ pair with $\epsilon \geq 0$ and $0 \leq \delta \leq 1$, if a mechanism $P_{Y \mid X}$ satisfies $(\epsilon, \delta)$-closedness then all $P_{\bar Y \mid X} \in \mathcal C(P_{Y \mid X})$ also satisfy $(\epsilon, \delta)$-closedness.
\end{prop}
\begin{IEEEproof}
Let the function $f: \mathcal C(P_{Y \mid X}) \times [0,1] \to \mathbb R^+$ be defined as  
\begin{equation*}
    f(P_{\bar Y \mid X} , \delta) = \sup_{P_{Z \mid \bar Y}} \max_{\substack{z \in \mathrm{supp}(P_Z) : \\ P_Z(z) \geq \delta}} \ell_{P_{XZ}}(X \to z),
\end{equation*}
that is, $f$ represents the largest PML over all outcomes $z$ of all post-processing channels with probability at least $\delta$. We argue that $f(\cdot, \delta)$ is constant on $\mathcal C(P_{Y \mid X})$ for all $0 \leq \delta\leq 1$. To see this, fix an arbitrary $P_{\bar Y \mid X} \in \mathcal C(P_{Y \mid X})$ and note that the Markov chain $X - \bar Y - Y_r$ holds, where $Y_r$ denotes the random variable induced by the reduced channel $P_{Y_r \mid X}$. By definition, $I(X;\bar Y) = I(X; Y_r)$, therefore $Y_r$ is a sufficient statistic of $\bar{Y}$ for $X$, and the Markov chain $X - Y_r - \bar Y$ also holds. Now, we write 
\begin{align*}
    f(P_{\bar Y \mid X} , \delta) &= \sup_{P_{Z \mid \bar Y}} \max_{\substack{z \in \mathrm{supp}(P_Z) : \\ P_Z(z) \geq \delta}} \ell_{P_{XZ}}(X \to z) \\
    &= \sup_{\substack{P_{Z \mid Y_r} : \\ P_{Z \mid Y_r} = P_{Z \mid \bar Y} \circ P_{\bar Y \mid Y_r}}} \max_{\substack{z \in \mathrm{supp}(P_Z) : \\ P_Z(z) \geq \delta}} \ell_{P_{XZ}}(X \to z)\\
    &\leq \sup_{{P_{Z \mid Y_r}}} \max_{\substack{z \in \mathrm{supp}(P_Z) : \\ P_Z(z) \geq \delta}} \ell_{P_{XZ}}(X \to z)\\
    &= f(P_{Y_r \mid X} , \delta). 
\end{align*}
Reversing the role of $\bar Y$ and $Y_r$, it can also be established that $f(P_{Y_r \mid X} , \delta) \leq f(P_{\bar Y \mid X}, \delta)$; hence, we obtain $f(P_{Y_r \mid X} , \delta) = f(P_{\bar Y \mid X}, \delta)$ for all $P_{\bar Y \mid X} \in \mathcal C(P_{Y \mid X})$ and $0 \leq \delta \leq 1$. Finally, if $P_{Y \mid X}$ satisfies $(\epsilon, \delta)$-closedness then $f(P_{Y \mid X}, \delta) \leq \epsilon$, which implies that $\sup_{P_{\bar Y \mid X} \in \mathcal C(P_{Y \mid X})} f(P_{\bar Y \mid X}, \delta) \leq \epsilon$.
\end{IEEEproof}

One last concept that we need to introduce is the notion of the maximal leakage associated with arbitrary events (that is, subsets) of $\mathrm{supp}(P_Y)$. We will call this new form of leakage \emph{event maximal leakage} (EML), which is defined fairly similarly to PML. That said, the real benefit of EML is in that it allows us to come up with an alternative formulation of Definition~\ref{def:stable_post_proc}.

\begin{definition}[Event maximal leakage (EML)]
Let $P_{XY}$ denote the joint distribution of $X$ and $Y$. Given an event $\mathcal{E} \subseteq \mathrm{supp}(P_{Y})$, the maximal leakage from $X$ to $\mathcal E$ is defined as 
\begin{align*}
    &\ell_{P_{XY}}(X \to \mathcal{E}) \coloneqq \log \max_{x \in \mathrm{supp}(P_X)} \frac{P_{Y \mid X=x}(\mathcal E)}{P_{Y}(\mathcal{E})}\\
    &= \log \max_{x \in \mathrm{supp}(P_X)} \frac{\sum_{y \in \mathcal{E}} P_{Y \mid X=x}(y)}{\sum_{y' \in \mathcal{E}} P_Y(y')}\\
    &= \log \max_{x \in \mathrm{supp}(P_X)} \sum_{y \in \mathcal{E}} \; \frac{P_{Y}(y)}{\sum_{y' \in \mathcal{E}} P_Y(y')} \; \exp \big( i_{P_{XY}}(x;y) \big).
\end{align*}
\end{definition}

EML essentially quantifies the information leaking from the outcomes of a deterministic function of $Y$. To see why, suppose $Z$ is a random variable produced by some deterministic post-processing of $Y$. Then, outcomes of $Z$ are either the re-labeled outcomes of $Y$ or they result from merging several outcomes of $Y$ into a single symbol. If $z \in \mathrm{supp}(P_{Z})$ is a re-labeling of $y \in \mathrm{supp}(P_{Y})$, then $\ell_{P_{XZ}} (X \to z ) = \ell_{P_{XY}} (X \to y)$. On the other hand, if $z' \in \mathrm{supp}(P_{Z})$ is a result of combining $y \in \mathcal E \subseteq \mathrm{supp}(P_{Y})$ into a single symbol, then $\ell_{P_{XZ}} (X \to z') = \ell_{P_{XY}} (X \to \mathcal E)$.

Having defined the maximal leakage associated with arbitrary subsets of $\mathrm{supp}(P_{Y})$, we are now ready to provide an alternative form for Definition~\ref{def:stable_post_proc}.

\begin{definition}
\label{def:eps_delta_eml}
Given an arbitrary but fixed prior $P_X$, we say that a privacy mechanism $P_{Y \mid X}$ satisfies $(\epsilon, \delta)$-EML with $\epsilon \geq 0$ and $0 \leq \delta \leq 1$ if for all $P_{\bar Y \mid X} \in \mathcal C(P_{Y \mid X})$ and all events $\mathcal E \subseteq \mathrm{supp}(P_{\bar Y})$, $P_{\bar Y}(\mathcal E) \geq \delta$ implies $\ell_{P_{X \bar Y}}(X \to \mathcal E) \leq \epsilon$.  
\end{definition}
Clearly, $(\epsilon, 0)$-EML and $\epsilon$-PML are equivalent. Furthermore, given an arbitrary but fixed prior $P_X$, if a channel $P_{Y \mid X}$ satisfies $(\epsilon, \delta)$-EML, then it also satisfies $(\epsilon', \delta')$-EML for all $\epsilon \leq \epsilon'$, and all $\delta \leq \delta' \leq 1$.

Next, we show that a privacy mechanism $P_{Y \mid X}$ satisfies $(\epsilon, \delta)$-closedness if and only if it satisfies $(\epsilon, \delta)$-EML. That is, Definitions~\ref{def:stable_post_proc} and~\ref{def:eps_delta_eml} are equivalent. 

\begin{theorem}
\label{thm:equivalence}
Given an arbitrary but fixed prior $P_X$, and a pair $(\epsilon, \delta)$ with $\epsilon \geq 0$ and $0 \leq \delta \leq 1$, a privacy mechanism $P_{Y \mid X}$ satisfies $(\epsilon, \delta)$-closedness if and only if  it satisfies $(\epsilon, \delta)$-EML.
\end{theorem}
\begin{IEEEproof}
Suppose without loss of generality that $P_X$ has full support. We first show that if a privacy mechanism $P_{Y \mid X}$ satisfies $(\epsilon, \delta)$-closedness, then it satisfies $(\epsilon, \delta)$-EML. Informally, this result follows from the fact that for all $P_{\bar{Y} \mid X} \in  \mathcal C(P_{Y \mid X})$, optimizing over the events in $\mathrm{supp}(P_{\bar Y})$ with probability at least $\delta$ is equivalent to optimizing over the outcomes of all deterministic mappings $P_{Z \mid \bar{Y}}$ with probability at least $\delta$. More concretely, suppose $P_{Z \mid \bar{Y}}$ is a deterministic channel, that is, the matrix form of $P_{Z \mid \bar{Y}}$ consists only of zeros and ones. Then, each outcome $z \in \mathrm{supp}(P_Z)$ corresponds to some event  $\mathcal E_z \subseteq \bar{\mathcal Y}$ such that $P_{Z \mid \bar{Y}=y}(z) = 1$ for all $y \in \mathcal E_z$, and $P_{\bar Y}(\mathcal E_z) = P_Z(z)$. Let $\mathcal D_{\bar{\mathcal Y}}$ denote the set of all deterministic mappings with domain $\bar{\mathcal{Y}}$. We can write
\begin{align*}
    \epsilon &\geq \sup_{P_{\bar Y \mid X} \in \mathcal C(P_{Y \mid X})} \sup_{P_{Z \mid \bar Y}} \max_{\substack{z \in \mathrm{supp}(P_Z) : \\ P_Z(z) \geq \delta}} \ell_{P_{XZ}}(X \to z)\\
    &\geq  \sup_{P_{\bar Y \mid X} \in \mathcal C(P_{Y \mid X})} \sup_{P_{Z \mid \bar Y} \in \mathcal D_{\bar{\mathcal Y}}} \max_{\substack{z \in \mathrm{supp}(P_Z) : \\ P_Z(z) \geq \delta}} \ell_{P_{XZ}}(X \to z) \\
    &= \sup_{P_{\bar Y \mid X} \in \mathcal C(P_{Y \mid X})} \sup_{P_{Z \mid \bar Y} \in \mathcal D_{\bar{\mathcal Y}}} \max_{\substack{z \in \mathrm{supp}(P_Z) : \\ P_Z(z) \geq \delta}} \log \; \max_x \frac{P_{Z \mid X=x}(z)}{P_Z(z)} \\
    &= \sup_{P_{\bar Y \mid X} \in \mathcal C(P_{Y \mid X})} \sup_{P_{Z \mid \bar Y} \in \mathcal D_{\bar {\mathcal Y}}} \max_{\substack{z \in \mathrm{supp}(P_Z) : \\ P_Z(z) \geq \delta}} \log \; \max_x \frac{P_{\bar Y \mid X=x}(\mathcal E_z)}{P_{\bar Y}(\mathcal E_z)} \\
    &= \sup_{P_{\bar Y \mid X} \in \mathcal C(P_{Y \mid X})} \max_{\substack{\mathcal E \subseteq \mathrm{supp}(P_{\bar Y}) : \\ P_{\bar Y}(\mathcal E) \geq \delta}} \log \; \max_x \frac{P_{\bar Y \mid X=x}(\mathcal E)}{P_{\bar Y}(\mathcal E)} \\
    &= \sup_{P_{\bar Y \mid X} \in \mathcal C(P_{Y \mid X})} \max_{\substack{\mathcal E \subseteq \mathrm{supp}(P_{\bar Y}) : \\ P_{\bar Y}(\mathcal E) \geq \delta}} \ell_{P_{X{\bar Y}}}(X \to \mathcal E),
\end{align*}
where the first inequality follows from Proposition~\ref{prop:one_for_all}. Thus, $P_{Y \mid X}$ satisfies $(\epsilon, \delta)$-EML. 

Now, we show that if a mechanism $P_{Y \mid X}$ satisfies $(\epsilon, \delta)$-EML, then it satisfies $(\epsilon, \delta)$-closedness. Let the function $h : \mathcal C(P_{Y \mid X}) \times [0,1] \to [1, \infty)$ be defined as
\begin{equation}
\label{eq:optim_0}
    h(P_{\bar Y \mid X} , \delta) = \sup_{P_{Z \mid \bar Y}} \max_{\substack{z \in \mathrm{supp}(P_Z) : \\ P_Z(z) \geq \delta}} \exp \Big( \ell_{P_{XZ}}(X \to z) \Big).
\end{equation}
We can write $h$ as 
\begin{equation*}
    h(P_{\bar Y \mid X}, \delta) = \max_x h_x(P_{\bar Y \mid X}, \delta),
\end{equation*}
where 
\begin{align}
\label{eq:optim_1}
    &h_x(P_{\bar Y \mid X}, \delta) \coloneqq \sup_{P_{Z \mid \bar Y}} \max_{\substack{z \in \mathrm{supp}(P_Z) : \\ P_Z(z) \geq \delta}} \frac{P_{Z \mid X=x}(z)}{P_Z(z)}\nonumber\\
    &= \sup_{P_{Z \mid \bar Y}} \max_{\substack{z \in \mathrm{supp}(P_Z) : \\ P_Z(z) \geq \delta}} \frac{\sum\limits_{y \in \mathrm{supp}(P_{\bar Y})} P_{Z \mid \bar Y=y}(z) P_{\bar Y \mid X=x}(y)}{\sum\limits_{y' \in \mathrm{supp}(P_{\bar Y})} P_{Z \mid \bar Y=y'}(z) P_{\bar Y}(y')}\nonumber\\
    &= \sup_{P_{Z \mid \bar Y}} \max_{\substack{z \in \mathrm{supp}(P_Z) : \\ P_Z(z) \geq \delta}} \\
    & \hspace{1.3em}\sum_{y \in \mathrm{supp}(P_{\bar Y})} \hspace{-0.2em} \frac{P_{Z \mid \bar Y=y}(z) P_{\bar Y}(y)}{\sum\limits_{y' \in \mathrm{supp}(P_{\bar Y})} \hspace{-1.5em} P_{Z \mid \bar Y=y'}(z) P_{\bar Y}(y')} \left( \frac{P_{\bar Y \mid X=x}(y)}{P_{\bar Y}(y)} \right).\nonumber
\end{align}

Using Proposition~\ref{prop:one_for_all}, it suffices to show that for each $x$, there exists $P_{\bar Y \mid X} \in \mathcal C(P_{Y \mid X})$ satisfying $h_x(P_{\bar Y \mid X}, \delta) \leq \exp(\epsilon)$. Hence, we solve the above optimization problem for the reduced channel associated with the class $\mathcal C(P_{Y \mid X})$, denoted by $P_{Y_r \mid X}$. 

Fix some $x \in \mathcal X$. Let $n_r \coloneqq \abs{\mathcal Y_r}$  denote the cardinality of $\mathcal Y_r$ (recall that $P_{Y_r}$ has full support). We re-write~\eqref{eq:optim_1} for $P_{Y_r \mid X}$ as 
\begin{align}
\label{eq:optim_reduced}
    &\max_{a_1, \ldots, a_{n_{r}}} \quad  \sum_{j=1}^{n_{r}} \; \frac{a_j P_{ Y_r}(y_j)}{\sum_{j'=1}^{n_{r}} a_{j'} P_{Y_r}(y_{j'})} \;  \exp \big(i_{P_{X Y_r}}(x;y_j)\big),\nonumber\\
    &\mathrm{subject\;to} \quad \sum_{j=1}^{n_{r}} a_j P_{Y_r}(y_j) \geq \delta,\\
    &\qquad \qquad \quad \, 0 \leq a_j \leq 1, \qquad \forall j \in [n_{r}],\nonumber
\end{align}
where $\{a_j\}$ specify $P_{Z \mid Y_r=y_j} (z)$ for the $z \in \mathrm{supp}(P_Z)$ with the largest PML which also satisfies $P_Z(z) \geq \delta$.

Suppose the elements in $\mathcal Y_r$ are labelled  such that $i_{P_{XY_r}}(x;y_1) \geq i_{P_{XY_r}}(x;y_2) \geq \ldots \geq i_{P_{XY_r}}(x;y_{n_r})$. Given an integer $k \in [n_r]$, let $\mathcal{F}_k \coloneqq \{y_1, \ldots, y_k \}$ be the set containing $k$ elements from $\mathcal Y_r$ that have the largest information density with $x$. Let $k^* \in [n_r]$ be the smallest integer such that $P_{Y_r}(\mathcal F_{k^*}) \geq \delta$. 
The objective function in problem~\eqref{eq:optim_reduced} is a linear-fractional function which is quasi-convex (in fact, quasi-linear)~\cite[Section 3.4]{boyd2004convex}, and the feasible region is a convex polytope. Therefore, the optimal solution is an extreme point of the feasible region given by
\begin{equation*}
    a^*_j = \begin{cases}
    1, & \text{if}\;\; j = 1, \ldots, k^* -1,\\
    \zeta, & \text{if}\;\; j = k^*,\\
    0, & \text{otherwise},
    \end{cases}
\end{equation*}
where the parameter $0 < \zeta \leq 1$ can be calculated by
\begin{equation}
\label{eq:zeta}
    P_{Y_r}(\mathcal F_{k^*-1}) + \zeta P_{Y_r}(y_{k^*}) = \delta.
\end{equation}
Hence, we obtain $h_x(P_{Y_r \mid X}, \delta)$ as 
\begin{equation}
\label{eq:kappa_x}
    h_x(P_{Y_r \mid X}, \delta) = \frac{1}{\delta} \Big( P_{Y_r \mid X=x}(\mathcal F_{k^* -1}) + \zeta P_{Y_r \mid X=x}(y_{k^*}) \Big).
\end{equation}
Since $k^*$ is the smallest integer such that $P_{Y_r}(\mathcal F_{k^*}) \geq \delta$, we need to consider the following two possibilities: We either have $P_{Y_r}(\mathcal F_{k^*}) = \delta$ or $P_{Y_r}(\mathcal F_{k^*}) > \delta$. First, suppose $\mathcal F_{k^*} = \delta$. In this case, the optimal parameters become
\begin{equation*}
\label{eq:opt_variables_deter}
    a^*_j = \begin{cases}
    1, & \text{if}\;\; j = 1, \ldots, k^*,\\
    0, & \text{otherwise},
    \end{cases}
\end{equation*}
that is, $\zeta = 1$ in~\eqref{eq:zeta}. Since $\{a^*_j\}$ consist of only zeros and ones, it in fact specifies a deterministic outcome $z^* \in \mathrm{supp}(P_{Z^*})$ for some channel $P_{Z^* \mid Y_r}$. This outcome corresponds to the event $\mathcal F_{k^*}$ in the sense that $P_{Z \mid {Y_r}=y}(z^*) = 1$ for all $y \in \mathcal F_{k^*}$, and $P_{Y_r}(\mathcal F_{k^*}) = P_Z(z^*)$. Thus, we get 
\begin{align}
\begin{split}
\label{eq:worst_case_is_deterministic}
    h_x ({P_{Y_r \mid X}}, \delta) &= \sup_{P_{Z \mid \bar Y}} \max_{\substack{z \in \mathrm{supp}(P_Z) : \\ P_Z(z) \geq \delta}} \frac{P_{Z \mid X=x}(z)}{P_Z(z)}\\
    &=  \frac{P_{Z^* \mid X=x}(z^*)}{P_{Z^*}(z^*)}\\
    &= \frac{P_{Y_r \mid X=x}(\mathcal F_{k^*})}{P_{Y_r}(\mathcal F_{k^*})}\\
    &\leq \exp(\epsilon),
\end{split}
\end{align}
where the last inequality follows from the fact that $P_{Y \mid X}$ satisfies $(\epsilon, \delta)$-EML. 

Now, suppose $P_{Y_r}(\mathcal F_{k^*}) > \delta$ which implies that $0 < \zeta < 1$ in~\eqref{eq:zeta}. In this case, we construct $P_{\hat Y \mid X} \in \mathcal C(P_{Y \mid X})$ whose columns are identical to the columns of $P_{Y_r \mid X}$, except that the $k^*$-th column of $P_{Y_r \mid X}$ is split into two corresponding columns in $P_{\hat Y \mid X}$ given by 
\begin{equation*}
    P_{\hat Y \mid X=x}(y_{k^*_{(1)}}) = \zeta P_{Y_r \mid X=x}(y_{k^*}),  
\end{equation*}
and
\begin{equation*}
    P_{\hat Y \mid X=x}(y_{k^*_{(2)}}) = (1- \zeta) P_{Y_r \mid X=x}(y_{k^*}),
\end{equation*}
for all $x \in \mathcal X$. Note that the outcomes $y_{k^*_{(1)}}, y_{k^*_{(2)}} \in \mathrm{supp}(P_{\hat Y})$ defined above are similar, and satisfy
\begin{equation*}
    i_{P_{X \hat Y}}(x;y_{k^*_{(1)}}) = i_{P_{X \hat Y}}(x;y_{k^*_{(2)}}) = i_{P_{X Y_r}}(x;y_{k^*}).
\end{equation*}
Now, we find $h_x (P_{\hat Y \mid X}, \delta)$. Forming the optimization problem~\eqref{eq:optim_1} for $P_{\hat Y \mid X}$, it is easy to see that the optimal parameters are
\begin{equation*}
    a^*_j = \begin{cases}
    1, & \text{if}\;\; j = 1, \ldots, k^* -1, k^*_{(1)}\\
    0, & \text{otherwise},
    \end{cases}
\end{equation*}
which, once again, specifies a deterministic outcome. Using arguments similar to~\eqref{eq:worst_case_is_deterministic}, we get $h_x(P_{\hat Y \mid X}, \delta) \leq \exp(\epsilon)$. 

Finally, as $x$ was chosen arbitrarily, we conclude that $h(P_{Y \mid X}, \delta) \leq \exp(\epsilon)$, that is, $P_{Y \mid X}$ satisfies $(\epsilon, \delta)$-closedness.
\end{IEEEproof}

\begin{remark}
\label{rem:equivalence_class}
The proof of Theorem~\ref{thm:equivalence} sheds light on the role of the class $\mathcal C(P_{Y \mid X})$: For each $0 \leq \delta \leq 1$, there exists $P_{\bar Y \mid X} \in \mathcal C(P_{Y \mid X})$ and a \say{least private} event $\mathcal E^* \subseteq \mathrm{supp}(P_{\bar Y})$ satisfying $P_{\bar Y}(\mathcal E^*) = \delta$. As such, without loss of generality, we unify the channels in the equivalence class $\mathcal C(P_{Y \mid X})$ and assume that $\mathcal E^* \subseteq \mathrm{supp}(P_{Y})$. Then, to show that $P_{Y \mid X}$ satisfies $(\epsilon, \delta)$-EML, it suffices to show that $\ell_{P_{XY}}(X \to \mathcal E^*) \leq \epsilon$.
\end{remark}

Now, recall that our motivation for introducing the notion of $(\epsilon, \delta)$-EML was to obtain a probabilistic privacy guarantee which is closed under both pre- and post-processing. The following result formally shows that this is indeed the case.
\begin{prop}
\label{prop:data_processing_EML}
Suppose the three random variables $X$, $Y$, and $Z$ form the Markov chain $X - Y - Z$. Given $\epsilon \geq 0$ and $0 \leq \delta \leq 1$, it holds that:
\begin{enumerate}
    \item (Pre-processing) If $P_{Z \mid Y}$ satisfies $(\epsilon, \delta)$-EML, then $P_{Z \mid X}$ satisfies $(\epsilon, \delta)$-EML. 
    
    \item (Post-processing) If $P_{Y \mid X}$ satisfies $(\epsilon, \delta)$-EML, then $P_{Z \mid X}$ satisfies $(\epsilon, \delta)$-EML. 
\end{enumerate}
\end{prop}
\begin{IEEEproof}
In both cases, we use Theorem~\ref{thm:equivalence} and verify the conditions of Definition~\ref{def:stable_post_proc}. Consider the Markov chain $X-Y-Z-T$. 
\begin{enumerate}
    \item Fix an arbitrary $P_{T \mid Z}$ and $t \in \mathrm{supp}(P_T)$ satisfying $P_{T}(t) \geq \delta$. Then, 
    \begin{align*}
        \ell(X \to t) \leq \ell(Y \to t) \leq \epsilon, 
    \end{align*}
    where the first inequality is due to Lemma~\ref{lemma:properties} and the second inequality follows by the assumption that $P_{Z \mid Y}$ satisfies $(\epsilon, \delta)$-EML. Thus, $P_{Z \mid X}$ satisfies $(\epsilon, \delta)$-EML. 

    \item The result follows directly by noticing that $T$ is a post-processing of $Y$ through the channel $P_{T \mid Y} = P_{T \mid Z} \circ P_{Z \mid Y}$. Hence, $P_{T}(t) \geq \delta$ implies $\ell(X \to t) \leq \epsilon$ with $t \in \mathrm{supp}(P_T)$ and $P_{Z \mid X}$ satisfies $(\epsilon, \delta)$-EML. 
\end{enumerate}
\end{IEEEproof}

Now, let us re-visit Example~\ref{ex:eps_delta_post_proc} and analyze it through the lens of event maximal leakage. 
\begin{example}
\label{ex:eml_post_proc}
Suppose $P_X$, $P_{Y \mid X}$ and, $P_{Z \mid Y}$ are defined as in Example~\ref{ex:eps_delta_post_proc}, and let $\delta = \frac{1}{6}$. Our goal is to find the smallest $\epsilon_1 \geq 0$ such that $P_{Y \mid X}$ satisfies $(\epsilon_1, \delta)$-EML, and the smallest $\epsilon_2 \geq 0$ such that $P_{Z \mid X}$ satisfies $(\epsilon_2, \delta)$-EML. First, note that the outcomes $y_3$ and $y_4$ are similar; hence, by merging them, we obtain the reduced channel $P_{Y_r \mid X}$ as 
\begin{align*}
    P_{Y_r \mid X} = \begin{bmatrix}
    0 & 0 & 1\\[0.5em]
    0 & 0 & 1\\[0.5em]
    0 & \frac{1}{3} & \frac{2}{3}\\[0.5em]
    \frac{1}{3} & 0 & \frac{2}{3}
    \end{bmatrix}.
\end{align*}
Now, for each $x$, we find $h_x(P_{Y_r \mid X}, \delta)$ defined in~\eqref{eq:kappa_x}:
\begin{gather*}
    h_{x_1}(P_{Y_r \mid X}, \delta) = h_{x_2}(P_{Y_r \mid X}, \delta) = \frac{6}{5}, \\ 
    h_{x_3}(P_{Y_r \mid X}, \delta) = h_{x_4}(P_{Y_r \mid X}, \delta) = \frac{12}{5},
\end{gather*}
which implies that $P_{Y \mid X}$ satisfies $(\epsilon_1, \delta)$-EML with 
\begin{equation*}
    \epsilon_1 = \log \max_x h_{x_1}(P_{Y_r \mid X}, \delta) = \log \frac{12}{5}. 
\end{equation*}
Furthermore, the channel $P_{Z \mid X}$ is given by 
\begin{equation*}
    P_{Z \mid X} = \begin{bmatrix}
    \frac{1}{2} & \frac{1}{2}\\[0.5em]
    \frac{1}{2} & \frac{1}{2}\\[0.5em]
    \frac{1}{3} & \frac{2}{3}\\[0.5em]
    \frac{2}{3} & \frac{1}{3}
    \end{bmatrix}.
\end{equation*}
Since $P_Z(z_1) = P_Z(z_2) = \frac{1}{2}$ and $\ell(X \to z_1) = \ell(X \to z_2) = \log \frac{4}{3}$, it follows that $P_{Z \mid X}$ satisfies $(\epsilon_2, \delta)$-EML with $\epsilon_2 = \log \frac{4}{3}$. Note that since $\epsilon_2 < \epsilon_1$, $P_{Z \mid X}$ also satisfies $(\epsilon_1, \delta)$-EML which was expected from Proposition~\ref{prop:data_processing_EML}.
\end{example}

\begin{remark}
The above example also describes the computational complexity of finding the smallest $\epsilon$ associated with a given $\delta$ in an $(\epsilon, \delta)$-EML privacy guarantee (or alternatively, finding the least private event with probability $\delta$). For each $x \in \mathcal X$, finding $h_x(P_{Y \mid X}, \delta)$ requires sorting the vector of information density $i(x;y_1), \ldots, i(x;y_{\abs{\mathcal Y}})$, which can be implemented with time complexity of $\mathcal O(\abs{\mathcal Y} \log \abs{\mathcal Y})$. Hence, the overall procedure can be implemented relatively efficiently.
\end{remark}

As the final topic in this section, we discuss the relationship between the $(\epsilon, \delta)$-EML and $(\epsilon, \delta)$-PML privacy guarantees. By Examples~\ref{ex:eps_delta_post_proc} and~\ref{ex:eml_post_proc}, it is clear that $(\epsilon, \delta)$-PML does not imply $(\epsilon, \delta)$-EML. In general, $(\epsilon, \delta)$-EML does not imply $(\epsilon, \delta)$-PML either. For example, consider the binary symmetric channel 
\begin{align*}
    P_{Y \mid X} = \begin{bmatrix}
    0.6 & 0.4 \\[0.5em]
    0.4 & 0.6
    \end{bmatrix},
\end{align*}
with uniform $P_X$. It is straightforward to verify that $P_{Y \mid X}$ satisfies $(\epsilon_1, \delta_1)$-EML with $\epsilon_1 = \log \frac{34}{30}$ and $\delta_1 = 0.6$. However, due to the symmetry of the channel, $P_{Y \mid X}$ satisfies $(\epsilon_2, \delta_2)$-PML with $\epsilon_2 = \log \frac{6}{5}$ and all $0 \leq \delta_2 < 1$. Note that for $\delta_1 = \delta_2 = 0.6$ we have $\epsilon_1 < \epsilon_2$. 

While in general $(\epsilon, \delta)$-EML does not imply $(\epsilon, \delta)$-PML, there exists a condition under which $(\epsilon, \delta)$-EML does in fact imply $(\epsilon, \delta)$-PML. This condition is discussed below. 
\begin{prop}
Given an arbitrary but fixed prior $P_X$, suppose the privacy mechanism $P_{Y \mid X}$ satisfies $(\epsilon, \delta)$-EML. Let $\mathcal A = \{y \in \mathrm{supp}(P_Y) \colon \ell(X \to y) > \epsilon \}$. If there exists $x^* \in \mathrm{supp}(P_X)$ satisfying $x^* \in \argmax_{x \in \mathrm{supp}(P_X)} i(x;y)$ for all $y \in \mathcal{A}$, then $P_{Y \mid X}$ satisfies $(\epsilon, \delta)$-PML. 
\end{prop}
\begin{IEEEproof}
We need to show that $P_Y(\mathcal A) \leq \delta$. We can write
\begin{subequations}
\begin{align}
    \ell(X \to \mathcal A) &= \log \max_{x \in \mathrm{supp}(P_X)} \frac{\sum_{y \in \mathcal{A}} P_{Y \mid X=x}(y)}{\sum_{y \in \mathcal{A}} P_Y(y)}\nonumber\\
    &= \log \frac{\sum_{y \in \mathcal{A}} \max_{x \in \mathrm{supp}(P_X)} P_{Y \mid X=x}(y)}{\sum_{y \in \mathcal{A}} P_Y(y)}\label{subeq:eml_implies_pml_1}\\
    &\geq \log \; \min_{y \in \mathcal A} \, \max_{x \in \mathrm{supp}(P_X)}\frac{P_{Y \mid X=x}(y)}{P_Y(y)}\nonumber\\
    &= \log \; \min_{y \in \mathcal A} \, \ell(X \to y)\nonumber\\
    &> \epsilon, \label{subeq:eml_implies_pml_2}
\end{align}
\end{subequations}
where~\eqref{subeq:eml_implies_pml_1} follows from the assumption that all $P_{Y \mid X=x}(y)$ are maximized at the same $x$, and~\eqref{subeq:eml_implies_pml_2} follows from the definition of the event $\mathcal A$. Since $P_{Y \mid X}$ satisfies $(\epsilon, \delta)$-EML, $\ell(X \to \mathcal A) > \epsilon$ implies that $P_Y(\mathcal A) < \delta$. 
\end{IEEEproof}
\subsection{Composition Properties}
\label{ssec:composition}
A second group of properties that are helpful while assessing the privacy levels of more complicated systems are composition properties. Let $X$ denote the sensitive data, and let $Y$ be the output of a channel $P_{Y \mid X}$. Suppose $X$ and $Y$ are fed into another channel $P_{Z \mid XY}$ inducing a random variable $Z$ in its output. What we are interested in is to find out how much information the overall channel $P_{ZY \mid X}$ leaks about $X$. 

The composition property stated in Lemma~\ref{lemma:properties} describes an upper bound on the PML resulting from composing the two channels $P_{Y \mid X}$ and $P_{Z \mid XY}$. In this section, our goal is to understand how different privacy guarantees, namely, $\epsilon$-PML, $(\epsilon, \delta)$-PML and $(\epsilon, \delta)$-EML are affected by adaptively composing two channels. We note that some related results have been derived in~\cite{wu2020strong}, where \emph{non-adaptive} composition is studied asymptotically for maximal leakage and Sibson mutual information~\cite{sibson1969information, verdu2015alpha}. The bounds we derive in this section differ from previous works in that here we study \emph{adaptive} composition, that is, we allow $Z$ to depend arbitrarily on both $X$ and $Y$. 

Naturally, one can formulate various problems by making different assumptions about the channels $P_{Y \mid X}$ and $P_{Z \mid XY}$. The following result contains several such problem formulations and results, and its proof is deferred to Appendix~\ref{sec:composition_lemma_proof}.
\begin{theorem}
\label{thm:composition}
Consider three random variables $X$, $Y$ and $Z$ where $X$ denotes the secret, $Y$ is the output of a channel $P_{Y \mid X}$, and $Z$ is the output of a channel $P_{Z \mid X Y}$.
\begin{enumerate}
    \item Suppose $P_{Y \mid X}$ satisfies $\epsilon_1$-PML, and for all $y \in \mathrm{supp}(P_Y)$, the channel $P_{Z \mid X,Y=y}$ satisfies $\epsilon_2$-PML with $\epsilon_1, \epsilon_2 \geq 0$. Then, the channel $P_{YZ \mid X}$ satisfies $\epsilon$-PML with $\epsilon = \epsilon_1 + \epsilon_2$.
    \item Suppose $P_{Y \mid X}$ satisfies $(\epsilon_1, \delta_1)$-PML, and for all $y \in \mathrm{supp}(P_Y)$, $P_{Z \mid X, Y=y}$ satisfies $(\epsilon_2, \delta_2)$-PML. Then, the channel $P_{YZ \mid X}$ satisfies $(\epsilon, \delta )$-PML with $\epsilon = \epsilon_1 + \epsilon_2$ and $\delta = \delta_1 + \delta_2 - \delta_1 \delta_2$. 
    \item Suppose $P_{Y \mid X}$ satisfies $(\epsilon_1, \delta_1)$-PML, and 
    \begin{equation*}
        \mathbb P_{(Y,Z) \sim P_{YZ}} \Big[ \ell(X \to Z \mid Y) \leq \epsilon_2 \Big] \geq 1 - \delta_2.
    \end{equation*}
    Then, the channel $P_{YZ \mid X}$ satisfies $(\epsilon, \delta )$-PML with $\epsilon = \epsilon_1 + \epsilon_2$ and $\delta = \delta_1 + \delta_2$. 
    \item Suppose $P_{Y \mid X}$ satisfies $(\epsilon_1, \delta_1)$-EML, and for all $y \in \mathrm{supp}(P_Y)$, $P_{Z \mid X, Y=y}$ satisfies $(\epsilon_2, \delta_2)$-EML. Given an event $\mathcal E \subseteq \mathrm{supp}(P_{YZ})$, define the sets
    \begin{gather*}
        \mathcal E_Y \coloneqq \{ y \in \mathrm{supp}(P_Y) \colon (y,z) \in \mathcal E , \;\, z \in \mathrm{supp}(P_Z) \}\\
        \mathcal E_Z(y) \coloneqq \{ z \in \mathrm{supp}(P_Z) \colon (y,z) \in \mathcal E\}.
    \end{gather*}
    If $0 \leq \delta_2 \leq \min\limits_{y \in \mathcal E_Y} P_{Z \mid Y=y}(\mathcal E_Z(y))$, then $P_{YZ}(\mathcal E) \geq \delta_1$ implies ${\ell(X \to \mathcal E) \leq \epsilon_1 + \epsilon_2}$. 
    
    Specifically, if $P_{Y \mid X}$ satisfies $(\epsilon_1, \delta_1)$-EML, and $P_{Z \mid X, Y=y}$ satisfies $\epsilon_2$-PML, then $P_{YZ \mid X}$ satisfies $(\epsilon_1 + \epsilon_2, \delta_1)$-EML.
    
    \item Suppose $P_{Y \mid X}$ satisfies $(\epsilon_1, \delta_1)$-EML, and for all $y \in \mathrm{supp}(P_Y)$, $P_{Z \mid X, Y=y}$ satisfies $(\epsilon_2, \delta_2)$-EML. Then, $P_{YZ \mid X}$ satisfies $(\epsilon, \delta)$-EML with
    \begin{equation*}
        \epsilon = \log \Big( \frac{\delta_2}{\delta_1 + \delta_2} \cdot \exp(\epsilon_\mathrm{max}) + \exp \big( \epsilon_1 + \epsilon_2 \big) \Big),
    \end{equation*}
    and $\delta = \delta_1 + \delta_2$, where  $\epsilon_{\mathrm{max}} \coloneqq -\log \min\limits_{x \in \mathrm{supp}(P_X)} P_X(x)$.
\end{enumerate}
\end{theorem}

\section{Relationship to other privacy/statistical notions}
\label{sec:comparisons}

In this section, we study how pointwise maximal leakage and the privacy guarantees defined in the previous section relate to several existing privacy/statistical notions from the literature. More specifically, we discuss max-information~\cite{dwork2015generalization,rogers2016max}, local differential privacy~\cite{kasiviswanathan2011can, duchi2013local}, local information privacy~\cite{du2012privacy, jiang2021context}, local differential identifiability~\cite{lee2012differential, wang2016relation}, mutual information, $f$-information~\cite{diaz2019robustness}, and total variation privacy~\cite{rassouli2019optimal}. The results of this section are summarized in Table~\ref{table:comparisons}.

\subsection{Max-information}
\label{ssec:max_information}
Max-information is a statistical quantity that was introduced as a tool for studying generalization in adaptive data analysis~\cite{dwork2015generalization,rogers2016max}. Note that while max-information has not been developed as a notion of privacy, it is defined similarly to pointwise maximal leakage, and therefore, their comparison is appropriate. Before we give the definition of (approximate) max-information, we need the following definition of \emph{approximate max-divergence} which is a weakening of \emph{max-divergence} (that is, Rényi divergence of order infinity~\cite{van2014renyi}). 

\begin{table*}[!t]
\renewcommand{\arraystretch}{2.3}
\caption{Summary of the Results of Section~\ref{sec:comparisons}}
\label{table:comparisons}
\centering
\begin{tabular}{|c||c|c|}
\hline
 Privacy/Statistical Notion & Relation/Bound & Ref.\\
\hline\hline
Max-information & $I_\infty (X ; Y) = \max_y \ell(X \to y)$ & Def.~\ref{def:max_info}\\
\hline
Approximate max-information &  $(\epsilon, \delta)\text{-PML} \implies I_\infty^\delta (X ; Y) \leq \epsilon$ & Prop.~\ref{prop:approx_max_info}\\
\hline
Local information privacy & $\epsilon\text{-LIP} \implies \epsilon\text{-PML}$ & Def.~\ref{def:lip} \\
\hline
Local differential privacy & $\epsilon\text{-LDP} \implies \log \frac{1}{p_\mathrm{min} + e^{-\epsilon} (1 - p_\mathrm{min})}$-PML & Prop.~\ref{prop:lip_ldp_ldi} \\
\hline
Local differential identifiability & 
    $\epsilon\text{-LDI} \implies \log \frac{1}{p_{\mathrm{min}} \big(1 + e^{-\epsilon}(\abs{\mathrm{supp}(P_X)} - 1) \big)}$-PML
    & Prop.~\ref{prop:lip_ldp_ldi}\\
\hline
Mutual information & $I(X ; Y) \leq \mathbb E_{Y \sim P_Y}[\ell(X \to Y)]$ & Prop.~\ref{prop:mutual_info}\\
\hline
$f$-information & $I_f(P_{XY}) \leq \mathbb E_{Y \sim P_Y} \Big[\max \left\{f\left(\exp(\ell(X \to Y))\right), f(0) \right\} \Big]$ & Prop.~\ref{prop:f_info_bound}\\
\hline
Total variation privacy & $T(X;Y) \leq \min \left\{ \frac{1}{2} \mathbb E_{Y\sim P_Y} \Big[\max \left \{\exp(\ell(X \to Y)) - 1, \; 1  \right \} \Big], \exp \left(\mathcal{L}(P_{Y \mid X}) \right) - 1 \right\}$ &
{\renewcommand{\arraystretch}{0.8}
    \begin{tabular}{@{}c@{}}
        Prop.~\ref{prop:f_info_bound}\\ Prop.~\ref{prop:total_var_privacy}
    \end{tabular}} \\
\hline
\end{tabular}
\end{table*}

\begin{definition}[Approximate max-divergence~\cite{dwork2015generalization}]
Let $P$ and $Q$ be probability distributions on a finite set $\Omega$ and suppose $P \ll Q$. Given $0 \leq \delta \leq 1$, the $\delta$-approximate max-divergence between $P$ and $Q$ is defined as 
\begin{equation*}
    D_\infty^\delta (P \Vert Q) = \log \max_{\mathcal{E} \subseteq \Omega , P(\mathcal{E}) \geq \delta } \frac{P(\mathcal{E}) - \delta}{Q(\mathcal E)}. 
\end{equation*}
\end{definition}
Note that if $\delta = 0$, then the above definition reduces to the max-divergence between $P$ and $Q$, denoted by $D_\infty(P \Vert Q)$.

\begin{definition}[(Approximate) max-information~\cite{dwork2015generalization}]
\label{def:max_info}
Suppose $V$ and $W$ are random variables supported on finite sets $\mathcal{V}$ and $\mathcal{W}$, respectively, and let $P_{V W}$ denote their joint distribution. The max-information between $V$ and $W$ is defined as 
\begin{align*}
    I_\infty (V ; W) &\coloneqq D_\infty(P_{V W} \Vert P_V P_W)\\
    &= \log \max_{v \in \mathcal{V}, w \in \mathcal{W}} \frac{P_{V W}(v,w)}{P_V(v) P_W(w)}.
\end{align*}
Similarly, the $\delta$-approximate max-information between $V$ and $W$ is defined as 
\begin{align*}
    I_\infty^\delta (V;W) &\coloneqq D_\infty^\delta (P_{V W} \Vert P_V P_W)\\
    &= \log \max_{\mathcal{E} \subseteq \mathcal V \times \mathcal W : P_{VW}(\mathcal{E}) \geq \delta } \frac{P_{V W}(\mathcal{E}) - \delta}{P_VP_W (\mathcal E)}, 
\end{align*}
where $P_VP_W(\mathcal E) = \sum_{(v,w) \in \mathcal E} P_V(v) P_{W} (w)$.

\end{definition}

It follows from the definition of max-information that, given a fixed prior $P_X$, a mechanism $P_{Y \mid X}$ satisfies $\epsilon$-PML if and only if $I_\infty(X ; Y) \leq \epsilon$. Therefore, below we examine how $(\epsilon, \delta)$-PML privacy compares with guarantees given in terms of the $\delta$-approximate max-information. To do this, first, we recall a lemma from~\cite{dwork2015generalization_arxiv}. As the proof of the lemma was omitted in~\cite{dwork2015generalization_arxiv}, here we also provide a proof.

\begin{lemma}[\hspace{-0.5pt}{\cite[Lemma 18]{dwork2015generalization_arxiv}}]
\label{lemma:high_prob_max_dvgc}
Let $P$ and $Q$ be probability distributions on a finite set $\Omega$ and suppose $P \ll Q$. Let the event $\mathcal{O} \subset \Omega$ be defined as 
\begin{equation*}
    \mathcal O \coloneqq \{\omega \in \Omega : \frac{P(\omega)}{Q(\omega)} > e^\epsilon\}.
\end{equation*}
If $P(\mathcal{O}) \leq \delta$, then $D_\infty^\delta (P \Vert Q) \leq \epsilon$. 
\end{lemma}
\begin{IEEEproof}
Let $\tilde{\mathcal O} \subseteq \Omega$ be any event satisfying $P(\Tilde{\mathcal O}) \geq \delta$. Then, we have 
\begin{align*}
    P(\Tilde{\mathcal O}) &= P(\Tilde{\mathcal O} \cap \mathcal O) + P(\Tilde{\mathcal O} \cap \mathcal O^c)\\
    &\leq P(\mathcal O) + Q(\Tilde{\mathcal O} \cap \mathcal O^c) e^\epsilon\\
    &\leq \delta + Q(\Tilde{\mathcal O}) e^\epsilon,
\end{align*}
which gives
\begin{equation*}
    \frac{P(\Tilde{\mathcal O}) - \delta }{Q(\Tilde{\mathcal O})} \leq e^\epsilon, \qquad \forall \Tilde{\mathcal O} \subseteq \Omega, P(\Tilde{\mathcal O}) \geq \delta. 
\end{equation*}
\end{IEEEproof}
Now, we use Lemma~\ref{lemma:high_prob_max_dvgc} to relate $\mathbb{P}_{Y \sim P_Y}[\ell(X\to Y) \leq \epsilon]$ and the $\delta$-approximate max-information.
\begin{prop}
\label{prop:approx_max_info}
Given an arbitrary but fixed $P_X$, if the channel $P_{Y \mid X}$ satisfies $(\epsilon, \delta)$-PML, then $I_\infty^\delta (X ; Y) \leq \epsilon$. 
\end{prop}
\begin{IEEEproof}
First, note that 
\begin{align}
\begin{split}
\label{eq:pml_max_dvgc_proof}
    &\mathbb{P}_{(X,Y) \sim P_{XY}} \left[\frac{P_{XY}(X,Y)}{P_X(X) P_Y(Y)} \leq e^\epsilon \right]\\
    &\geq \mathbb{P}_{Y \sim P_Y} \left[ \max_{x \in \mathrm{supp}(P_X)} \frac{P_{XY}(x,Y)}{P_X(x) P_Y(Y)} \leq e^\epsilon \right].
\end{split}
\end{align}
Now, since the privacy mechanism $P_{Y \mid X}$ satisfies $(\epsilon, \delta)$-PML, we have 
\begin{align*}
     &\mathbb{P}_{Y \sim P_Y} \Bigg[ \max_{x \in \mathrm{supp}(P_X)} \frac{P_{XY}(x,Y)}{P_X(x) P_Y(Y)} \leq e^\epsilon \Bigg]\\
     &= \mathbb{P}_{Y \sim P_Y}[\ell(X\to Y) \leq \epsilon]\\
     &\geq 1 - \delta, 
\end{align*}
which combined with~\eqref{eq:pml_max_dvgc_proof} yields 
\begin{equation*}
    \mathbb{P}_{(X,Y) \sim P_{XY}} \left[\frac{P_{XY}(X,Y)}{P_X(X) P_Y(Y)} \leq e^\epsilon \right] \geq 1 - \delta, 
\end{equation*}
or equivalently, 
\begin{equation*}
    \mathbb{P}_{(X,Y) \sim P_{XY}} \left[\frac{P_{XY}(X,Y)}{P_X(X) P_Y(Y)} > e^\epsilon \right] \leq \delta.
\end{equation*}
Finally, using the above inequality and Lemma~\ref{lemma:high_prob_max_dvgc} we conclude that 
\begin{equation*}
    I_\infty^\delta (X ; Y) = D_\infty^\delta (P_{XY} \Vert P_X P_Y) \leq \epsilon.  
\end{equation*}
\end{IEEEproof}

The previous result shows that $(\epsilon,\delta)$-PML is a stronger guarantee compared to $I_\infty^\delta(X ; Y) \leq \epsilon$. Roughly speaking, this is because under a $I_\infty^\delta(X ; Y) \leq \epsilon$ guarantee, the \say{good} $y$'s are those that have small information density $i(x ; y)$ \emph{with high probability} over the $x$'s. However, under an $(\epsilon, \delta)$-PML guarantee, the \say{good} $y$'s need to have small $i(x ; y)$ for \emph{all} $x$'s in $\mathrm{supp}(P_X)$, that is, $i(x ; y)$ must be small \emph{with probability one} over the $x$'s. In addition, note that $I_\infty^\delta(X ; Y)$ treats random variables $X$ and $Y$ symmetrically and the probability of a good event is calculated according to $P_{XY}$. On the other hand, under $(\epsilon, \delta)$-PML, the probability of a good event (i.e., low leakage) is calculated according to $P_Y$ over the $y$'s.

\subsection{LDP, LIP, and LDI}
Now, we discuss the relationship between PML-based guarantees and three other notions of privacy, namely, local differential privacy (LDP)~\cite{kasiviswanathan2011can, duchi2013local}, local information privacy (LIP)~\cite{du2012privacy,jiang2020local,jiang2021context} and local differential identifiability (LDI)~\cite{lee2012differential, wang2016relation}. First, we recall their definitions.

\begin{definition}[Local differential privacy{~\cite{kasiviswanathan2011can, duchi2013local}}]
A privacy mechanism $P_{Y \mid X}$ satisfies $\epsilon$-LDP with $\epsilon \geq 0$ if for all $y \in \mathrm{supp}(P_Y)$, $x,x' \in \mathrm{supp}(P_X)$ it holds that
\begin{equation*}
    \frac{P_{Y \mid X=x} (y)}{P_{Y \mid X=x'}(y)} \leq e^\epsilon.
\end{equation*}
\end{definition}
Note that the above definition depends only on the channel $P_{Y \mid X}$, so an $\epsilon$-LDP guarantee is valid for all priors $P_X$. 

\begin{definition}[Local information privacy{~\cite{jiang2021context}}]
\label{def:lip}
Given an arbitrary but fixed prior $P_X$, we say that a privacy mechanism $P_{Y \mid X}$ satisfies $\epsilon$-LIP with $\epsilon \geq 0$ if for all $y \in \mathrm{supp}(P_Y)$ and $x \in \mathrm{supp}(P_X)$  it holds that
\begin{equation*}
\label{eq:lip}
    e^{-\epsilon} \leq \frac{P_{X \mid Y=y}(x)}{P_X(x)} \leq e^\epsilon.
\end{equation*}
\end{definition}
While the definition of $\epsilon$-LIP privacy is similar to $\epsilon$-PML privacy, it differs from $\epsilon$-PML in that it requires an additional lower bound on the information density, that is, we must also have 
\begin{equation*}
    \frac{P_X(x)}{P_{X \mid Y=y}(x)} \leq e^\epsilon.
\end{equation*}
The above bound has no clear operational interpretation in our current framework, and may be superfluous. To see why, consider the following simple example. Suppose $X$ is a uniformly distributed binary random variable. Assume $P_{X \mid Y=y}(0) = p$ for some $y \in \mathrm{supp}(P_Y)$, where $p > 0$ is a small positive number. Then, $P_{X \mid Y=y}(1) = 1 - p$ with $1 - p$ close to one, and $\ell(X \to y)$ is 
\begin{equation*}
    \ell(X \to y) = \log \max_{x \in \{0, 1\}} \frac{P_{X \mid Y=y}(x)}{P_X(x)} = \log \big( 2 (1-p) \big),
\end{equation*}
which is close to $\epsilon_\mathrm{max} = \log 2$. Hence, the outcome $y$ has large information leakage. This example shows that small values of the ratio $\frac{P_{X \mid Y=y}(x)}{P_X(x)}$ can increase the pointwise maximal leakage simply because we must have $\sum_x P_{X \mid Y=y}(x) = 1$ for all $y \in \mathrm{supp}(P_X)$. As such, it may not necessary to impose a lower bound on the information density as a separate constraint; a privacy guarantee defined based on an upper bound on information density will be automatically penalized for small values of the ratio $\frac{P_{X \mid Y=y}(x)}{P_X(x)}$. 

\begin{definition}[Local differential identifiability{~\cite{lee2012differential, wang2016relation}}]
Given an arbitrary but fixed prior $P_X$, we say that a privacy mechanism $P_{Y \mid X}$ satisfies $\epsilon$-LDI with $\epsilon \geq 0$ if for all $y \in \mathrm{supp}(P_Y)$, $x,x' \in \mathrm{supp}(P_X)$ it holds that 
\begin{equation*}
    \frac{P_{X \mid Y=y} (x)}{P_{X \mid Y=y}(x')} \leq e^\epsilon.
\end{equation*}
\end{definition}
Note that the notion of identifiability has been previously considered in centralized settings~\cite{lee2012differential, wang2016relation}, where $x$ and $x'$ denote neighboring databases. Here, we have given a local version of the definition, where the ratio $\frac{P_{X \mid Y=y} (x)}{P_{X \mid Y=y}(x')}$ must be bounded by $e^\epsilon$ for all $x,x' \in \mathrm{supp}(P_X)$.

What the above three notions of privacy have in common is that they are strictly intolerant of zero-probability assignments in the channel $P_{Y \mid X}$, that is, the existence of a single input-output pair $(x,y)$ such that $P_{Y \mid X=x}(y) = 0$ (where $x \in \mathrm{supp}(P_X)$ and $y \in \mathrm{supp}(P_Y)$) immediately implies that the channel does not satisfy any of the above notions of privacy. On the other hand, $\ell(X \to y)$ is always bounded by $\epsilon_\mathrm{max}$, so naturally, a guarantee given in terms of pointwise maximal leakage may not satisfy any of the above notions. 

To see why zero-probability assignments in the channel $P_{Y \mid X}$ do not necessarily imply \say{bad privacy}, consider the following simple example. Suppose $X$ and $Y$ are random variables defined on sets with cardinality $n$ and assume that $X$ is uniformly distributed. Consider the following channel:
\begin{align*}
    P_{Y \mid X=x_1} (y_i)= \begin{cases}
    0, & \text{if} \; i=1,\\
    \frac{1}{n-1} & \; \text{otherwise},
    \end{cases}
\end{align*}
and $P_{Y \mid X=x_j}(y_i) = \frac{1}{n}$ with $i \in \{1, \ldots, n\}$ and $j \in \{2, \ldots, n\}$. Intuitively, if $n$ is larger, then $P_{Y \mid X}$ leaks very little information which also becomes apparent by calculating the leakage:
\begin{gather*}
    \ell(X \to y_1) = \log \frac{n}{n-1},\\
    \ell(X \to y_i) = \log \frac{n^2}{n^2-n+1}, \quad i=2, \ldots, n.
\end{gather*}
However, under LDP/LIP/LDI, no matter how large $n$ is, the above channel is considered to be equally non-private as a deterministic mapping from $X$ to $Y$. This is clearly an overly pessimistic assessment. 

That being said, it is straightforward to verify that $\epsilon$-LDP/LIP/LDI guarantees imply a guarantee based on pointwise maximal leakage. According to Definition~\ref{def:lip}, $\epsilon$-LIP implies $\epsilon$-PML. We also have the following relations. 

\begin{prop}
\label{prop:lip_ldp_ldi}
Given an arbitrary but fixed prior $P_X$, let $p_\mathrm{min} \coloneqq \min_{x \in \mathrm{supp}(P_X)} P_X(x)$. If $P_{Y \mid X}$ satisfies $\epsilon$-LDP, then it also satisfies $\epsilon'$-PML with
\begin{equation}
\label{eq:ldp_pml_translation}
    \epsilon' \geq - \log \Bigg(p_\mathrm{min} + e^{-\epsilon} (1 - p_\mathrm{min}) \Bigg),
\end{equation}
Moreover, if $P_{Y \mid X}$ satisfies $\epsilon$-LDI, then it also satisfies $\tilde \epsilon$-PML with
\begin{align*}
    \tilde \epsilon \geq - \log \Bigg(p_{\mathrm{min}} \big(1 + e^{-\epsilon}(\abs{\mathrm{supp}(P_X)} - 1) \big) \Bigg).
\end{align*}
\end{prop}
\begin{IEEEproof}
The first statement regarding LDP is an intermediate step in the proof of~\cite[Thm. 2]{jiang2021context}. To see the second statement, note that $P_{Y \mid X}$ satisfies $\epsilon$-LDI if 
\begin{equation*}
    \frac{P_{Y \mid X=x}(y) P_X(x)}{P_{Y \mid X=x'}(y) P_X(x')} \leq e^\epsilon,
\end{equation*}
for all $y \in \mathrm{supp}(P_Y)$, $x,x' \in \mathrm{supp}(P_X)$. Fix $y \in \mathrm{supp}(P_Y)$. We write 
\begin{align*}
    &\max_{x \in \mathrm{supp}(P_X)} \frac{P_{Y \mid X=x}(y)}{P_Y(y)}\\
    &= \max_{x \in \mathrm{supp}(P_X)} \frac{P_{Y \mid X=x}(y)}{\sum_{x'} P_{Y \mid X=x}(y) P_X(x')}\\
    &= \max_{x \in \mathrm{supp}(P_X)} \frac{P_{Y \mid X=x}(y)}{P_{Y \mid X=x}(y) P_X(x) + \sum\limits_{x' \neq x} P_{Y \mid X=x'}(y) P_X(x')}\\
    &\leq \max_{x \in \mathrm{supp}(P_X)} \!\frac{P_{Y \mid X=x}(y)}{P_{Y \mid X=x}(y) P_X(x) + \!\!\sum\limits_{x' \neq x} P_{Y \mid X=x}(y) P_X(x) e^{-\epsilon}}\\
    &= \max_{x \in \mathrm{supp}(P_X)} \frac{1}{P_X(x) 
    \big(1 + e^{-\epsilon}(\abs{\mathrm{supp}(P_X)} - 1) \big)}\\
    &= \frac{1}{p_{\mathrm{min}} \big(1 + e^{-\epsilon}(\abs{\mathrm{supp}(P_X)} - 1) \big)}.
\end{align*}
\end{IEEEproof}

By Proposition~\ref{prop:lip_ldp_ldi}, mechanisms satisfying $\epsilon$-LDP can be deployed to achieve $\epsilon$-PML. Therefore, given an arbitrary prior, the class of $\epsilon$-LDP mechanisms is a subset of the class of $\epsilon$-PML mechanisms, which in turn, implies that PML can be used to achieve better privacy-utility tradeoffs compared to LDP. Below, we study the \emph{randomized response} (RR) mechanism~\cite{warner1965randomized,kairouz2014extremal} which is one of the most common implementations of LDP. Interestingly, the RR mechanism attains the translation bound of~\eqref{eq:ldp_pml_translation} between LDP and PML. 

\begin{example}[Randomized response~\cite{warner1965randomized,kairouz2014extremal}]
\label{ex:randomized_response}
Suppose $\mathcal X = \mathcal Y = [n]$. Given $\epsilon > 0$, the RR mechanism is expressed as  
\begin{equation*}
    P_{Y \mid X=x}(y) = \begin{cases}
    {\displaystyle \frac{e^\epsilon}{n-1 + e^\epsilon}} & \mathrm{if} \; x=y,\\[0.5em]
    {\displaystyle \frac{1}{n-1 + e^\epsilon}} & \mathrm{if} \; x\neq y.
    \end{cases}
\end{equation*}
The RR mechanism satisfies $\epsilon$-LDP and $\epsilon'$-PML with $\epsilon' = - \log \Big(p_\mathrm{min} + e^{-\epsilon} (1 - p_\mathrm{min}) \Big)$. In~\cite{kairouz2014extremal}, \citeauthor{kairouz2014extremal} show that the RR mechanism is optimal in the low privacy regime for a broad class of utility measures including mutual information. The RR mechanism is, however, sub-optimal when the considered notion of privacy is PML. To demonstrate this, we compare the mutual information achieved by the RR mechanism with the mutual information of an optimized PML mechanism calculated numerically by implementing the following constrained non-convex program: 
\begin{gather*}
    \underset{P_{Y \mid X}}{\text{maximize}} \quad I(X ; Y) = H(X) - H(X \mid Y),\\
    \hspace{1em}\text{subject to} \quad \frac{\max_x P_{Y \mid X=x}(y)}{P_Y(y)} \leq e^\epsilon, \quad \forall y \in \mathcal Y,
\end{gather*}
where $H(\cdot)$ denotes Shannon entropy. Numerical results are depicted in \figurename~\ref{fig:mi_rr_vs_opt} in which the achievable mutual information is plotted against $\epsilon$ when $\abs{\mathcal X} = 3$ with priors $P_X = [0.05 \; 0.05 \; 0.9]$, $P_X = [0.1 \; 0.3 \; 0.6]$, and $P_X = [\frac{1}{3} \; \frac{1}{3} \; \frac{1}{3}]$. Note that for each $P_X$, we are interested in the region $\epsilon \leq \epsilon_{\mathrm{max}} = -\log p_\mathrm{min}$, which is marked by the vertical dashed line. If $\epsilon >  \epsilon_{\mathrm{max}}$, then $Y$ can be a deterministic function of $X$ and $I(X;Y) = H(X)$ trivially. As expected, the optimized PML mechanism achieves higher mutual information compared to the RR mechanism even in the low privacy regime where RR is optimal under LDP. 

\begin{figure*}
    \centering
    \subfloat[{$P_X = [0.05 \;\, 0.05 \;\, 0.9]$}]{\includegraphics[scale=0.15]{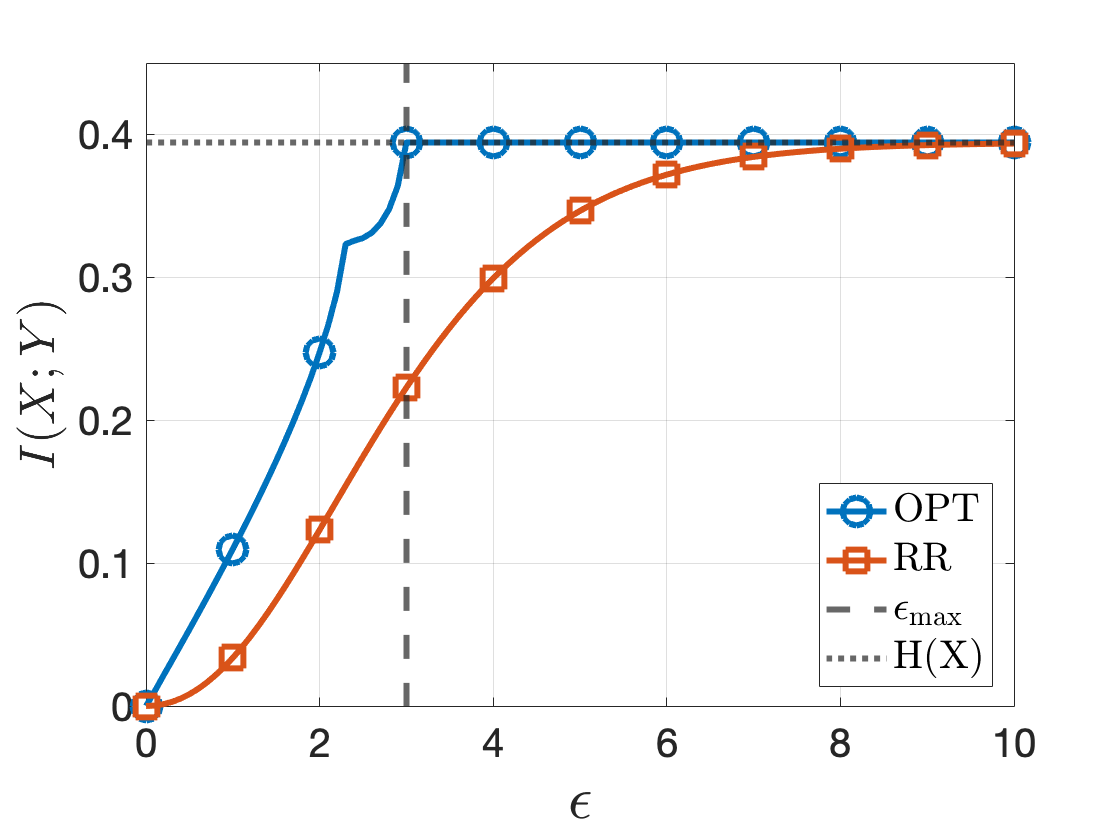}}
    \hfil
    \subfloat[{$P_X = [0.1 \;\, 0.3 \;\, 0.6]$}]{\includegraphics[scale=0.15]{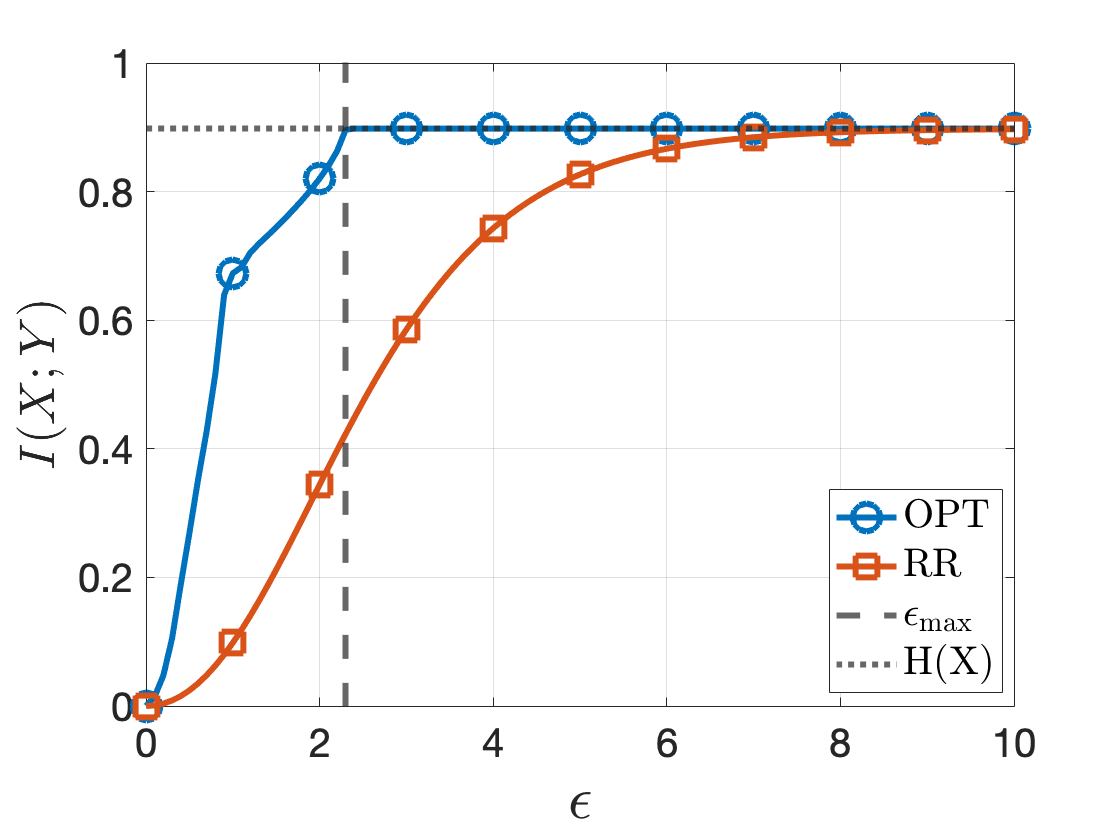}}
    \hfil
    \subfloat[{$P_X = [\frac{1}{3} \;\, \frac{1}{3}  \;\, \frac{1}{3}]$}]{\includegraphics[scale=0.15]{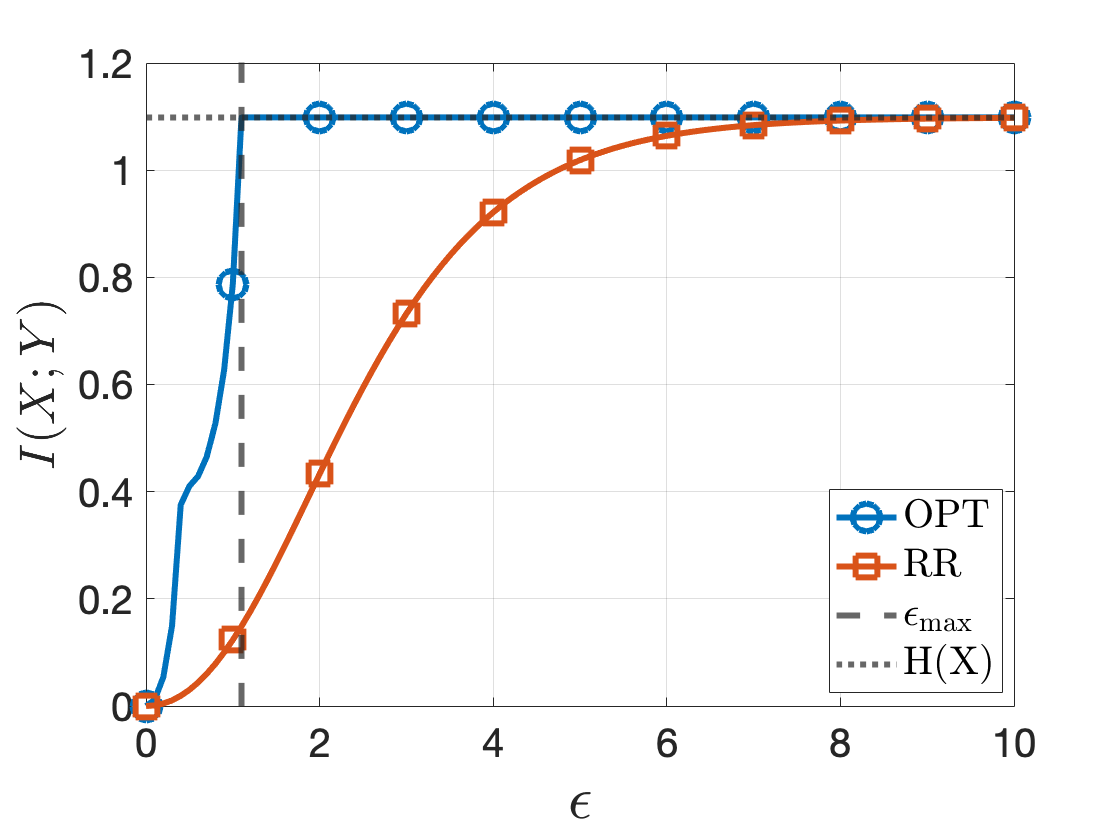}}
    \caption{Comparing the mutual information (in nats) of the RR mechanism and a locally optimal mechanism found by implementing a constrained non-linear program. The RR mechanism is strictly sub-optimal in all three cases.}
    \label{fig:mi_rr_vs_opt}
\end{figure*}
\end{example}

\subsection{Mutual Information}
Mutual information has been studied as a privacy measure in a number of works, e.g.,~\cite{asoodeh2016information,asoodeh2015maximal, wang2016relation, makhdoumi2014information}, even though it has been argued that in some cases mutual information underestimates the information leakage~\cite{issa2019operational}. Previous works have shown that maximal leakage upper bounds mutual information, and in fact, no scalar multiple of mutual information can upper bound maximal leakage~\cite{alvim2012measuring,issa2019operational}. Here, we show how mutual information can be bounded in terms of $\ell(X \to Y)$. 
\begin{prop}
\label{prop:mutual_info}
Given a joint distribution $P_{XY}$, it holds that $I(X ; Y) \leq \mathbb E_{Y \sim P_Y}[\ell(X \to Y)]$ with equality if and only if $P_{Y \mid X=x}(y) = P_{Y \mid X=x'}(y)$ for all $x,x' \in \mathcal{X}$ and $y \in \mathcal{Y}$ such that $P_{XY}(x,y) > 0$ and $P_{XY}(x',y) > 0$. Hence, given an arbitrary but fixed prior $P_X$,
\begin{enumerate}
    \item if $P_{Y \mid X}$ satisfies $\epsilon$-PML, then $I(X;Y) \leq \epsilon$, and \label{itm:mutual_info_1}
    \item if $P_{Y \mid X}$ satisfies $(\epsilon, \delta)$-PML, then $I(X;Y) \leq \epsilon + \delta \cdot \epsilon_\mathrm{max}$.\label{itm:mutual_info_2}
\end{enumerate}
\end{prop}

\begin{IEEEproof}
\begin{align*}
    I(X;Y) &= \mathbb E_{Y \sim P_Y}\left[\mathbb E_{X \sim P_{X \mid Y}}\left[\log \frac{P_{XY}(X,Y)}{P_X(X) P_Y(Y)} \right] \right]\\
    &\leq \mathbb E_{Y \sim P_Y}\left[\max_{x \in \mathrm{supp}(P_X)} \log \frac{P_{XY}(x,Y)}{P_X(x) P_{Y}(Y)} \right]\\
    &= \mathbb E_{Y \sim P_Y}[\ell(X \to Y)],
\end{align*}
where the inequality holds with equality if and only if for all $x,x',y$ such that $P_{XY}(x,y) > 0$ and $P_{XY}(x',y) > 0$ we have $i(x ; y) = i(x' ; y)$, or equivalently, $P_{Y \mid X=x}(y) = P_{Y \mid X=x'}(y)$ (the condition for equality has also been noted in~\cite[Lemma 2]{issa2019operational}). The remaining statements then follow directly from the definitions of $\epsilon$-PML and $(\epsilon, \delta)$-PML, and the above inequality. 
\end{IEEEproof}

\subsection{\texorpdfstring{$f$}{f}-information}
$f$-information~\cite{diaz2019robustness} refers to a class of information measures that are defined based on $f$-divergences~\cite{csiszar1967information}. First, we recall the definition of $f$-divergence. 
\begin{definition}[$f$-divergence] 
Let $f : (0, \infty) \to \mathbb R$ be a convex function satisfying $f(1) = 0$. Let $P$ and $Q$ be probability distributions defined on a finite set $\Omega$, and suppose $P \ll Q$. The $f$-divergence between $P$ and $Q$ is defined as 
\begin{equation*}
    D_f(P \Vert Q) \coloneqq \mathbb{E}_Q \left[ f \left(\frac{P}{Q}\right) \right] = \sum_{\omega \in \Omega} Q(\omega) f\left(\frac{P(\omega)}{Q(\omega)}\right).
\end{equation*}
\end{definition}
Common $f$-divergences include KL-divergence, total variation distance, and $\chi^2$-divergence. Now, we can define $f$-information as the $f$-divergence between the joint distribution and the product of the marginals of two random variables. 
\begin{definition}[{$f$-information~\cite[Def. 7]{diaz2019robustness}}]
Let $f : (0, \infty) \to \mathbb R$ be a convex function satisfying $f(1) = 0$. Given a joint distribution $P_{VW}$ on the finite set $\mathcal V \times \mathcal W$, the $f$-information of $P_{VW}$ is defined as 
\begin{align*}
    I_f(P_{VW}) &\coloneqq D_f(P_{VW} \Vert P_V P_W)\\
    &= \mathbb E_{(V,W) \sim P_V P_W} \left[ f\left(\frac{P_{VW}(V,W)}{P_V(V) P_W(W)}\right) \right].
\end{align*}
\end{definition}
\citet{diaz2019robustness} justify the definition of $f$-information as an information/privacy measure on account of its operational meaning for specific choices of the function $f$, such as mutual information (associated with KL-divergence) and $\chi^2$-information~\cite{du2017principal, wang2017estimation} (associated with $\chi^2$-divergence). Below, we show how PML can be used to upper bound $f$-information.
\begin{prop}
\label{prop:f_info_bound}
 Let $f : (0, \infty) \to \mathbb R$ be a convex function satisfying $f(1) = 0$, and suppose $\lim_{t \to 0^+} f(t) < \infty$. Then,
 \begin{equation*}
     I_f(P_{XY}) \leq \mathbb E_{Y \sim P_Y} \Big[\max \left\{f\left(\exp(\ell(X \to Y))\right), f(0) \right\} \Big],
 \end{equation*}
 where $f(0)$ is defined by continuity as $f(0) \coloneqq f(0^+)$. 
\end{prop}
\begin{IEEEproof}
\begin{subequations}
\begin{align}
    &I_f(P_{XY}) = \mathbb E_{Y \sim P_Y} \left[ \mathbb E_{X \sim P_X} \left[ f\left(\frac{P_{XY}(X,Y)}{P_X(X) P_Y(Y)}\right) \right] \right]\nonumber\\
    &\leq \mathbb E_{Y \sim P_Y} \left[\max_{x \in \mathrm{supp}(P_X)} f\left(\frac{P_{XY}(x,Y)}{P_X(x) P_Y(Y)}\right) \right]\nonumber\\
    &= \mathbb E\Bigg[\max \Bigg\{f\left(\max_{x \in \mathrm{supp}(P_X)} \frac{P_{XY}(x,Y)}{P_X(x) P_Y(Y)}\right),\label{eq:f_dvgc_1}\\
    &\hspace{8em} f\left(\min_{x \in \mathrm{supp}(P_X)} \frac{P_{XY}(x,Y)}{P_X(x) P_Y(Y)}\right) \Bigg\} \Bigg]\nonumber\\
    &\leq \mathbb E\left[\max \left\{f\big(\max_{x \in \mathrm{supp}(P_X)} \frac{P_{XY}(x,Y)}{P_X(x) P_Y(Y)}\big), f(0) \right\} \right]\label{eq:f_dvgc_2}\\
    &= \mathbb E \Big[\max \Big\{f\Big(\exp(\ell(X \to Y))\Big), f(0) \Big\} \Big],\nonumber
\end{align}
\end{subequations}
where~\eqref{eq:f_dvgc_1} follows from the fact that the maximum of a convex function is attained at an extreme point, and~\eqref{eq:f_dvgc_2} follows from $\min_{x \in \mathrm{supp}(P_X)} \frac{P_{X \mid Y=y}(x)}{P_X(x)} \geq 0$ for all $y \in \mathrm{supp}(P_Y)$.
\end{IEEEproof}

\subsection{Total variation privacy}
Let $P$ and $Q$ be probability distributions on the finite set $\Omega$. The total variation distance between $P$ and $Q$ is defined as 
\begin{equation*}
    \mathrm{TV}(P,Q) \coloneqq \frac{1}{2} \mathbb{E}_Q \left[ \left \lvert \frac{P}{Q} -1 \right \rvert \right] = \frac{1}{2} \sum_{\omega \in \Omega} \left \lvert P(\omega) - Q(\omega) \right \rvert,
\end{equation*}
which is an $f$-divergence with $f(x) = \frac{1}{2} \abs{x-1}$. Total variation privacy~\cite{rassouli2019optimal} is a privacy measure that is defined as the expected total variation distance between the posterior distribution $P_{X \mid Y}$ and the prior $P_X$, given by 
\begin{align}
\begin{split}
\label{eq:totalvar_privacy}
    T(X;Y) &\coloneqq \mathbb E_{Y\sim P_Y} \left[\mathrm{TV}(P_{X \mid Y}, P_X)\right] \\
    &= \sum_{y \in \mathcal{Y}} P_Y(y)\, \mathrm{TV}(P_{X \mid Y=y}, P_X).
\end{split}
\end{align}
\citet{rassouli2019optimal} motivate the use of total variation distance as a privacy measure by arguing that $T(X;Y)$ is closed under pre- and post-processing, and by showing that controlling $T(X;Y)$ restricts the inference quality of an adversary optimizing an additive gain function, described in~\cite{du2012privacy}. Furthermore, the following bound between $T(X;Y)$ and maximal leakage is derived: 
\begin{equation}
\label{eq:totalvar_stupid_bound}
    T(X;Y) \leq (\abs{\mathcal{X}} - 1) \cdot \max_{x \in \mathcal{X}} P_X(x) \cdot \left( \exp(\mathcal{L}(P_{Y \mid X})) - 1\right),
\end{equation}
which is a rather loose bound as it depends on the cardinality of $\mathcal{X}$ (considering that $0 \leq T(X;Y) \leq 1$). Note that $T(X;Y)$ is actually the $f$-information associated with the total variation distance, so by applying Proposition~\ref{prop:f_info_bound} with $f(x) = \frac{1}{2} \abs{x-1}$ we get 
\begin{equation}
\label{eq:totalvar_bound_1}
    T(X;Y) \leq \frac{1}{2} \mathbb E_{Y\sim P_Y} \Big[\max \left \{\exp(\ell(X \to Y)) - 1, \; 1  \right \} \Big],
\end{equation}
which is tighter than~\eqref{eq:totalvar_stupid_bound}. In the following result, we derive another upper bound on $T(X;Y)$ in terms of maximal leakage $\mathcal L(P_{Y \mid X})$, and also show how $(\epsilon, \delta)$-PML privacy guarantee constrains $T(X;Y)$.

\begin{prop}
\label{prop:total_var_privacy}
The following relationship holds between $T(X;Y)$ and $\mathcal L(P_{Y \mid X})$:
\begin{equation*}
    T(X ; Y) \leq \exp \left(\mathcal{L}(P_{Y \mid X}) \right) - 1. 
\end{equation*}
Furthermore, given an arbitrary but fixed prior $P_X$, if the channel $P_{Y\mid X}$ satisfies $(\epsilon, \delta)$-PML, then $T(X;Y)$ is bounded as follows:
\begin{enumerate}
    \item if $\epsilon \leq \log \frac{3}{2}$, then
    \begin{equation*}
        T(X;Y) \leq e^\epsilon - 1 + \frac{\delta}{2} \, \left( e^{\epsilon_\mathrm{max}} -1 \right),
    \end{equation*}

    \item if $\log \frac{3}{2} \leq \epsilon \leq \log 2$, then
    \begin{equation*}
        T(X;Y) \leq \frac{1}{2} + \frac{\delta}{2} \, \left( e^{\epsilon_\mathrm{max}} -1 \right), 
    \end{equation*}

    \item if $\epsilon \geq \log 2$, then
    \begin{equation*}
        T(X;Y) \leq \frac{1}{2} \, \left(e^\epsilon - 1 \right) + \frac{\delta}{2} \, \left( e^{\epsilon_\mathrm{max}} -1 \right).
    \end{equation*}
\end{enumerate}
\end{prop}
\begin{IEEEproof}
Fix some $y \in \mathrm{supp}(P_Y)$ and define the set $\mathcal{A}_y \coloneqq \{x \in \mathrm{supp}(P_X) : P_{X \mid Y=y}(x) \geq P_{X}(x)\}$. We can write
\begin{align}
\label{eq:tv_bound2}
     \mathrm{TV}(P_{X \mid Y=y}, P_X) &= \frac{1}{2} \sum_{x \in \mathrm{supp}(P_X)} \left \lvert P_{X \mid Y=y}(x) - P_X(x) \right \rvert \nonumber\\
    &= \sum_{x \in \mathcal{A}_y} P_{X \mid Y=y}(x) - P_X(x) \nonumber\\
    &= \sum_{x \in \mathcal{A}_y} \left( \frac{P_{X \mid Y=y}(x)}{P_X(x)} - 1 \right) P_X(x)\nonumber\\
    &\leq \max_{x \in \mathcal{A}_y} \left( \frac{P_{X \mid Y=y}(x)}{P_X(x)} - 1 \right) \sum_{x \in \mathcal{A}_y} P_X(x) \nonumber\\
    &\leq \exp\left(\ell(X \to y)\right) - 1.
\end{align}
Taking the expectation of the above expression over $y$, we get 
\begin{equation*}
    T(X;Y) \leq \exp\left(\mathcal{L}(P_{Y\mid X}) \right) -1.
\end{equation*}
Now, define the function $\eta(y) \coloneqq \exp \left(\ell(X \to y) \right) -1$, $y \in \mathrm{supp}(P_Y)$. By~\eqref{eq:totalvar_bound_1} and~\eqref{eq:tv_bound2}, we obtain
\begin{equation*}
     T(X;Y) \leq \mathbb E_{Y\sim P_Y} \left[ \min \left\{\eta(Y), \max \left\{\frac{1}{2}\eta(Y), \frac{1}{2} \right\} \right\}\right].
\end{equation*}

Suppose $P_X$ is arbitrary but fixed, and the mechanism $P_{Y \mid X}$ satisfies $(\epsilon, \delta)$-PML. Define $\eta_\mathrm{max} \coloneqq e^{\epsilon_\mathrm{max}} -1$. Using the fact that $\epsilon_\mathrm{max} \geq \log 2$, we conclude that with probability smaller than $\delta$ over $Y$, we have 
\begin{equation}
\label{eq:tv_bad_prob}
    \mathrm{TV}(P_{X \mid Y=y}, P_X) \leq \frac{1}{2} \, \eta_\mathrm{max} = \frac{1}{2} \, \left( e^{\epsilon_\mathrm{max}} -1 \right).
\end{equation}
We need to consider the following three cases for $\epsilon$:
\begin{enumerate}
    \item $\epsilon \leq \log \frac{3}{2}$: With probability at least $1-\delta$ we have $\mathrm{TV}(P_{X \mid Y=y}, P_X) \leq \eta(y) \leq e^\epsilon - 1$, which implies that 
    \begin{equation*}
        T(X;Y) \leq e^\epsilon - 1 + \frac{\delta}{2} \, \left( e^{\epsilon_\mathrm{max}} -1 \right). 
    \end{equation*}
    \item $\log \frac{3}{2} \leq \epsilon \leq \log 2$: With probability at least $1-\delta$ we have $\mathrm{TV}(P_{X \mid Y=y}, P_X) \leq \frac{1}{2}$, which implies that 
    \begin{equation*}
        T(X;Y) \leq \frac{1}{2} + \frac{\delta}{2} \, \left( e^{\epsilon_\mathrm{max}} -1 \right). 
    \end{equation*}
    \item $\epsilon \geq \log 2$: With probability at least $1-\delta$ we have $\mathrm{TV}(P_{X \mid Y=y}, P_X) \leq \frac{1}{2} \, \eta(y) \leq \frac{1}{2} \, \left(e^\epsilon - 1 \right)$, which implies that 
    \begin{equation*}
        T(X;Y) \leq \frac{1}{2} \, \left(e^\epsilon - 1 \right) + \frac{\delta}{2} \, \left( e^{\epsilon_\mathrm{max}} -1 \right). 
    \end{equation*}
\end{enumerate}
\end{IEEEproof}
\section{Conclusions}
\label{sec:conclusions}

In this paper, we have introduced a new privacy measure called pointwise maximal leakage extending the pre-existing notion of maximal leakage, which quantifies the amount of information leaking about a secret $X$ by disclosing a single outcome of a (randomized) function calculated on $X$. Our results demonstrate that a framework centered around PML can be used to reason about privacy in a wide range of problems: First, we make no assumptions about the nature of the sensitive data, e.g., $X$ can represent an entire database or a single data point collected from an individual. Then, PML is operationally meaningful in the sense that it is obtained by analyzing threat models in which all assumptions about adversaries are made explicit. Moreover, PML is a robust measure of privacy since it is meaningful when considering any adversary whose objective can be described by a gain function. In addition, PML satisfies useful properties, for example, it behaves well under composition, enables us to model side information, and satisfies data-processing (i.e., pre- and post-processing) inequalities. Last but not least, pointwise maximal leakage allows us to view privacy leakage as a random variable; consequently, we have the freedom to define different types of privacy guarantees based on the particular requirements of different applications, both in terms of privacy and utility.

\appendices
\section{Proof of Theorem~\ref{thm:gain_u_equivalence}}
\label{sec:eqvlnc_thm_proof}
Suppose without loss of generality that $P_X$ has full support. Given an arbitrary gain function $g$, the $g$-leakage of $X$ can be written as 
\begin{align}
\label{eq:g_representation}
    &\ell_g(X \to y) = \log \frac{\sup_{P_{\hat{X} \mid Y}} \mathbb{E} \left[g(X,\hat{X}) \mid Y=y \right]}{\max_{\hat{x} \in \hat{\mathcal X}} \mathbb E\left[g(X, \hat{x})\right]}\nonumber\\[0.3em]
    &= \log \frac{\sup\limits_{P_{\hat{X} \mid Y}} \sum\limits_{x \in \mathcal X} \sum\limits_{\hat{x} \in \hat{\mathcal{X}}} g(x,\hat{x}) P_{X \mid Y=y}(x) P_{\hat{X}\mid Y=y}(\hat{x})}{\max_{\hat{x} \in \hat{\mathcal{X}}} \sum_{x \in \mathcal X} g(x,\hat{x}) P_{X}(x)}\nonumber\\[0.3em]
    &= \log \frac{\max_{\hat{x} \in \hat{\mathcal{X}}} \sum_{x \in \mathcal X} g(x,\hat{x}) P_{X \mid Y=y}(x)}{\max_{\hat{x} \in \hat{\mathcal{X}}} \sum_{x \in \mathcal X} g(x,\hat{x}) P_{X}(x)},
\end{align}
where the last equality follows by plugging in $P^*_{\hat{X} \mid Y}$ satisfying
\begin{align*}
    &P^*_{\hat{X} \mid Y=y}(\hat{x}) \coloneqq\\
    &\hspace{2em} \begin{cases}
    1 & \mathrm{for\;some} \; \hat{x} \in \argmax\limits_{\hat{x}}  \sum\limits_{x \in \mathcal X} g(x,\hat{x}) P_{X \mid Y=y}(x) \\
    0 & \mathrm{otherwise},
    \end{cases}
\end{align*}
in the numerator. Furthermore, as shown in the proof of Theorem~\ref{thm:privacy_leakage_rv}, given a randomized function of $X$ denoted by $U$, the $U$-leakage of $X$ can be expressed as 
\begin{align}
\label{eq:u_representation}
    &\ell_U(X \to y) = \log \frac{\max_{u \in \mathcal U} P_{U \mid Y=y}(u)}{\max_{u} P_U(u)} \nonumber\\[0.5em]
    &= \log \frac{\max_{u \in \mathcal U} \sum_{x \in \mathcal X} P_{U \mid X=x}(u) \; P_{X \mid Y=y}(x)}{\max_{u \in \mathcal{U}} \sum_{x \in \mathcal X} P_{U \mid X=x}(u) \; P_X(x)} .
\end{align}

To show the equivalence, first, we prove the simpler direction by showing that each $U$-leakage can be written as a $g$-leakage. Given an arbitrary randomized function of $X$ denoted by $U$, define $\hat{\mathcal{X}}_{_U} \coloneqq \mathcal U$ such that each $u \in \mathcal U$ corresponds uniquely to some $\hat{x}_u \in \hat{\mathcal{X}}_{_U}$, and let $g_{_U}(x, \hat{x}_u) \coloneqq P_{U \mid X=x} (u)$ for all $x \in \mathcal{X}$ and $u \in \mathcal{U}$. By computing expressions~\eqref{eq:g_representation} and~\eqref{eq:u_representation}, it is easy to see that $\ell_U(X \to y)  = \ell_{g_{_U}} (X \to y)$. This construction implies that a randomized function of $X$ is simply a gain function that satisfies $\sum_{\hat{x}} g_{_U}(x, \hat{x}) = 1$, for all $x \in \mathcal X$, that is, the total gain associated with each secret $x \in \mathcal{X}$ is a constant. 

Now, we show that each $g$-leakage can be written as a $U$-leakage. Fix an arbitrary gain function $g$. Without loss of generality, suppose $g(x, \hat{x}) \leq 1$ for all $x \in \mathcal {X}$ and $\hat{x} \in \hat{\mathcal X}$ (this can be achieved by normalizing the gain function by $\max_{x , \hat{x}} g(x, \hat{x})$). In what follows, we construct a randomized function of $X$ using a channel that generalizes the shattering channel of Definition~\ref{def:shattering}. We need to consider the following two cases:

\textbf{Case 1: The same $\hat{x}$ maximizes the numerator and the denominator in~\eqref{eq:g_representation}}.\\
Here, we will construct a randomized function of $X$, denoted by $V$, which is described by the kernel $P_{V \mid X}$ and satisfies $\ell_V(X \to y) = \ell_g(X \to y)$. Let $\hat{x}_V \in \argmax_{\hat{x}} \sum_x g(x, \hat{x}) P_{X \mid Y=y}(x)$ which (following from the definition of Case 1) also satisfies $\hat{x}_V \in \argmax_{\hat{x}} \sum_x g(x,\hat{x}) P_X(x)$. Informally, $\hat{x}_V$ denotes the adversary's best guess both after observing $Y=y$, and with no observations. 

Define the set $\mathcal{X}_V \coloneqq \{x \in \mathcal X \colon g(x, \hat{x}_V) > 0\}$. For now, let us assume that $\mathcal{X}_V = \mathcal{X}$ but later we will discuss how the proof can be adapted if $\mathcal{X}_V$ is a proper subset of $\mathcal{X}$. For each $x \in \mathcal{X}_V$, let $k_V(x) \coloneqq 1/{g(x, \hat{x}_V)}$, and define $k_V \coloneqq \max_{x \in \mathcal{X}_V} k_V(x)$. Roughly speaking, $k_V(x)$ is the cardinality of the support of $P_{V \mid X=x}$ while $k_V$ is the cardinality of the support of $P_V$. 

Now, we construct a random variable $V$ taking values in an alphabet $\mathcal V$ such that $\abs{\mathcal V} = \ceil{k_V}$. For all $x \in \mathcal{X}_V$, the kernel $P_{V \mid X=x}$ is defined by
\begin{align*}
    &P_{V \mid X=x}(v_i)\coloneqq\\
    &\hspace{1em} \begin{cases}
    g(x, \hat{x}_V) & \text{if}\; 1 \leq i \leq \floor{k_V(x)},\\
    1 - \floor{k_V(x)} \, g(x, \hat{x}_V) & \text{if}\; i = \ceil{k_V(x)},\\
    0 &  \text{if}\; \ceil{k_V(x)} + 1 \leq i \leq \ceil{k_V}.
    \end{cases}
\end{align*}
Informally, for each $x$, the above kernel allocates chunks of probability equal to $g(x, \hat{x}_V)$ to the first $\floor{k_V(x)}$ letters, and the $\ceil{k_V(x)}$-th letter is used to contain the remaining probability $1 - \floor{k_V(x)} \, g(x, \hat{x}_V)$. Note that the above kernel indeed satisfies
\begin{equation*}
    \sum_{i=1}^{\ceil{k_V}} P_{V \mid X=x}(v_i) = 1,
\end{equation*}
for all $x \in \mathcal{X}_V$. 

Renaming $V$ to $U_g$, we verify that the random variable constructed above satisfies $\ell_{U_g}(X \to y) = \ell_g(X \to y)$:
\begin{align*}
    &\max_{u \in \mathcal{U}_g} \sum_x P_{U_g \mid X=x}(u) P_{X \mid Y=y}(x) = \sum_{x} g(x, \hat{x}_{U_g}) P_{X \mid Y=y}(x)\\
    &= \max_{\hat{x}} \sum_{x} g(x, \hat{x}) P_{X \mid Y=y}(x), 
\end{align*}
where the last equality follows from the definition of $\hat{x}_{U_g}$. Similarly, we have 
\begin{align*}
    \max_{u \in \mathcal{U}_g} \sum_x P_{U_g \mid X=x}(u) P_{X}(x) &= \sum_{x} g(x, \hat{x}_{U_g}) P_{X}(x)\\
    &= \max_{\hat{x}} \sum_{x} g(x, \hat{x}) P_{X}(x).
\end{align*}

\textbf{Case 2: The maximizing $\hat{x}$'s in the numerator and the denominator of~\eqref{eq:g_representation} are different.}\\
We will construct two randomized functions of $X$ denoted by $V$ and $W$, one for the numerator of~\eqref{eq:g_representation} and one for the denominator. Let
\begin{gather*}
    \hat{x}_V \in \argmax_{\hat{x}} \sum_x g(x, \hat{x}) P_{X \mid Y=y}(x),\\
    \hat{x}_W \in \argmax_{\hat{x}} \sum_x g(x,\hat{x}) P_X(x),
\end{gather*}
and define
\begin{gather*}
    \mathcal{X}_V \coloneqq \{x \in \mathcal X \colon g(x, \hat{x}_V) > 0\},\\
    \mathcal{X}_W \coloneqq \{x \in \mathcal X \colon g(x, \hat{x}_W) > 0\},
\end{gather*}
where $\hat{x}_V$ denotes the adversary's best guess having observed $Y=y$, and $\hat{x}_W$ denotes the adversary's best guess without an observation. We need to consider the following two cases:

\textbf{Case 2.1: $\mathcal{X}_V = \mathcal{X}_W$}. \\
Let $\mathcal{X}_{U_g} \coloneqq \mathcal{X}_V = \mathcal{X}_W$. Once again, let us assume that $\mathcal{X}_{U_g} = \mathcal{X}$. Similarly to what we had in Case 1, for all $x \in \mathcal{X}_{U_g}$ we define
\begin{gather*}
    k_V(x) \coloneqq 1/{g(x, \hat{x}_V)}, \qquad k_V \coloneqq \max_{x \in \mathcal{X}_{U_g}} k_V(x),\\
    k_W(x) \coloneqq 1/{g(x, \hat{x}_W)}, \qquad k_W \coloneqq \max_{x \in \mathcal{X}_{U_g}} k_W(x).
\end{gather*}
Let $\mathcal{V}$ denote the support set of random variable $V$, and $\mathcal{W}$ denote the support set of random variable $W$, where $\abs{\mathcal V} = \ceil{k_V}$ and $\abs{\mathcal W} = \ceil{k_W}$. For all $x \in \mathcal{X}_{U_g}$, the kernels $P_{V \mid X=x}$ and $P_{W \mid X=x}$ are defined as
\begin{align*}
    &P_{V \mid X=x}(v_i) \coloneqq\\
    &\hspace{1em}\begin{cases}
    g(x, \hat{x}_V) & \text{if}\; 1 \leq i \leq \floor{k_V(x)},\\
    1 - \floor{k_V(x)} \, g(x, \hat{x}_V) & \text{if}\; i = \ceil{k_V(x)},\\
    0 & \text{if}\; \ceil{k_V(x)} + 1 \leq i \leq \ceil{k_V},
    \end{cases}
\end{align*}
and
\begin{align*}
    &P_{W \mid X=x}(w_j) \coloneqq\\
    &\hspace{1em}\begin{cases}
    g(x, \hat{x}_W) & \text{if} \; 1 \leq j \leq \floor{k_W(x)},\\
    1 - \floor{k_W(x)} \, g(x, \hat{x}_W) & \text{if}\; j = \ceil{k_W(x)},\\
    0 & \text{if}\; \ceil{k_W(x)} + 1 \leq j \leq \ceil{k_W}.
    \end{cases}
\end{align*}
Finally, we define the random variable $U_g$ as the Bernoulli mixture of $V$ and $W$. Let $\mathcal{U}_g \coloneqq \mathcal{V} \cup \mathcal{W}$ denote the alphabet of $U_g$. For all $x \in \mathcal{X}_{U_g}$, we define $P_{U_g \mid X=x}(u) \coloneqq \frac{1}{2} P_{V \mid X=x}(u) + \frac{1}{2} P_{W \mid X=x}(u)$, where $P_{V \mid X=x}(u) = 0$ for $u \in \mathcal{W}$ and $P_{W \mid X=x}(u) = 0$ for $u \in \mathcal{V}$.\footnote{This is a slight abuse of notation. Strictly speaking, $P_{V \mid X=x}(u)$ is defined only for $u \in  \mathcal V$ and $P_{W \mid X=x}(u)$ is defined only for $u \in \mathcal W$.} Let us verify that $U_g$ satisfies $\ell_{U_g}(X \to y) = \ell_g(X \to y)$:  
\begin{align*}
    &\max_{u \in \mathcal{U}_g} \sum_x P_{U_g \mid X=x}(u) P_{X \mid Y=y}(x) = \sum_{x} \frac{1}{2} \, g(x, \hat{x}_{V}) P_{X \mid Y=y}(x)\\
    &= \frac{1}{2} \max_{\hat{x}} \sum_{x} g(x, \hat{x}) P_{X \mid Y=y}(x), 
\end{align*}
and also, 
\begin{align*}
    \max_{u \in \mathcal{U}_g} \sum_x P_{U_g \mid X=x}(u) P_{X}(x) &= \sum_{x} \frac{1}{2} \, g(x, \hat{x}_{W}) P_{X}(x)\\
    &= \frac{1}{2} \max_{\hat{x}} \sum_{x} g(x, \hat{x}) P_{X}(x).
\end{align*}
Thus, we have
\begin{align*}
    \ell_{U_g}(X \to y) &= \log \frac{\max_{u \in \mathcal{U}_g} \sum_x P_{U_g \mid X=x}(u) P_{X \mid Y=y}(x)}{\max_{u \in \mathcal{U}_g} \sum_x P_{U_g \mid X=x}(u) P_{X}(x)}\\[0.5em]
    &= \log \frac{\frac{1}{2} \max_{\hat{x}} \sum_{x} g(x, \hat{x}) P_{X \mid Y=y}(x)}{\frac{1}{2} \max_{\hat{x}} \sum_{x} g(x, \hat{x}) P_{X}(x)}\\[0.5em]
    &= \ell_g(X \to y).
\end{align*}
\begin{figure*}
    \centering
    \includegraphics[scale=0.128]{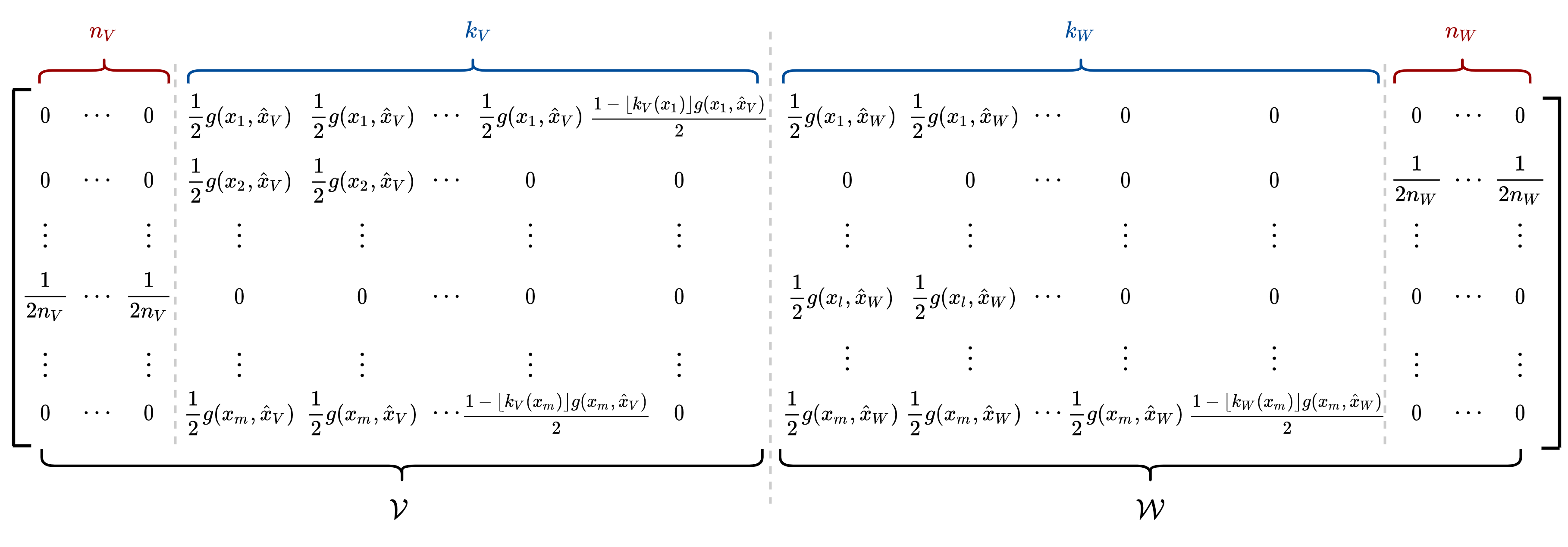}
    \caption{Illustration of the channel $P_{U_g \mid X}$ constructed in Case 2.2. We assume $\mathcal{X} = \mathcal{X}_V \cup \mathcal{X}_W$, where $\mathcal{X}$ contains $m$ elements. Each row of the above matrix corresponds to an $x$ from $\mathcal{X}$, and the columns correspond to letters from $\mathcal{U}_g$. In this example, we have $k_V(x_1) = k_V$ and $k_W(x_m) = k_W$. Also, we are assuming $x_2 \in \mathcal{X}_V \setminus \mathcal{X}_W$ and $x_l \in \mathcal{X}_W \setminus \mathcal{X}_V$, where $2 < l < m$.}
    \label{fig:shattering}
\end{figure*}

\textbf{Case 2.2: $\mathcal{X}_V \neq \mathcal{X}_W$}.\\
Let $n_V$ and $n_W$ be positive integers. Here, the idea is that we increase the sizes of the sets $\mathcal V$ and $\mathcal W$ by $n_V$ and $n_W$, respectively, where these extra letters are used to support those $x$'s for which we either have $g(x, \hat{x}_V) = 0$ or $g(x, \hat{x}_W) = 0$. We need to distinguish between three types of $x$'s: 
\begin{enumerate}
\item For $x \in \mathcal{X}_V \cap \mathcal{X}_W$ we define
\begin{align*}
    &P_{V \mid X=x}(v_i)\coloneqq \\
    &\hspace{1em}\begin{cases}
    g(x, \hat{x}_V) \hspace{8em} \text{if}\; 1 \leq i \leq \floor{k_V(x)},\\
    1 - \floor{k_V(x)} \, g(x, \hat{x}_V) \hspace{4.5em} \text{if}\; i = \ceil{k_V(x)},\\
    0 \hspace{5.2em} \text{if}\; \ceil{k_V(x)} + 1 \leq i \leq \ceil{k_V} + n_V,
\end{cases}
\end{align*}
and
\begin{align*}
    &P_{W \mid X=x}(w_j) \coloneqq\\
    &\hspace{1em}\begin{cases}
    g(x, \hat{x}_W) \hspace{7.8em} \text{if}\; 1 \leq j \leq \floor{k_W(x)},\\
    1 - \floor{k_W(x)} \, g(x, \hat{x}_W) \hspace{4em} \text{if}\; j = \ceil{k_W(x)},\\
    0 \hspace{4.7em} \text{if}\; \ceil{k_W(x)} + 1 \leq j \leq \ceil{k_W} + n_W.
    \end{cases}
\end{align*}
\item For $x \in \mathcal{X}_V \setminus \mathcal{X}_W$ we let
\begin{align*}
    &P_{V \mid X=x}(v_i) \coloneqq \\
    &\hspace{1em}\begin{cases}
    g(x, \hat{x}_V) \hspace{8em} \text{if}\; 1 \leq i \leq \floor{k_V(x)},\\
    1 - \floor{k_V(x)} \, g(x, \hat{x}_V) \hspace{4.5em} \text{if}\; i = \ceil{k_V(x)},\\
    0 \hspace{5.15em} \text{if}\; \ceil{k_V(x)} + 1 \leq i \leq \ceil{k_V} + n_V,
    \end{cases}
\end{align*}
and
\begin{align*}
    &P_{W \mid X=x}(w_j) \coloneqq\\
    &\hspace{3em}\begin{cases}
    0 & \text{if}\; 1 \leq j \leq \ceil{k_W},\\
    \frac{1}{n_W} & \text{if}\; \ceil{k_W}+1 \leq j \leq \ceil{k_W}+n_W.
    \end{cases}
\end{align*}
\item For $x \in \mathcal{X}_W \setminus \mathcal{X}_V$ we let
\begin{equation*}
P_{V \mid X=x}(v_i) \coloneqq \begin{cases}
0 & \text{if}\; 1 \leq i \leq \ceil{k_V},\\
\frac{1}{n_V} & \text{if}\; \ceil{k_V} + 1 \leq i \leq \ceil{k_V} + n_V,
\end{cases}
\end{equation*}
and
\begin{align*}
    &P_{W \mid X=x}(w_j) \coloneqq \\
    &\hspace{1em}\begin{cases}
    g(x, \hat{x}_W) \hspace{7.8em} \text{if}\; 1 \leq j \leq \floor{k_W(x)},\\
    1 - \floor{k_W(x)} \, g(x, \hat{x}_W) \hspace{4em} \text{if}\; j = \ceil{k_W(x)},\\
    0 \hspace{4.7em} \text{if}\; \ceil{k_W(x)} + 1 \leq j \leq \ceil{k_W} + n_W.
    \end{cases}
\end{align*}
\end{enumerate}
Now, we define $U_g$ as before. Suppose $\mathcal{U}_g = \mathcal{V} \cup \mathcal{W}$ is the alphabet of $U_g$. For $x \in \mathcal{X}_V \cup \mathcal{X}_W$ (where we are assuming $\mathcal{X} =  \mathcal{X}_V \cup \mathcal{X}_W$), let $P_{U_g \mid X=x} (u) = \frac{1}{2} P_{V \mid X=x}(u) + \frac{1}{2} P_{W \mid X=x}(u)$, where $P_{V \mid X=x}(u) = 0$ for $u \in \mathcal{W}$ and $P_{W \mid X=x}(u) = 0$ for $u \in \mathcal{V}$. Then, we can write 
\begin{align*}
    &\max_{u \in \mathcal{U}_g} \sum_x P_{U_g \mid X=x}(u) P_{X \mid Y=y}(x) = \\
    &\max \Bigg\{  \hspace{-0.2em} \sum_{x \in \mathcal{X}_V}   \frac{g(x, \hat{x}_{V})}{2} P_{X \mid Y=y}(x), \hspace{-1em}\sum_{x \in \mathcal{X}_W \setminus \mathcal{X}_V}  \!\!\frac{1}{2 n_V} P_{X \mid Y=y}(x) \Bigg\}.
\end{align*}
By taking $n_V$ to be large enough, we can ensure that
\begin{align*}
    \sum_{x \in \mathcal{X}_W \setminus \mathcal{X}_V} \frac{1}{2 n_V} P_{X \mid Y=y}(x) \leq \sum_{x \in \mathcal{X}_V}  \frac{1}{2} \, g(x, \hat{x}_{V}) P_{X \mid Y=y}(x),
\end{align*}
which yields 
\begin{align*}
    &\max_{u \in \mathcal{U}_g} \sum_x P_{U_g \mid X=x}(u) P_{X \mid Y=y}(x)\\
    &= \sum_{x \in \mathcal{X}_V} \frac{1}{2} \, g(x, \hat{x}_{V}) P_{X \mid Y=y}(x)\\
    &= \frac{1}{2} \max_{\hat{x}} \sum_{x \in \mathcal{X}} g(x, \hat{x}) P_{X \mid Y=y}(x).
\end{align*}
Similarly, for $n_W$ large enough we have 
\begin{align*}
    &\max_{u \in \mathcal{U}_g} \sum_x P_{U_g \mid X=x}(u) P_{X}(x)\\
    &= \max \left\{ \sum_{x \in \mathcal{X}_W} \frac{1}{2} \, g(x, \hat{x}_{W}) P_{X}(x), \sum_{x \in \mathcal{X}_V \setminus \mathcal{X}_W} \frac{1}{2 n_W} P_X(x) \right\}\\
    &= \sum_{x \in \mathcal{X}_W} \frac{1}{2} \, g(x, \hat{x}_{W}) P_{X}(x)\\
    &= \frac{1}{2} \max_{\hat{x}} \sum_{x \in \mathcal{X}} g(x, \hat{x}) P_{X}(x).
\end{align*}
Hence, we conclude that $\ell_{U_g}(X \to y) = \ell_g (X \to y)$.

The only point left to discuss is regarding the case where $\mathcal{X}_V \cup \mathcal{X}_W$ is a proper subset of $\mathcal{X}$. Let $n_O$ be a positive integer. Once again, we increase the size of the alphabet $\mathcal{U}_g$ by $n_O$ letters, where these extra letters are used to support the $x$'s in $\mathcal{X} \setminus (\mathcal{X}_V \cup \mathcal{X}_W)$. Hence, we let $\mathcal{U}_g = \mathcal{V} \cup \mathcal{W} \cup \mathcal{O}$, where $\mathcal{O}$ is a finite set containing $n_O$ elements. For $x \in \mathcal{X} \setminus (\mathcal{X}_V \cup \mathcal{X}_W)$ we define the channel $P_{U_g \mid X=x}$ as 
\begin{equation*}
    P_{U_g \mid X=x}(u) = \begin{cases}
    \frac{1}{n_O} & \text{if} \; u \in \mathcal{O},\\
    0 & \text{otherwise}. 
    \end{cases}
\end{equation*}
For $x \in \mathcal{X}_V \cup \mathcal{X}_W$, we let $P_{U_g \mid X=x}(u) = 0$ when $u \in \mathcal{O}$; otherwise $P_{U_g \mid X=x}(u)$ is defined as in Case 2.2. It is straightforward to verify that for $n_O$ large enough ($\frac{1}{n_O}$ small enough), the values of the numerator and the denominator in the expression of $\ell_{U_g}(X \to y)$ remain as before, from which we may conclude $\ell_{U_g}(X \to y) = \ell_g (X \to y)$.
$\hfill \blacksquare$

\section{Proof of Lemma~\ref{lemma:properties}}
\label{sec:porperties_lemma_proof}
\begin{enumerate}
\item \underline{Upper bound}: 
\begin{align*}
    \ell(X \to y) &= \log \max_{x \in \mathrm{supp}(P_X)} \frac{P_{X \mid Y=y}(x)}{P_{X}(x)}\\
    &\leq \log \max_{x \in \mathrm{supp}(P_X)} \frac{1}{P_X(x)}\\
    &= \log \frac{1}{\min_{x \in \mathrm{supp}(P_X)} P_X(x)},
\end{align*}
where the inequality holds with equality if an only if $P_{X \mid Y=y}(x^*) = 1$ with $x^* \in \argmin_{ x \in \mathrm{supp}(P_X)} P_X(x)$.

\underline{Lower bound}: Here, the idea is that since both $P_{X \mid Y=y}$ and $P_X$ are probability distributions over $\mathrm{supp}(P_X)$, then for any fixed $y \in \mathrm{supp}(P_Y)$, there exists at least one $x \in \mathrm{supp}(P_X)$ such that $P_{X \mid Y=y}(x) \geq P_{X}(x)$. Suppose to the contrary that for all $x \in \mathcal{X}$, $P_{X \mid Y=y}(x) < P_{X}(x)$. Then, $1 = \sum_{x} P_{X \mid Y=y}(x) < \sum_{x} P_X(x) = 1$ which is a contradiction. Therefore, 
\begin{equation*}
    \ell(X \to y) = \log \max_{x \in \mathrm{supp}(P_X)} \frac{P_{X \mid Y=y}(x)}{P_{X}(x)} \geq \log 1 = 0.
\end{equation*}
The above inequality holds with equality if and only if $\max_x P_{Y \mid X=x}(y) = P_Y(y) = \sum_x P_{Y \mid X=x}(y) P_X(x)$ which holds whenever $P_{Y \mid X=x}(y) = P_{Y \mid X=x'}(y)$ for all $x,x' \in \mathrm{supp}P_X$. 

\item Both statements follow directly from the definition. 

\item Let $x^* \in \argmax_x P_{Z \mid X=x}(z)$. Then, 
\begin{align*}
    &\ell(X \to z) = \log \max_x \frac{P_{Z \mid X=x}(z)}{P_{Z}(z)}\\
    &= \log \frac{\sum_{y \in \mathrm{supp}(P_{Y\mid X=x^*})}P_{Z \mid Y=y}(z) P_{Y \mid X=x^*}(y)}{P_{Z}(z)}\\
    &\leq \log \max_{y' \in \mathrm{supp}(P_{Y \mid X=x^*})} \frac{P_{Z \mid Y=y'}(z)}{P_Z(z)} \sum_{y} P_{Y \mid X=x^*}(y)\\
    &\leq \log \max_{y' \in \mathrm{supp}(P_Y)} \frac{P_{Z \mid Y=y'}(z)}{P_Z(z)} = \ell(Y \to z),
\end{align*}
where the first inequality holds with equality if $P_{Z \mid Y=y}(z) = P_{Z \mid Y=y'}(z)$ for all $y, y' \in \mathrm{supp}(P_{Y \mid X=x^*})$, and the second inequality holds with equality if $\max_{y \in \mathrm{supp}(P_Y)} i(y;z)$ is attained at some $y \in \mathrm{supp}(P_{Y \mid X=x^*})$. 

\item 
\begin{align*}
    &\ell(X \to z) = \log \max_x \frac{P_{Z \mid X=x}(z)}{P_{Z}(z)}\\
    &= \log \max_x \frac{\sum_{y \in \mathrm{supp}(P_Y)}P_{Z \mid Y=y}(z) P_{Y \mid X=x}(y)}{\sum_{y \in \mathrm{supp}(P_Y)}P_{Z \mid Y=y}(z) P_Y(y)}\\
    &\leq \log \max_x \max_{y \in \mathrm{supp}(P_Y)} \frac{P_{Y \mid X=x}(y)}{P_Y(y)}\\
    &= \max_{y \in \mathrm{supp}(P_Y)} \ell(X \to y).
\end{align*}
Now, if $X$ and $Y$ are independent then $\ell(X \to z) = \ell(X \to y) = 0$ for all $y,z$, and the inequality holds with equality. Furthermore, if the distribution $P_{Y \mid Z=z}$ is degenerate, then $z$ is mapped uniquely to some $y_z \in \mathrm{supp}(P_Y)$. This implies that $P_{Z \mid Y=y}(z) = 0$ for $y \neq y_z$, hence, we have
\begin{align*}
    \ell(X \to z) &= \log \max_x \frac{P_{Z \mid X=x}(z)}{P_{Z}(z)}\\ 
    &= \log \max_x \frac{P_{Y \mid X=x}(y_z)}{P_{Y}(y_z)}\\ 
    &= \ell(X \to y_z)\\
    &\leq \max_{y \in \mathrm{supp}(P_Y)} \ell(X \to y),
\end{align*}
with equality if $\ell(X \to y_z) = \max_{y \in \mathrm{supp}(P_Y)} \ell(X \to y)$.

\item 
\begin{align*}
    &\ell(X \to y \mid z) = \log \; \max_x \frac{P_{Y \mid X=x,Z=z}(y)}{P_{Y \mid Z=z}(y)}\\
    &= \log \; \max_x \frac{P_{Y \mid X = x}(y) P_{Y}(y)}{P_{Y}(y) P_{Y \mid Z=z}(y)}\\
    &= \log \; \max_x \frac{P_{Y \mid X = x}(y)}{P_{Y}(y)} + \log \frac{P_{Y}(y)}{P_{Y \mid Z=z}(y)}\\
    &= \ell(X \to y) - i(y;z).
\end{align*}

\item 
\begin{align*}
    &\ell(X \to y,z) = \log \; \max_{x} \frac{P_{YZ \mid X=x}(y,z)}{P_{YZ}(y,z)}\\
    &= \log \; \max_{x} \frac{P_{Y \mid X=x,Z=z}(y) P_{Z \mid X=x}(z)}{P_{Y \mid Z=z}(y) P_Z(z)}\\
    &\leq \log \; \max_x \frac{P_{Y \mid X=x,Z=z}(y)}{P_{Y \mid Z=z}(y)} + \log \; \max_x \frac{P_{Z \mid X=x}(z)}{P_Z(z)}\\
    &= \ell(X \to y \mid z) + \ell(X \to z),
\end{align*}
\end{enumerate}
with equality if and only if there exists $x^* \in \mathrm{supp}(P_X)$ maximizing both $i(x;y \mid z)$ and $i(x;z)$. 
$\hfill \blacksquare$

\section{Proof of Theorem~\ref{thm:composition}}
\label{sec:composition_lemma_proof}
\begin{enumerate}
    \item This result is an immediate consequence of the composition property given in Lemma~\ref{lemma:properties}. For all $y \in \mathrm{supp}(P_Y)$ and all $z \in \mathrm{supp}(P_Z)$ we have 
    \begin{align*}
        &\ell(X \to y,z) \leq \ell(X \to y) + \ell(X \to z \mid y)\\
        & \leq \max_{y \in \mathrm{supp}(P_Y)} \ell(X \to y) + \max_{\substack{y \in \mathrm{supp}(P_Y), \\ z \in \mathrm{supp}(P_Z)}} \ell(X \to z \mid y)\\
        &\leq \epsilon_1 + \epsilon_2.
    \end{align*}
    Therefore, $P_{YZ \mid X}$ satisfies $\epsilon_1 + \epsilon_2$-PML. 
    \item Since $\ell(X \to y,z) \leq \ell(X \to y) + \ell(X \to z \mid y)$ for all $(y,z) \in \mathrm{supp}(P_Y) \times \mathrm{supp}(P_Z)$, we can write 
    \begin{align*}
        &\mathbb P\Big[ \ell(X \to Y,Z) >\epsilon_1 + \epsilon_2 \Big] \\
        &\leq \mathbb P\Big[ \ell(X \to Y) + \ell(X \to Z \mid Y)  >\epsilon_1 + \epsilon_2 \Big]\\
        &= 1 - \mathbb P\Big[ \ell(X \to Y) + \ell(X \to Z \mid Y)  \leq\epsilon_1 + \epsilon_2 \Big].
    \end{align*}
    We define the following \say{good} events:
    \begin{gather*}
        \mathcal G \coloneqq \{(y,z) \in \mathrm{supp}(P_Y) \times \mathrm{supp}(P_Z) \colon \\
        \hspace{4em}\ell(X \to y) \leq \epsilon_1 \;\; \text{and} \; \; \ell(X \to z \mid y) \leq \epsilon_2 \},\\[0.5em]
        \mathcal G_Y \coloneqq \{y \in \mathrm{supp}(P_Y) \colon (y,z) \in \mathcal G, z \in \mathrm{supp}(P_Z)\},\\[0.5em]
        \mathcal G_Z(y) \coloneqq \{z \in \mathrm{supp}(P_Z) \colon (y,z) \in \mathcal G\}.
    \end{gather*}
    Our goal is to lower bound the probability of event $\mathcal G$. We can write 
    \begin{subequations}
    \begin{align}
        P_{YZ} (\mathcal G) &= \sum_{(y,z) \in \mathcal G} P_{Y}(y) P_{Z \mid Y=y}(z)\nonumber\\
        &= \sum_{y \in \mathcal G_Y} P_Y(y) \; P_{Z \mid Y=y}(\mathcal G_Z(y)) \nonumber\\
        &\geq (1 - \delta_2) \sum_{y \in \mathcal G_Y} P_Y(y) \label{subeq:comp_1}\\
        &\geq (1 - \delta_2) (1 - \delta_1),\label{subeq:comp_2}
    \end{align}
    \end{subequations}
    where 
    \begin{itemize}
        \item \eqref{subeq:comp_1} follows from the fact that for all $y \in \mathrm{supp}(P_Y)$, $P_{Z \mid X, Y=y}$ satisfies $(\epsilon_2, \delta_2)$-PML which implies that 
        \begin{align*}
            P_{Z \mid Y=y} (\mathcal G_Z(y)) &= \mathbb P_{Z \sim P_{Z \mid Y=y}} \Big[\ell(X \to Z \mid y)\leq \epsilon_2 \Big]\\
            &\geq 1- \delta_2,
        \end{align*}
        
        \item and \eqref{subeq:comp_2} follows from the fact that $P_{Y \mid X}$ satisfies $(\epsilon_1, \delta_1)$-PML, that is, 
        \begin{equation*}
            P_Y(\mathcal G_Y) = \mathbb P_{Y \sim P_Y} \Big[\ell(X \to Y) \leq \epsilon_1 \Big] \geq 1 - \delta_1.
        \end{equation*}
    \end{itemize}
    It follows that
    \begin{align*}
        &\mathbb P_{(Y,Z) \sim P_{YZ}} \Big[ \ell(X \to y) + \ell(X \to z \mid y)  \leq\epsilon_1 + \epsilon_2 \Big]\\
        &\geq P_{YZ} (\mathcal G)\\
        &\geq (1 - \delta_2) (1 - \delta_1),
    \end{align*}
    which yields 
    \begin{equation*}
        \mathbb P_{(Y,Z) \sim P_{YZ}} \Big[ \ell(X \to Y,Z) >\epsilon_1 + \epsilon_2 \Big] \leq \delta_1 + \delta_2 - \delta_1 \delta_2.
    \end{equation*}
    \item Define the event 
    \begin{align*}
        \mathcal A_Y = \{y \in \mathrm{supp}(P_Y) \colon \ell(X \to y) \leq \epsilon_1 \}.
    \end{align*}
    As $P_{Y \mid X}$ satisfies $(\epsilon_1, \delta_1)$-PML, we have 
    \begin{align*}
        1 - \delta_1 &\leq P_Y(\mathcal{A}_Y) = P_{YZ} (\mathcal A),
    \end{align*}
    where $\mathcal A \coloneqq \mathcal A_Y \times \mathrm{supp}(P_Z)$. Moreover, we define the event
    \begin{gather*}
        \mathcal B \coloneqq \{(y,z) \in \mathrm{supp}(P_Y) \times \mathrm{supp}(P_Z) \colon\\
        \hspace{14em}\ell(X \to z \mid y) \leq \epsilon_2 \}. 
    \end{gather*}
    By assumption, $P_{YZ} (\mathcal B) \geq 1 - \delta_2$. Therefore,
    \begin{align*}
        &\mathbb P_{(Y,Z) \sim P_{YZ}} \Big[ \ell(X \to Y,Z) \leq \epsilon_1 + \epsilon_2 \Big]\\
        &\geq \mathbb P_{(Y,Z) \sim P_{YZ}} \Big[ \ell(X \to Y) +  \ell(X \to Z \mid Y) \leq \epsilon_1 + \epsilon_2 \Big]\\
        &\geq P_{YZ} (\mathcal A \cap \mathcal B)\\
        &= 1 - P_{YZ} (\mathcal A^c \cup \mathcal B^c)\\
        &\geq 1 - \delta_1 - \delta_2.
    \end{align*}
    
    \item Let $\mathcal E \subseteq \mathrm{supp}(P_{YZ})$ be an event satisfying $P_{YZ}(\mathcal{E}) \geq \delta_1$ and $0 \leq \delta_2 \leq \min_{y \in \mathcal E_Y} P_{Z \mid Y=y}(\mathcal E_Z(y))$. Since $P_{Z \mid X,Y=y}$ satisfies $(\epsilon_2, \delta_2)$-EML for all $y \in \mathcal E_Y$, we have
    \begin{equation}
    \label{eq:comp_EML_special_event}
        \max_{x \in \mathrm{supp}(P_X)}\frac{P_{Z \mid Y=y, X=x}(\mathcal E_Z(y))}{P_{Z \mid Y=y}(\mathcal E_Z(y))} \leq \exp(\epsilon_2).
    \end{equation}
    Now, we write
    \begin{subequations}
    \begin{align}
        &\exp \big( \ell(X \to \mathcal E) \big) = \max_{x \in \mathrm{supp}(P_X)} \frac{P_{YZ \mid X=x} (\mathcal E)}{P_{YZ} (\mathcal E)}\nonumber\\
        &= \max_{x \in \mathrm{supp}(P_X)} \frac{\sum_{y \in \mathcal E_Y} \sum_{z \in \mathcal E_Z(y)} P_{YZ \mid X=x}(y,z)}{\sum_{y \in \mathcal E_Y} \sum_{z \in \mathcal E_Z(y)} P_{YZ}(y,z)}\nonumber\\[0.5em]
        &= \max_{x \in \mathrm{supp}(P_X)} \!\!\frac{\sum\limits_{y \in \mathcal E_Y} P_{Y\mid X=x}(y) \!\!\sum\limits_{z \in \mathcal E_Z(y)} P_{Z \mid Y=y, X=x}(z)}{\sum\limits_{y \in \mathcal E_Y} P_Y(y) \sum\limits_{z \in \mathcal E_Z(y)} P_{Z \mid Y=y}(z)}\nonumber\\[0.5em]
        &= \max_{x \in \mathrm{supp}(P_X)} \sum_{y \in \mathcal E_Y} \frac{ P_Y(y) P_{Z \mid Y=y}(\mathcal E_Z(y))}{\sum\limits_{y' \in \mathcal E_Y} P_Y(y') P_{Z \mid Y=y'}(\mathcal E_Z(y'))} \cdot \nonumber\\
        &\hspace{5em}\left(\frac{ P_{Y\mid X=x}(y)}{P_Y(y)} \right) \cdot \left(\frac{P_{Z \mid Y=y, X=x}(\mathcal E_Z(y))}{P_{Z \mid Y=y}(\mathcal E_Z(y))} \right)\nonumber\\
        &\leq e^{\epsilon_2} \cdot \max_{x \in \mathrm{supp}(P_X)} \!\sum_{y \in \mathcal E_Y} \!\!\frac{ P_Y(y) P_{Z \mid Y=y}(\mathcal E_Z(y))}{\sum\limits_{y' \in \mathcal E_Y} \!\!P_Y(y') P_{Z \mid Y=y'}(\mathcal E_Z(y'))} \cdot \nonumber\\
        &\hspace{13em}\left(\frac{ P_{Y\mid X=x}(y)}{P_Y(y)} \right) \label{subeq:comp_EML_1}\\
        &\leq e^{\epsilon_2} \cdot \max_{x \in \mathrm{supp}(P_X)} h_x(P_{Y \mid X=x}, \delta_1) \label{subeq:comp_EML_2}\\
        &\leq e^{\epsilon_2 + \epsilon_1},\label{subeq:comp_EML_3}
    \end{align}
    \end{subequations}
    where 
    \begin{itemize}
        \item \eqref{subeq:comp_EML_1} follows from inequality~\eqref{eq:comp_EML_special_event},
        \item the function $h_x$ in~\eqref{subeq:comp_EML_2} is defined in~\eqref{eq:optim_1},
        \item and \eqref{subeq:comp_EML_3} follows from the fact that $P_{Y \mid X}$ satisfies $(\epsilon_1, \delta_1)$-EML.
    \end{itemize}
    
    \item Let $\mathcal E \subseteq \mathrm{supp}(P_{YZ})$ be an event satisfying $P_{YZ}(\mathcal{E}) \geq \delta_1 + \delta_2$. We define the following \say{bad} sets
    \begin{gather*}
        \mathcal B_Y \coloneqq \{y \in \mathcal E_Y \colon P_Z(\mathcal E_Z(y)) < \delta_2 \},\\
        \mathcal B \coloneqq \{(y,z) \in \mathcal E \colon y \in \mathcal B_Y\},
    \end{gather*}
    and the \say{good} sets $\mathcal G_Y = \mathcal E_Y \setminus \mathcal B_Y$ and $\mathcal G = \mathcal E \setminus \mathcal B$.
    Note that 
    \begin{align*}
        P_{YZ}(\mathcal B) = \sum_{y \in \mathcal B_Y} P_Y(y) P_{Z \mid Y=y} (\mathcal E_Z(y)) < \delta_2, 
    \end{align*}
    which implies that $P_{YZ}(\mathcal G) = P_{YZ}(\mathcal E) - P_{YZ}(\mathcal B) > \delta_1$. Now, similarly to the previous part, we write
    \begin{subequations}
    \begin{align}
        &\exp \big( \ell(X \to \mathcal E) \big) \nonumber\\
        &= \max_{x \in \mathrm{supp}(P_X)} \frac{\sum_{(y,z) \in \mathcal E} P_{YZ \mid X=x}(y,z)}{ P_{YZ}(\mathcal E)} \nonumber\\[0.5em]
        &\leq\max_{x \in \mathrm{supp}(P_X)} \frac{\sum\limits_{(y,z) \in \mathcal B} P_{YZ}(y,z) \left( \frac{P_{YZ \mid X=x}(y,z)}{P_{YZ}(y,z)} \right)}{P_{YZ}(\mathcal E)} + \nonumber\\
        &\hspace{8em}\max_{x \in \mathrm{supp}(P_X)} \frac{\sum_{(y,z) \in \mathcal G} P_{YZ \mid X=x}(y,z)}{P_{YZ}(\mathcal E)} \nonumber\\[0.5em]
        &\leq \frac{\delta_2}{\delta_1 + \delta_2} e^{\epsilon_\mathrm{max}} + \max_{x \in \mathrm{supp}(P_X)} \frac{\sum\limits_{(y,z) \in \mathcal G} P_{YZ \mid X=x}(y,z)}{P_{YZ}(\mathcal G)} \label{subeq:comp_EML_sad_1}\\[0.5em]
        &\leq \frac{\delta_2}{\delta_1 + \delta_2} e^{\epsilon_\mathrm{max}} + e^{\epsilon_1 + \epsilon_2}, \label{subeq:comp_EML_sad_2}
    \end{align}
    where 
    \begin{itemize}
        \item \eqref{subeq:comp_EML_sad_1} follows from the fact that $P_{YZ}(\mathcal B) < \delta_2$, $P_{YZ}(\mathcal E) \geq \delta_1 + \delta_2$, and for all $(y,z) \in \mathrm{supp}(P_{YZ})$,
        \begin{equation*}
            \max_{x \in \mathrm{supp}(P_X)}\frac{P_{YZ \mid X=x}(y,z)}{P_{YZ}(y,z)} \leq \exp(\epsilon_\mathrm{max}),
        \end{equation*}
        
        \item and \eqref{subeq:comp_EML_sad_2} follows from the definition of the set $\mathcal G$ and the previous part.
    \end{itemize}
    \end{subequations}
\end{enumerate}
$\hfill \blacksquare$

\section*{Acknowledgement}
The authors would like to thank the anonymous reviewers for their careful reading of the paper and for making suggestions leading to a shorter proof for Proposition~\ref{prop:one_for_all}.

\bibliographystyle{IEEEtranN}
{\footnotesize
\bibliography{IEEEabrv, main}}

\balance

\begin{IEEEbiographynophoto}{Sara Saeidian}
(M’20) received the B.S. in Electrical Engineering from the University of Tehran in 2017 and the M.S. in Information and Network Engineering from KTH Royal Institute of Technology in 2019. She is currently pursuing the Ph.D. degree at the Division of Information Science and Engineering at KTH Royal Institute of Technology. Her research interests include privacy and information theory. 
\end{IEEEbiographynophoto}

\begin{IEEEbiographynophoto}{Giulia Cervia}
(S’15-M’19) received the degree in Mathematics from the University of Pisa, Italy, in 2014 and the Ph.D. degree in 2018 from Université Paris Seine (ETIS Laboratory – UMR 8051, CY Université, ENSEA, CNRS). From 2019 to 2020, she was a postdoctoral researcher in Information Theory at the Information Science and Engineering Division, School of Electrical Engineering and Computer Science, KTH – Royal Institute of Technology, Stockholm, Sweden. Since 2020, she is Associate Professor at  IMT Nord Europe, Institut Mines-Télécom, Univ. Lille, CERI SN - Centre for Digital Systems, Lille, France.

Her main research interests are in the areas of information theory, cooperative communications, error-control coding, privacy and physical layer security.
\end{IEEEbiographynophoto}

\begin{IEEEbiographynophoto}{Tobias J. Oechtering}
(S’01-M’08-SM’12) received his Dipl-Ing degree in Electrical Engineering and Information Technology in 2002 from RWTH Aachen University, Germany, his Dr-Ing degree in Electrical Engineering in 2007 from the Technische Universit\"at Berlin, Germany. In 2008 he joined KTH Royal Institute of Technology, Stockholm, Sweden and has been a Professor since 2018. In 2009, he received the ``F\"orderpreis 2009” from the Vodafone Foundation.

Dr. Oechtering is currently Senior Editor of IEEE Transactions on Information Forensic and Security since May 2020 and served previously as Associate Editor for the same journal since June 2016, and IEEE Communications Letters during 2012-2015. He has served on numerous technical program committees for IEEE sponsored conferences, and he was general co-chair for IEEE ITW 2019. His research interests include information theory, privacy and physical layer security, statistical learning and signal processing, communication, as well as networked control.
\end{IEEEbiographynophoto}

\begin{IEEEbiographynophoto}{Mikael Skoglund} 
(S'93-M'97-SM'04-F'19) received the Ph.D. degree in
1997 from Chalmers University of Technology, Sweden.  In 1997, he
joined the Royal Institute of Technology (KTH), Stockholm, Sweden,
where he was appointed to the Chair in Communication Theory in 2003.
At KTH, he heads the Division of Information Science and Engineering,
as well as the Department of Intelligent Systems.

Dr. Skoglund has worked on problems in source-channel coding, coding
and transmission for wireless communications, Shannon theory,
information-theoretic security, information theory for statistics and
learning, information and control, and signal processing. He has
authored and co-authored around 200 journal and more than 400
conference
papers.

Dr. Skoglund is a Fellow of the IEEE. During 2003--08 he was an
associate editor for the IEEE Transactions on Communications.  In the
interval 2008--12 he was on the editorial board for the IEEE
Transactions on Information Theory and starting in the Fall of 2021 he
he joined it once again. He has served on numerous technical program
committees for IEEE sponsored conferences, he was general co-chair for
IEEE ITW 2019, and in 2022 he is one of three TPC co-chairs for IEEE
ISIT 2022. Starting Jan. 1st 2023 he is an elected member of the IEEE
Information Theory Society Board of Governors.
\end{IEEEbiographynophoto}

\end{document}